\documentclass[3p,onecolumn]{elsarticle}
\pdfoutput=1
\usepackage{graphicx}
\usepackage{latexsym}
\usepackage{amsfonts}
\usepackage{amsmath,amssymb}
\usepackage{slashed}
\usepackage{feynmp}
\usepackage{hyperref}
\usepackage{url}
\usepackage{color}
\usepackage{cancel}
\usepackage{multirow,array}
\usepackage{float}
\usepackage{subfigure}
\usepackage{bm}
\usepackage{comment}
\usepackage{multirow}
\usepackage{natbib}

\begin{document}

\newcommand{\mwimp}{$m_\chi$}
\newcommand{\kcut}{$k_{\mathrm{cut}}$}
\newcommand{\mhalo}{$M_{\text{halo}}$}
\newcommand{\msun}{$M_{\odot}$}
\newcommand{\mvir}{$M_{\text{vir}}$}

\title{Gravitational probes of dark matter physics}

\author[mrb]{Matthew R.~Buckley}
\ead{mbuckley@physics.rutgers.edu}

\author[apphys,apastro]{Annika H.~G.~Peter\corref{cor1}}
\ead{apeter@physics.osu.edu}

\address[mrb]{\mbox{Department of Physics and Astronomy, Rutgers University, Piscataway, NJ 08854, USA}}
\address[apphys]{\mbox{CCAPP and Department of Physics, The Ohio State University, 191 W. Woodruff Ave., Columbus, OH 43210, USA}}
\address[apastro]{\mbox{Department of Astronomy, The Ohio State University, 140 W. 18th Ave., Columbus, OH 43210, USA}}
\cortext[cor1]{Principal corresponding author}

\begin{abstract}
The nature of dark matter is one of the most pressing questions in physics. Yet all our present knowledge of the dark sector to date comes from its gravitational interactions with astrophysical systems.  Moreover, astronomical results still have immense potential to constrain the particle properties of dark matter in the near future. We introduce a simple 2D parameter space which classifies models in terms of a particle physics interaction strength and a characteristic astrophysical scale on which new physics appears, in order to facilitate communication between the fields of particle physics and astronomy. We survey the known astrophysical anomalies that are suggestive of non-trivial dark matter particle physics, and present a theoretical and observational program for future astrophysical measurements that will shed light on the nature of dark matter. 

\end{abstract}


\maketitle

\section{Introduction}

For particle physicists, dark matter is clear evidence of the existence of new physics beyond the Standard Model. Significant theoretical effort has been applied to construct viable models of particle dark matter, and the resulting model-space is enormous. While many of these models contain particles which are ``minimal'' dark matter candidates in the cosmological sense (i.e., cold, functionally collisionless, non-baryonic matter), many others contain dark matter candidates that deviate at some level from the cold dark matter (CDM) paradigm \cite{Zwicky:1933gu,blumenthal1982,frenk1983}. Well known examples include warm dark matter (WDM) \cite{Hogan:2000bv,Dalcanton:2000hn} and self-interacting dark matter (SIDM) \cite{Spergel:1999mh}. Given the variety of models available, experimental input is necessary to narrow down the possibilities.  While many experiments have searched for signals of particle dark matter using a variety of traditional particle physics techniques \cite{Bertone:2004pz,Feng:2010gw,Patrignani:2016xqp} (e.g., collider production \cite{Bai:2010hh,Abercrombie:2015wmb,Penning:2017tmb}, direct detection \cite{Goodman:1984dc,Wasserman:1986hh,vanBibber:1988ge,Graham:2015ouw,Baudis:2016qwx}, and indirect detection \cite{Stecker:1978du,Silk:1985ax,Slatyer:2017sev}), so far all results have been negative.

Consequentially, the primary source of positive statements about the microphysical properties of dark matter continues to be the astrophysical and cosmological study of dark matter's gravitational interactions. This is perhaps not surprising. After all, it was through gravity that dark matter was discovered \cite{Zwicky:1933gu,1939LicOB..19...41B,Rubin:1980zd}, and its status as a non-relativistic (``cold,'' or at least no more than ``warm'') component of the Universe at the time of structure-formation verified \cite{blumenthal1984,davis1985,bahcall1999}. As astronomical surveys and galaxy simulations continue to improve, so too will our ability to explore the distribution of dark matter in the Universe and potentially discover deviations from the CDM paradigm, as predicted by many of the theoretical dark matter models. In many cases these astrophysics techniques are sensitive to dark matter properties orthogonal to those measured by particle physics experiments. Specifically, traditional particle physics techniques probe the microphysical interactions between dark matter and the Standard Model.  However, the distribution of dark matter in the Universe may be governed by additional interactions within a ``dark sector'' that are secluded from the Standard Model, in addition to interactions between dark matter and the Standard Model. Such interactions would not appear in experiments probing the Standard Model-dark matter scattering, but would appear in astrophysical measurements by altering the spatial distribution of dark matter in the Universe with respect to the CDM prediction.

The microphysical properties of dark matter --- particle mass, number of species, interaction strength of each particle with Standard Model and dark particles --- can imprint themselves on the macroscopic distribution of dark matter in the Universe.  To illustrate the power of astrophysics to reveal the microphysical properties of dark matter, it is useful to consider the decidedly non-minimal Standard Model. For example, because the Standard Model includes charged particles and photons, gas can cool.  This enables the formation of stars and galaxies. One might imagine that hypothetical researchers in the dark sector would be able to identify at least some of the major components and non-gravitational forces of the Standard Model based on gravitational evidence of the clustering of Standard Model particles. We explore this thought experiment in more detail in Section~\ref{sec:darkscience} of this work.  As this conceptual exercise will demonstrate, astronomical observations constitute \emph{measurements} of dark matter microphysics. 

There is immense potential for discovery in the intersection of particle and astrophysical measurements of dark matter. Positive signals of deviations from the gravitational predictions of CDM would give theoretical physicists much-needed experimental guidance about parameters that are not easily measured in particle physics experiments. If, on the other hand, all astrophysical studies of dark matter were to find correspondence with the CDM predictions, the improved knowledge of dark matter distributions would be extremely useful for reducing major sources of theoretical uncertainties in the particle physics experiments \cite{Buckley:2015doa,Caputo:2016ryl,Sloane:2016kyi,green2017,benito2017}. Likewise, results from particle experiments can either restrict the possible model space relevant to novel astrophysical signals, or suggest specific deviations from the CDM paradigm which can be confirmed by later surveys.

Though the connection between astrophysical measurements and particle physics models of dark matter is recognized for particular models (notably, sterile neutrinos \cite{Dodelson:1993je,Shi:1998km,Dolgov:2000ew,Abazajian:2001nj,Abazajian:2001vt}), the full potential is perhaps not appreciated in the broader context. This is in part due to the separate languages used by particle physicists and astrophysicists to describe dark matter, and in part due to the fact that astrophysical and particle-physics measurements of dark matter often depend on very different subsets of the parameter space of dark matter microphysics.  Finally, the ``natural" set of metrics used to constrain or define dark matter depend strongly on the type of experiment or observation used to find dark matter.

\bigskip

We have two primary goals for this paper.  First, we want to present a compact parameter space to capture the phenomenology of dark matter models relevant to both particle physicists and astrophysicists. To that end, we propose a simple two-dimensional parameter space, with one axis closely tied to particle dark matter searches and the other closely tied to astronomical searches, to orient dark-matter hunters in both fields.  We show where models of current interest lie in this space, and where potential discrepancies between observation and CDM predictions are located.  Second, we advocate a long-term plan for sharpening astrophysical constraints on dark-matter microphysics. This project will require close collaboration between physicists and astronomers (many of whom may not have dark matter as a primary area of research), which motivates our simple parameter space which translates between the two fields.

The two-dimensional description is a simple mapping which can be applied to a wide range of models and makes clear the sensitivity of both particle- and astrophysical-probes to the dark sector. By considering where specific models fall in this parameter space, particle physicists may identify new constraints or observables from astrophysics, while astrophysicists may find applications for their results of which they were previously unaware.

The two parameters we adopt are: 
\begin{enumerate}
\item The characteristic interaction strength between dark matter and the Standard Model, $\Lambda^{-1}$; and 
\item The characteristic largest dark matter halo mass where significant deviations from CDM predictions should arise, \mhalo.  The halo is the fundamental unit of cosmological dark matter on non-linear scales, a self-gravitating collection of dark-matter particles in equilibrium formed when a perturbation in the homogeneous universe detaches from the Hubble flow, and the structure within which galaxies live.
\end{enumerate}
The former parameter (defined to have dimensions of inverse energy) gives an estimate of the ability of particle physics experiments to discover dark matter through interactions with baryonic (used in the astronomical sense: all Standard Model particles but neutrinos and photons) matter. The latter gives an estimate of the type of astrophysical system whose dynamics must be understood in order to discover the presence of unique dark matter physics. 

We emphasize that these two parameters are not intended to encompass the full behavior from a particular model of dark matter. Given the breadth of theoretical work on this topic, many well-motivated models of dark matter can have unique phenomenology that would be difficult to capture in any low-dimensional set of parameters. However, flattening the model space down to these two parameters allows for direct comparison of a variety of models, which is particularly useful to scientists in one community (particle physics or astrophysics) looking to make connections with the results in the other. 

As part of this exercise, we highlight current challenges and new opportunities in dark matter astrophysics.  We show what the current theory systematics are in mapping between dark matter microphysics and \mhalo, and where the current observational constraints on deviations from CDM lie.  Importantly, we outline a systematic plan for tightening theory systematics and improving observational constraints. Some ideas have been discussed in the literature, but have not been unified into a comprehensive approach. We also identify a range of $M_{\rm halo} \lesssim 10^5\,M_\odot$ which is a critical test of the CDM paradigm, but is not expected to be probed by any present or near-future set of observations. This calls for new ideas, and the community must rise to meet this challenge. 

We begin our paper in Section~\ref{sec:darkscience} with our thought experiment, considering what a dark-matter scientist could discover of the particle physics in the Standard Model, using only evidence from our gravitational imprint. Having demonstrated the interplay between particle physics and astronomy, in Section~\ref{sec:fom}, we define our two figures of merit, and demonstrate their applicability to various dark matter models. In Section~\ref{sec:observations}, we discuss a series of observations that suggest that some non-trivial physics may be at play in the dark sector. These observations are at this point just hints, and in Section~\ref{sec:future}, we discuss near-term astrophysical probes and theoretical improvements that can increase the existing sensitivity to the astrophysical figure of merit, as well as more accurately defining \mhalo\, given known astrophysical systematics. This program will conclusively answer the question whether these hints are of dark matter origin, as well as bringing more dark matter models into experimental reach. We conclude with a discussion of the prospects for the joint astrophysical-particle physics investigation of dark matter model-space, and the important lessons about the physics of the dark sector that can be learned even in the absence of positive results. 

\section{Dark Matter Discovery of the Standard Model \label{sec:darkscience}}

Before developing a set of parameters to characterize the physics of dark matter models, it is interesting and informative to consider a thought experiment which makes clear the power of astrophysical studies to uncover novel particle physics in the dark sector. While we do not know if dark matter has non-trivial internal particle physics associated with it, our own Standard Model is decidedly non-trivial, with multiple particles and interactions. If the Standard Model sector were as invisible to a hypothetical dark matter observer as the dark sector is to us, one might assume that none of this complexity could be inferred. However, this is not the case, as we will show. The goal of the following thought experiment is two-fold.  First, and most importantly, we illustrate the power of astrophysics to unveil some major aspects of dark-matter physics.  Second, we show which aspects of a secluded model will be most challenging to discover with astrophysics.

We imagine a hypothetical ``dark-matter scientist,'' capable of perceiving the dark matter in the Universe but not the baryonic matter.\footnote{Following the usual nomenclature we will use ``dark'' to denote the sector of physics invisible to us but visible to our hypothetical dark-matter scientist, while ``visible'' corresponds to the baryonic sector which is detectable to us but invisible to the dark-matter scientist.} 
For the purposes of the thought experiment, this implies some massless or very low mass photon equivalent, i.e., a ``dark photon'' \cite{Holdom:1985ag,Ackerman:mha,Bjorken:2009mm}, which has nontrivial constraints from data \cite{CyrRacine:2012fz,Cyr-Racine:2013fsa,Boddy:2016bbu}. To avoid the issues of extra light degrees of freedom in the Cosmic Microwave Background (CMB), we assume the dark sector decoupled from the visible sector early on, and was colder than the Standard Model bath at the time of recombination \cite{Feng:2008mu,Ackerman:mha}. We assume that this dark photon provides the dark-matter scientist with a ``dark CMB'' which allows them precision measurements of the Universe's energy budget. Imagine that, just as we visible-sector scientists discovered dark matter through a combination of astronomical and cosmological measurements (through, e.g., rotation curves \cite{Rubin:1980zd} or the CMB), the dark-matter scientist discovered hints of the visible sector in the sky.

Having divined its existence, what else can our dark-matter scientist learn of this mysterious matter? Particle physics searches would prove difficult.  That is true even if the dark-matter scientist lives in a dark matter halo that also contains a significant number of baryons, rather than in a baryon-poor filament or void.  This is because the visible galaxy, composed of baryons, inhabits the very center of the halo --- the radius enclosing half the mass of the baryonic component is at best a few percent of the virial radius for dark matter \cite{kravtsov2013,huang2017}. The dark-matter scientist would have to be deep inside the potential well of their halo in order to have a chance at detecting baryons using the dark matter equivalent of direct detection experiments \cite{Goodman:1984dc,Wasserman:1986hh,Baudis:2016qwx}, assuming a weak-scale coupling between the two sectors.  Indirect detection of dark sector byproducts from baryonic self-annihilation would also be ineffective, as protons are not their own antiparticle, and the Universe is overwhelmingly matter-dominated \cite{steigman1976}. Thus, there would be little annihilation at the center of any halo.  A dark-matter scientist could probe  interactions between the sectors at or below the strength of the weak nuclear force only in the dark-matter equivalent of a particle accelerator (where they would face the equivalent of all the experimental difficulties inherent in searching for dark matter at the LHC). 

Yet, from astrophysics and cosmology, our dark-matter scientist may infer much about the structure of the Standard Model. With the dark CMB and measurements of the clustering of dark-matter halos, they would discover that the dark matter could not account for all of the non-relativistic energy, and that some missing matter exists, with $\Omega_b \sim 0.05$.  Moreover, this missing matter must have many light degrees of freedom, in order to explain the redshift of matter-radiation equality.  In fact, the light degrees of freedom from the visible sector would overwhelmingly dominate cosmology over those from the dark sector. 

Depending on the redshift of dark decoupling relative to matter-radiation equality, the dark-matter scientist could obtain precision measurements of the number of light degrees of freedom in the Standard Model (including evidence for the neutrinos, though this would be difficult to disentangle from a non-abelian unbroken gauge sector), as well as the temperature of the visible sector.

From the two-point correlation function of dark matter halos, the dark-matter scientist would see the imprint of baryon acoustic oscillations (BAO) \cite{Anderson:2012sa}, which would indicate that, whatever this mysterious matter was, it must be strongly self-coupled. If the scientist then turned their attention to the inner structure of dark matter halos, they would see gravitational evidence for a surprising deviation from the spherical density distribution they are accustomed to finding at large radii, replaced by some kind of disky structure. 

The combination of evidence of a baryonic disk, matter-radiation equality, and BAO would tell our dark-matter scientist that the visible matter must not just be strongly self-coupled, but that it must have some light force carrier capable of radiating away the excess kinetic energy after the virialization of the halo. A reasonable conclusion would be that the light degrees of freedom, inferred from the scale factor of matter-radiation equality, are related to these light force carriers.  This would be suggestive of some unbroken (or nearly unbroken) gauge group (assuming that dark-matter scientists are at least as familiar with quantum field theory as we are). A non-abelian gauge group for this light force carrier would be a possibility in addition to the actual solution of the electromagnetic $U(1)$, but in either case the dark-matter scientist would recognize the need for the visible matter to be overall neutral under this force. This would suggest at least two particles in the baryonic sector with opposite charges: $H$ (heavy) and $L$ (light).\footnote{Alternatively, $N$ particles charged under the fundamental of some non-abelian $SU(N)$, as with the three quarks of $SU(3)_C$.} 

At this point, the dark-matter scientist would not be able to tell whether $H$ and $L$ are two different particles with different masses, or just antiparticles of each other (in which case, of course, $m_H=m_L$). However, as we will describe below, the observation that the baryonic matter has undergone gravitational collapse requires an efficient mechanism to radiate away energy, which is suggestive of at least one light charged particle. As we will see, if the dark-matter scientists consider the possibility that the kinetic energy of the baryons during gravitational collapse is much larger than the binding energy (so that free-free emission is possible), this would lead to the conclusion that $m_H \gg m_L$. Thus, while other possibilities might lead to self-consistent astrophysics, it will be reasonable for them to consider this limit.

Now the dark-matter scientist would want to determine the coupling between the $H$ and $L$ --- what we know as the electromagnetic fine structure constant $\alpha$. There are two options: either the $H$ and $L$ form bound states in the baryonic halo, or they are ionized. The former possibility requires binding energies much higher than the typical kinetic energy of the particles in the disk ($B \gg k T_{\mathrm{vir}}$, where $T_\text{vir}$ is the temperature of particles in a halo of mass $M_\text{vir}$ --- see Section~\ref{sec:primer} for precise definitions). Cooling would proceed through the collisional excitation \cite{Thoul:1994ir}. At minimum, cooling could be facilitated by hyperfine de-excitation, which would only be possible if the virial temperature of the halo were sufficiently larger than the hyperfine transition energy ($ k T_{\mathrm{vir}} \sim G m_H M_{\rm vir}/R_{\mathrm{vir}} \gg E_{hf}$), or
\begin{eqnarray}\label{eq:bound_go_condition}
\alpha_{\rm bound~states} < 10^{-1/2} \left(\frac{M_{\mathrm{vir}}}{10^{12}\,M_\odot}\right)^{1/6} \left(\frac{m_H/m_L}{m_p/m_e}\right)^{1/2}, 
\end{eqnarray}
where \mvir\ is the virial mass of a halo \cite{Boddy:2016bbu}.  This is the requirement on $\alpha$ for the existence of bound states and the possibility that cooling may proceed through hyperfine emission.  
In general, though, the resulting radiation transport and cooling functions are complicated \cite{Katz:1995up,Ferland:1998id}, especially if dark nuclear physics exists.  It also is more important for low-mass halos rather than high-mass halos, and generally the range of $\alpha$ and $m_H/m_L$ that allow bound states to cool efficiently is small.  For example, if hyperfine transitions are the dominant cooling mode, we require
\begin{eqnarray}\label{eq:bound_limit}
\alpha_{\rm HF~cooling} \gtrsim 0.1 \left(\frac{m_H}{m_p}\right)^{3/2} \left(\frac{M_{\rm vir}}{10^{12}\,M_\odot}\right)^{1/6}
\end{eqnarray}
if $m_H/m_L \gg 1$. This is the requirement on $\alpha$, that it lie between the go-criterion limit of Eq. (\ref{eq:bound_go_condition}) and the requirement that the baryonic matter cool in a Hubble time (Eq. (\ref{eq:bound_limit})), if the matter exists as bound states.  This scenario is an attractive possibility for low-mass dark-matter halos if one prefers $\alpha$ to remain in the perturbative limit, and is one that is realized in the visible sector on dwarf scales before reionization \cite{Thoul:1994ir}.

Turning to the possibility that the binding energy of the $H-L$ system is much lower than the average kinetic energy of particles in the halo, which is possible if
\begin{eqnarray}\label{eq:unbound_go_criterion}
	\alpha_{\rm ionized~states} \lesssim 10^{-2} \left( \frac{M_{\rm vir}}{10^{12}\,M_\odot}\right)^{1/3} \left( \frac{m_H/m_L}{m_p/m_e} \right)^{1/2}
\end{eqnarray}
then cooling would proceed through thermal bremsstrahlung. This limit on $\alpha$ is what is required for the gas to be ionized for a halo of mass $M_{\mathrm{vir}}$.  An argument of this sort was used in Ref.~\cite{Silk:1977wz} to argue for the typical mass scale of galaxies.  In the case that $m_H \gg m_L$, then the dark-matter scientist would determine the volumetric energy loss rate is \cite{radiativeproc}
\begin{equation}
\frac{d^2E}{dV dt} \approx \left(\frac{2\pi k_B T_{\rm vir}}{3m_L} \right)^{1/2} \frac{2^4 \hbar^2 \alpha^3}{3m_L}n_L n_H,
\end{equation}
where $n_L$ and $n_H$ are the number densities of $L$ and $H$.  Because Coulomb interactions keep the $L$ and $H$ species in kinetic equilibrium, $n_L = n_H$ under the assumption that the galaxy is neutral under whatever force is allowing the cooling of the $H-L$ mixture.  The total energy radiated away over a Hubble time is approximately equal to the kinetic energy in the baryonic sector at the initial virialization of the halo:
\begin{equation}
E_{\rm tot} \sim \frac{fM_{\rm vir}}{m_H}k_B T_{\rm vir},
\end{equation}
where $f = \Omega_b / \Omega_{\rm DM}$. Setting the volumetric loss rate times the lifespan of the Galaxy over the entire virial radius of the baryonic disk equal to the total energy lost, and using the virial density $\bar{\rho} = \Delta_c \rho_c$ (where $\rho_c$ is the critical density of the Universe and $\Delta_c \approx 200$ \cite{blumenthal1984}; Section~\ref{sec:primer}) to determine $n_H = n_L = f\bar{\rho}/m_H$, the dark physicist would finally come to the conclusion that
\begin{equation}\label{eq:unbound_limit}
\alpha_{\rm brems} \gtrsim 10^{-7/3} \left( \frac{M_{\rm vir}}{10^{12}\,M_\odot}\right)^{1/9} \left(\frac{m_H/m_L}{m_p/m_e} \right)^{1/2} \left(\frac{m_L}{m_e}\right).
\end{equation}
For visible matter to exist as unbound states and cool in halos in a Hubble time, $\alpha$ must lie between the limits of Eqs. (\ref{eq:unbound_go_criterion}) and (\ref{eq:unbound_limit}).

Interestingly, applied to Milky Way-mass halos and using the actual masses of the Standard Model particles, this minimum bound on $\alpha$ from the thermal bremsstrahlung estimate is close to the actual value of $\alpha \sim 1/137$.   Note that increasing the possible mass of the ``unknown'' baryonic particles would eventually lead the dark physicist to a non-perturbative theory. 

From this rough calculation, the dark-matter scientist would also be able to conclude that visible matter is not composed of a thermal bath of particle-antiparticle, as the implied $\alpha$ leads to a scattering rate for a bath of matter-antimatter partners which is much to large to allow a significant relic abundance to survive into this late era of the Universe's life \cite{2011PhRvD..84d3510B}. Given that $H$ and $L$ are not antiparticles, the possibility that $m_H$ and $m_L$ are parametrically different would be an attractive avenue of research --- resulting in self-consistent results for the dark-matter scientist.

Having now concluded that baryons are not a thermal relic, the dark-matter scientist could infer that there must be some form of baryogenesis at work in the early Universe, and therefore that the baryonic sector contains CP violation in some form. They would be however hard-pressed to come up with any theory about the origin of that CP violation --- a situation that we in the visible sector, with far more information about our own particle content, can certainly sympathize with.

Let us summarize what the dark-matter scientist would find in our thought experiment, and what open questions would still exist.  Via traditional cosmological and astronomical measurements, the dark-matter scientist would find that a new sector of matter must exist, and could quantify what fraction of the Universe's energy budget it must contain.  They could determine that this sector must have a number of light degrees of freedom, and that the sector must have been tightly coupled early in the Universe's history but not today.  Assuming that at least a few of these light degrees of freedom might be the force carriers that allow this sector to be tightly coupled, they might assume that there is something like a $U(1)$ symmetry (electromagnetism, or akin to it) in the visible sector.  Their suspicions would be strengthened by observing that halos often have dense pancake-like features at the center, indicative of a light force carrier being radiated away through collisional cooling.  They would be able to constrain $\alpha$ and two particle masses from cooling rate considerations.  Between this and calculations of relic densities in the early Universe relic densities as well as  baryogenesis, the dark-matter scientist would conclude that the two charged particles would likely not be anti-particles, but instead two species with opposite charges.  

This set of conclusions is non-trivial.  However, it is also useful to consider the parts of the Standard Model which might be harder for the dark-matter scientist to infer.  Would the dark-matter scientist be able to discover the three generations of lepton we know exist in the visible sector, and not just the electron?  Would the dark-matter scientist discover nuclear physics, revealing that our heavy particle, the proton, is a composite particle?  Would they discover three generations of quarks?  Would the extra light degrees of freedom be identified with neutrinos, or as additional $U(1)$ forces?  For some of these questions, the answer is a qualified yes.  For example, stars in the visible sector might act as microlenses for dark matter astronomers, as they do for visible-sector applications \cite{Wambsganss:2006nj}.  The abundance and stability of such objects may hint at nuclear fusion-type processes.  But for some of these questions --- e.g., three generations of quarks and leptons --- the answer might be no.

This thought experiment is not an attempt to argue that a dark-matter scientist could immediately determine the unique properties of the baryons.  As we demonstrated, there are several possible branches that our hypothetical researcher may wander down; it is not clear that all other options would not also lead to self-consistent results. Indeed, one might expect the same level of vigorous debate as to the nature of this mysterious extra component of the Universe as found among our baryonic theoretical physicists.  However, the dark-matter scientist would be able to map out some of the most important features of the Standard Model, like electromagnetism, which were also the first Standard Model features that were described by modern theory by visible-sector scientists. 

\section{Metrics for Dark Matter Models \label{sec:fom}}

\begin{figure*}[t]
\begin{center}
\includegraphics[width=0.75\textwidth]{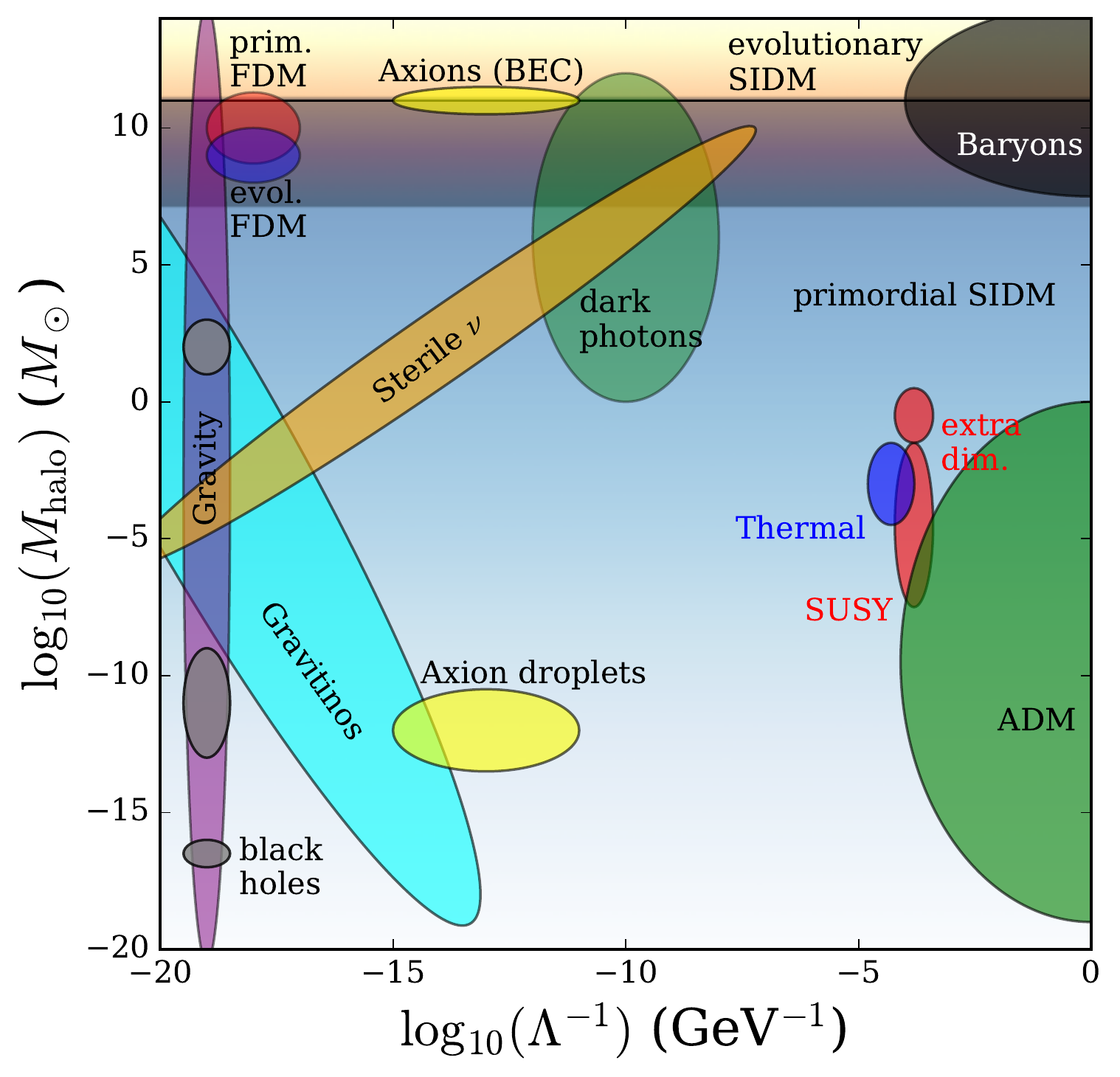}
\end{center}
\caption{Estimates for the range of particle physics and astrophysics figures of merit ($\Lambda^{-1}$ and $M_{\rm halo}$) for a variety of dark matter models. The range of $M_{\rm halo}$ covered by ``evolutionary'' and ``primordial'' self-interacting dark matter models (SIDM) are overlapping. The former covers the range $10^{6}-10^{15}\,M_\odot$, and the latter the range below $10^{11}\,M_\odot$. See text for further details.  \label{fig:fom}}
\end{figure*}

As our thought experiment demonstrates, much may be learned about the complicated Standard Model particle physics through measurements of the gravitational imprint of baryons if we were dark-matter scientists surveying the Universe. We can uncover non-trivial dark matter physics in the same manner. A comprehensive characterization of dark matter microphysics requires a combination of approaches: laboratory-based particle physics searches for interactions with the Standard Model, and the astronomical searches for interactions within a dark sector and also (as we will see) with the Standard Model.  To organize these searches, we need a compact space in which to classify models in terms of their observability in the laboratory and in the sky.  Our goal with this section is to motivate a specific choice for this space, and to show how particle dark matter models inhabit it.  The space is designed to be well-matched to the ways particle physicists and astronomers think about dark matter, making the mapping between the particle and astronomical spaces transparent and straight-forward, and compact but informative enough so that one might define ``figures of merit'' to quantify how well future experiments and observations will constrain dark matter models.

We classify dark matter models by their interaction strength with the Standard Model, $\Lambda^{-1}$, and the cosmological scale at which we expect to see a deviation from the Cold Dark Matter (CDM) paradigm, $M_{\rm halo}$. The former defines the sensitivity of particle physics detectors and the latter defines the largest size of the systems that must be understood in order to discover model-specific dark matter structures. We consider each axis of this parameter space in turn, and classify some well-known dark matter models by where they fall in the resulting two-dimensional parameter space. We summarize our estimates for these models in Figure~\ref{fig:fom}, with details given in the text.  Because one of our goals is to enable better communication between particle physicists and astronomers, we take a more pedagogical approach in defining parameters than is standard for either field.

For a more thorough review of the particle physics of many (though not all) dark matter models, see~Refs.~\cite{Herrero:1998eq,Bertone:2004pz,ProfumoBook}.

\subsection{Standard Model Interaction Strength}\label{sec:standardmodel}

\begin{figure}[t]
\begin{center}
\includegraphics[width=0.75\textwidth]{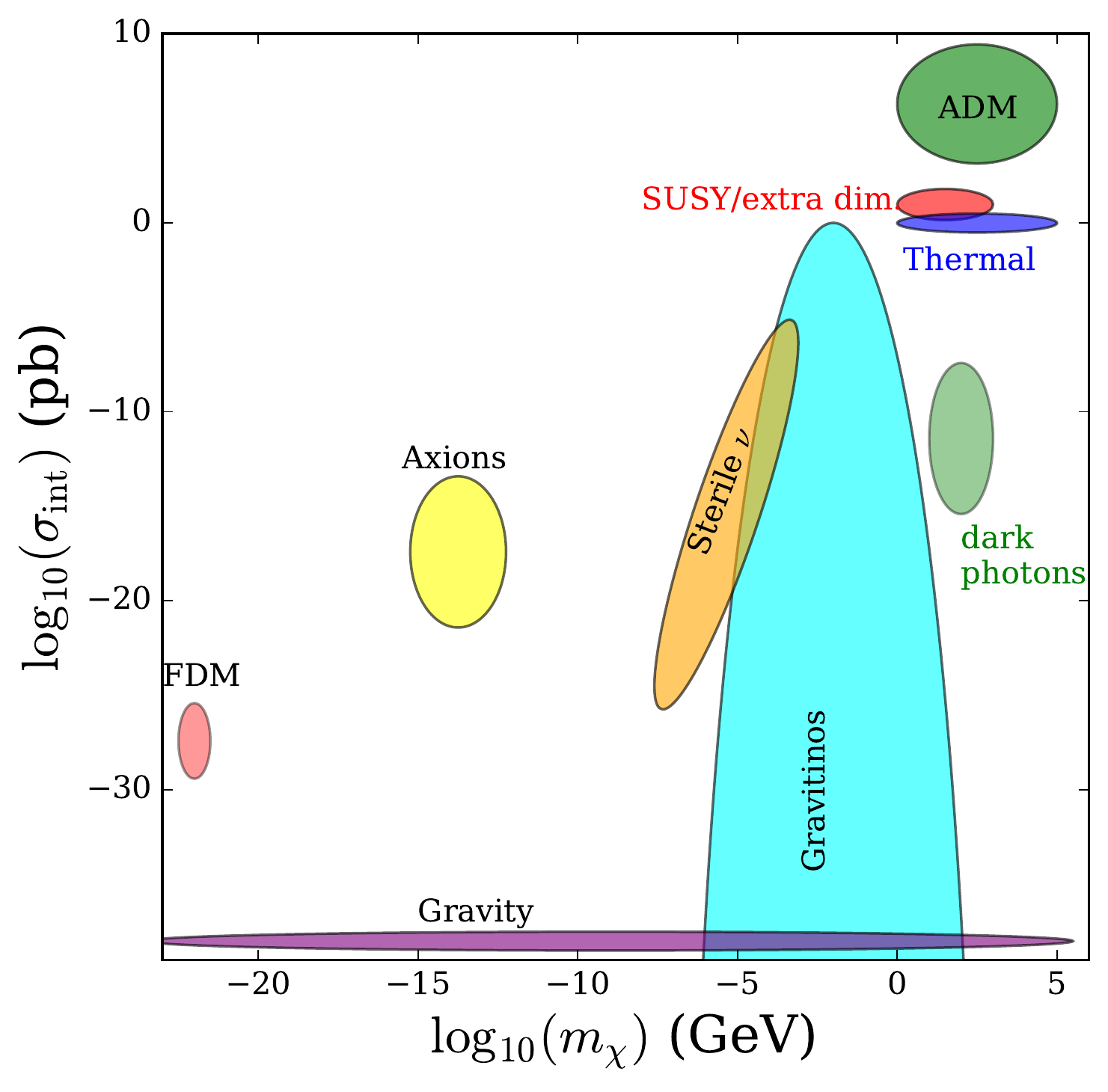}
\end{center}
\caption{\label{fig:dm_pspace}Representative example of the types of plots used to classify dark matter candidates in a parameter space useful to particle physicists. Particle masses are given in energy units (GeV), instead of mass units such as grams or GeV/c$^2$, because $c=1$ by convention in particle physics (i.e., mass $=$ energy).  The cross section $\sigma_\mathrm{int}$ is defined as the interaction cross section between dark matter and a Standard Model particle, where the relevant particle is different for different dark matter models, depending on which interaction is tested for a specific model in a specific experiment.  A pb is a picobarn, or $10^{-36}\hbox{ cm}^2$.}
\end{figure}

Defining a single parameter describing the interaction of dark matter with the Standard Model which can be applied to the wide variety of dark matter models is not straightforward, and any approach is bound to obscure some information about the models. A common approach in particle physics circles is to classify dark matter models in a two-dimensional parameter space, but even defining this simplified parameter space capturing the phenomenology of {\em particle physics} is challenging. We plot one commonly-used set of parameters in Figure~\ref{fig:dm_pspace}.  Typically, one dimension of the particle physics space is the particle mass, which is a well-motivated and straightforward choice.  From an experimental perspective, the particle mass conveys information about the energy scale (kinetic energy in the case of WIMP direct detection, total available energy in the case of creation in the lab or indirect detection in the sky) required to find  dark matter particle.  The second parameter is typically chosen to be an interaction cross section (as in Figure~\ref{fig:dm_pspace}) between dark matter and whatever Standard Model particle is most relevant for a particular dark matter model. This cross section is not uniquely defined between models, because it depends on the type of Standard Model particle involved in the interaction, the energy scale, and quantum numbers. Despite these limitations and ambiguities, this axis provides a quantity that captures some sense of how easy or difficult it would be to find a particular type of dark matter. As demonstrated in Figure~\ref{fig:dm_pspace}, the volume of the space is enormous, and it is not obvious how to compare the distinguishing power of different types of experiment fairly in this space. Moreover, this particle-physics parametrization fails to capture any of the experimental constraints on dark matter coming from astrophysics and cosmology.

Bearing these issues with the traditional two-dimensional space in mind, our goal is to quantify the interaction of dark matter with the Standard Model with a single parameter with a compact volume that retains distinguishing power among dark matter models.  As a case study, let us consider two of the best-known dark matter models: weakly interacting dark matter (WIMPs) \cite{Steigman:1984ac}, and axion dark matter \cite{Peccei:1977hh}: 

WIMP dark matter --- plotted in Figure~\ref{fig:dm_pspace} as ``supersymmetry/extra dimensions'' --- is relatively heavy (on the scale of the electroweak breaking scale, 246~GeV), and generally has interaction cross sections with Standard Model of approximately the same scale as neutrinos. The high mass of WIMP dark matter allows for direct and indirect detection searches with signals well-separated in energy from backgrounds, while the weak-scale interactions allow for the possibility of dark matter production at the Large Hadron Collider (LHC; see Ref.~\cite{Lisanti:2016jxe} for a review). However, for any given implementation of WIMP dark matter, including even the constrained version found in the Minimal Supersymmetric Standard Model (MSSM), the relevant cross section for these three detection techniques can vary significantly, depending on the details of the theory.

Axion models, in contrast, have phenomenology largely driven by a single parameter: the mass scale $f_a$. This single scale sets both the mass of the axion and the interaction with the Standard Model, both being proportional to inverse powers of $f_a$ (though specific implementations of axion models have different ${\cal O}(1)$ numbers which modify these relations). Viable models of axion dark matter require very high values for $f_a \gtrsim 10^{10}$~GeV, which in turn implies very small dark matter mass $\sim \mu$eV (see Ref.~\cite{Kim:2008hd} for a review). 

To describe models of dark matter as different as WIMPs and axions in a single variable, we define a particle physics parameter combining the dimensionless coupling constant ($\lambda$) that controls the strength of the interaction between the visible and dark particles with the relevant mass (energy) scale $M$ for the interaction. This scale $M$ is typically the mass scale of the particle(s) mediating the interaction (like the $Z$ boson mediating electron-neutrino elastic scattering in the Standard Model), rather than the mass of the dark matter itself. Larger interaction rates are the result of larger $\lambda$ or smaller $M$, all else being equal. Therefore, we define our figure of merit as
\begin{equation}
\Lambda^{-1} \equiv \lambda^2/4\pi M,
\end{equation}
where the extra factor of $4\pi$ is inserted by convention to account for phase space; this factor typically appears alongside $\lambda^2$ in cross section calculations. Larger values of this parameter correspond to stronger interactions with the Standard Model, while smaller values correspond to dark matter models which are less coupled to visible matter. 

However, it is important to recognize that, due to the difference in techniques that can be applied to search for different classes of dark matter, even though smaller $\Lambda^{-1}$ corresponds to weaker coupling to the Standard Model, the experimental reach in $\Lambda^{-1}$ is not the same for all types of dark matter. If axions and WIMP dark matter interacted identically with the Standard Model fields, then one would expect that any experiment that could probe one class of dark matter model would (at least in principle) be able to probe the other, albeit with an overall rate that differed by some eighteen orders of magnitude (the squared ratio between the scale $f_a$ and the electroweak breaking scale). This is not at all the case. Axion dark matter contains an interaction between axions and two photons, allowing for axion-photon conversion. Such interactions are completely absent in WIMP models. In contrast, WIMP dark matter is heavy enough to impart significant energy in a WIMP-nucleon scattering event, while axions are far too light. Thus, low-background resonance cavities allow searches for axions with low interaction rates (i.e., high $f_a$) in a $\Lambda^{-1}$ range inaccessible to WIMP dark matter searches, and low-background nuclear recoil detectors allow WIMP searches that are insensitive to axions.

Thus, while it is in general true that lower values of $\Lambda^{-1}$ are more difficult to experimentally probe, one cannot place a universal experimental lower limit on $\Lambda^{-1}$ for {\it all} models of dark matter. With this caveat, comparing different models via their $\Lambda^{-1}$ does allow useful heuristic for characterizing different types of dark matter. In general, experiments using a particular technique sensitive to dark matter-Standard Model interactions of a particular form and dark matter masses in a particular range will find smaller $\Lambda^{-1}$ harder to detect.

We now estimate $\Lambda^{-1}$ for various dark matter models.

\subsubsection*{Cold dark matter}
The minimum possible interaction a dark matter can have with the Standard Model is a gravitational-strength interaction, and is the closest we can come to a pure non-relativistic, non-interacting CDM paradigm (as described further in the next subsection). Note that there is no mechanism for production of dark matter in this pure-gravity model --- only an unspecified appeal to Planck-scale physics. As this is the most minimal model of dark matter (at least as it pertains to dark matter after it is produced), and the behavior of all other dark matter candidates in the evolving Universe can be thought of as departures from this model, we discuss it first.  Here, the mediation scale $M$ is the Planck scale, ($10^{19}$~GeV), and so
\begin{equation}
\Lambda_{\rm gravity}^{-1} \sim 1/m_{\rm Pl} \sim 10^{-19}~\mbox{GeV}^{-1}.
\end{equation}
Note that several of the models which we will discuss later have $\Lambda^{-1} < \Lambda_{\rm gravity}^{-1}$ (the result of dark matter-Standard Model interactions being mediating by a series of interactions, all of which are very weak). In such cases, the non-gravitational interaction would be subdominant, and the most important dark matter-Standard Model interaction would be the one mediated by the Planck scale. Given the incredibly small coupling that this implies, the difference is somewhat academic, as both would imply interaction strengths far below any potential particle physics detection strategy.

\subsubsection*{WIMPs}

Turning to dark matter models that have a plausible production mechanism, WIMP dark matter should have interactions mediated by electroweak-scale particles, with $M$ on the order of the $W$ or the Higgs mass. Couplings are expected to be on the order of the gauge couplings $g \sim 0.65$ or $g' \sim 0.35$. Thus, while specific implementations of WIMP dark matter might have additional mixing angles and tunings reducing the interaction rate, we expect
\begin{eqnarray}
\Lambda_{\rm WIMP}^{-1} & \sim & g'^2/(4\pi m_H)-g^2/(4\pi m_W) \\ 
& \sim & (6-40) \times 10^{-5}~\mbox{GeV}^{-1}. \nonumber
\end{eqnarray}
This estimate applies to many models of WIMP dark matter, including two of the most commonly discussed: supersymmetry (where the dark matter is a linear combination of the superpartners of the photon, the $Z$ boson, and the Higgs) and extra-dimensions (where the dark matter is a Kaluza-Klein excitation of the photon or $Z$). These specific iterations of WIMP dark matter are shown in Figure~\ref{fig:fom}.

\subsubsection*{Thermal relic dark matter}
WIMP dark matter is a specific example of dark matter produced through thermal freeze-out of non-relativistic particles in the early Universe. A necessary component of this is a thermally averaged annihilation cross section in the early Universe of $\langle \sigma v\rangle \approx 3\times 10^{-26}$~cm$^3/$s \cite{Steigman:2012nb}. In the case of WIMP dark matter, this annihilation is mediated by the weak nuclear force, however production of thermal dark matter requires only a sufficiently large cross section, not that it proceeds through the weak force. Assuming this annihilation is between dark matter and the Standard Model, then in general the cross section should be proportional to $\Lambda^{-2}$, giving
\begin{equation}
\Lambda_{\rm thermal}^{-1} \sim 5 \times 10^{-5}~\mbox{GeV}^{-1}.
\end{equation}
Note this is somewhat below the low-end of the WIMP $\Lambda^{-1}$ range; this is not a contradiction, as usually additional ${\cal O}(1)$ factors appear in any detailed calculation of $\langle \sigma v\rangle$ that one would perform in a given WIMP model.

\subsubsection*{Dark Photons}

For freeze-out to occur while the dark matter is non-relativistic, thermal dark matter is typically assumed to have a mass between a few GeV and a few tens of TeV. The particle that mediates the interaction between this relatively massive dark matter and the Standard Model is often of a similar mass. For example, in true WIMP dark matter, the mediating particle would be a $W$, $Z$, or Higgs boson, all with masses of ${\cal O}(100~\mbox{GeV})$. However, the mediator could potentially be much lighter, and have escaped detection in Standard Model experiments through highly-suppressed couplings.  Such scenarios are interesting because light mediators result in long-range forces which can significantly alter the distribution of dark matter on astrophysical scales, as we will explore in more detail in the next subsection. 

As an example, we consider here ``dark photons,'' (see Ref.~\cite{Alexander:2016aln} for a review of the relevant particle physics) where the mediating particle connecting dark matter to the Standard Model is a photon analog with a mass below $\sim 1$~GeV. This would include the possibility that the dark photon is massless.  A ``dark photon" is a quantum superposition of the mediator and a (slight admixture) of the Standard Model photon.  The amount of ``photon" contained in the dark photon is the mixing parameter $\epsilon \ll 1$.  The particle physics constraints on dark photons themselves are a subject of much active research in a wide variety of experiments (see Ref.~\cite{Battaglieri:2017aum} for a status update).

In these dark photon models, the dark matter can achieve a thermal cross section though pair annihilation into pairs of dark photons, followed by later decay of the dark photons into pairs of Standard Model particles \cite{Pospelov:2007mp,Feng:2009hw}. This divorces the thermal cross section from the interaction rate in a present day experiment, which must connect dark matter to the Standard Model through the dark photon. Thus, we might estimate
\begin{equation}
\Lambda_{\rm dark~photon}^{-1} \sim \frac{\epsilon g_D e}{\mbox{max}(m_{A'}^2/m_\chi,m_\chi)},
\end{equation}
where the requirement of thermal freeze-out sets the ratio of dark photon coupling to dark matter mass to $g_D^2/m_\chi \sim \Lambda_{\rm thermal}^{-1}$. Taking $\epsilon \sim 10^{-7}-10^{-3}$ as required by null search results \cite{Battaglieri:2017aum} and $g_D \sim e$ as suggested --- although not required --- by many Beyond the Standard Model scenarios results in 
\begin{equation}
\Lambda_{\rm dark~photon}^{-1} \sim 10^{-12}-10^{-8}~\mbox{GeV}^{-1}.
\end{equation}
We note that the concept of a ``dark photon'' encompasses a wide variety of models (some of which do not assume dark matter is a thermal relic), and the relevant couplings can likewise vary even more than in this rough estimate.

\subsubsection*{Asymmetric dark matter}

We can apply the thermal result to a non-thermal model of dark matter. Asymmetric dark matter (ADM) \cite{Barr:1990ca,Barr:1991qn,Kaplan:1991ah,Thomas:1995ze,Hooper:2004dc,Agashe:2004bm,Kitano:2004sv,Cosme:2005sb,Farrar:2005zd,Suematsu:2005kp,Banks:2006xr,Kitano:2008tk,Kribs:2009fy,An:2009vq,Cohen:2009fz,Kaplan:2009ag,Cohen:2010kn,Shelton:2010ta,Davoudiasl:2010am,Haba:2010bm,Buckley:2010ui} produces dark matter through some mechanism similar to baryogenesis (or possibly through a shared mechanism, see Ref.~\cite{Zurek:2013wia} for a comprehensive review). However, just as the asymmetry in the number of protons and antiprotons is only relevant in the Universe today because the number density of thermally produced proton-antiproton pairs is very small (a factor of $10^{10}$ less than the asymmetric proton density), the asymmetric density of dark matter can only be relevant if the symmetric component has been annihilated away. This requires annihilation cross sections in the early Universe at least as large as those of a thermal relic \cite{2011PhRvD..84d3510B}, and so
\begin{equation}
\Lambda_{\rm ADM}^{-1} \gtrsim 5 \times 10^{-5}~\mbox{GeV}^{-1}.
\end{equation}

\subsubsection*{Axions, QCD and otherwise}

Switching to dark matter that was never in thermal equilibrium with the Standard Model plasma, axions (as previously mentioned) can be produced if the axion field in the early Universe is not located at the minimum of the potential generated by interactions with colored particles. If such a ``misalignment angle'' occurs after the Universe cools through the QCD phase transition, the axion field will oscillate around the potential minimum. As it settles to the minimum, non-relativistic axions will be produced, providing a dark matter candidate.   Assuming the misalignment angle is ${\cal O}(1)$ and not unnaturally small, the observed cold dark matter relic abundance realized when $f_a \sim 10^{10} - 10^{12}$ GeV \cite{Kim:2008hd}. Larger $f_a$ are possible if the misalignment angle is suppressed for some reason. 
The coupling between axions and the Standard Model of relevance for experiments such as ADMX \cite{Asztalos:2009yp} is the axion-photon coupling, which is $\lambda^2/4\pi = \alpha$ (times ${\cal O}(1)$ numbers set by details of the axion models at very high energies). As a result,
\begin{equation}
\Lambda_{\rm axion}^{-1} \sim e^2/(4\pi f_a) \sim 10^{-11}-10^{-15}~\mbox{GeV}^{-1}.
\end{equation}

A true QCD axion has a very tight relationship between the coupling of the axion to Standard Model particles and the mass of the axion itself, both being dependent on the single new physics parameter $f_a$ to various powers (times known Standard Model parameters). However, the idea of dark matter as the result of a very low mass field oscillating coherently in the early Universe can be abstracted away from the QCD axion (though in doing so one loses the nice connection with a solution to the Strong CP problem). For reasons that will be made clear when we discuss astrophysical scales, such dark matter is often called ``fuzzy'' dark matter, and can have very low masses, down to $\sim 10^{-21}-10^{-22}$~GeV \cite{PhysRevLett.64.1084,Sin:1992bg,Sahni:1999qe,Hu:2000ke,Goodman:2000tg,Peebles:2000yy,Amendola:2005ad,Schive:2014dra,Hui:2016ltb}. The interactions with Standard Model particles are very model-dependent, but one might expect
\begin{equation}
\Lambda^{-1}_{\rm FDM} \sim 10^{-17}-10^{-19}~\mbox{GeV}^{-1}.
\end{equation}
However, it must be emphasized that this estimate is very rough, and is highly sensitive to (usually unspecified) details of a Fuzzy Dark Matter model. 

\subsubsection*{Sterile neutrinos}

Sterile neutrinos are a non-thermally produced dark matter candidate, created in the early Universe through a small mixing angle $\sin\theta$ with the active neutrino species. Due to this mixing, this type of dark matter is unstable, with a lifetime of 
\begin{equation}
\tau_{\rm sterile-\nu} \sim \left(10^{22}~\mbox{s}\right)\sin^{-2}\theta\left( \frac{m_\chi}{\mbox{keV}}\right)^{-5}.
\end{equation}
From null searches for the resulting monochromatic gamma-ray, the mixing angle is constrained to be $\theta^2 \lesssim 10^{-5}(m_\chi/\mbox{keV})^{-5}$ \cite{Abazajian:2001vt,Dolgov:2000ew,Boyarsky:2009ix}. The mass of sterile neutrino dark matter is constrained to be above $\sim 0.4$~keV by phase space arguments \cite{Boyarsky:2008ju} --- tighter constraints arise for specific production mechanisms in the context of astrophysical constraints, as discussed in the next section. The interaction figure of merit in the Universe today then is
\begin{equation}
\Lambda_{\rm sterile-\nu}^{-1} \sim e^2\sin\theta/(4\pi m_W) \lesssim 10^{-7}~\mbox{GeV}^{-1}.
\end{equation}

\subsubsection*{Gravitinos}

Supersymmetric dark matter is often considered to be a thermally produced neutralino, with $\Lambda^{-1}$ therefore given by the WIMP values. However, another possibility is gravitino dark matter from supergravity \cite{Pagels:1981ke,Ellis:1984er,Berezinsky:1991kf}, produced through decay of heavier superpartners. The present-day interaction of gravitinos with the Standard Model would then occur through interactions mediated by particles at the supersymmetric breaking scale $m_F$, which could be anywhere from $10^6-10^{11}$~GeV up to the scale of Grand Unification ($10^{16}$~GeV). For production of gravitinos at TeV-scale colliders (say, at the LHC), the effective coupling of the gravitinos with colliding Standard Model particles will be generated through these mediators with mass $m_F$. The coupling will increase with collision energy, but will be suppressed by $m_F^2$, resulting in a figure of merit of

\begin{equation}
\Lambda_{\rm gravitino}^{-1} \sim \mbox{TeV}/m_F^2 \lesssim 10^{-29}-10^{-9}~\mbox{GeV}^{-1}.
\end{equation}
As previously mentioned, for values of $\Lambda^{-1}$ below the scale relevant for purely gravitationally mediated dark matter, the gravitational mediation dominates.

\subsubsection*{Primordial Black Holes}

Finally, dark matter might be a relic population of primordial black holes, created early in the history of the Universe \cite{ZeldovichNovikov1,ZeldovichNovikov2,HawkingBH,Carr:2017jsz}. Black holes lighter than $10^{-18}\,M_\odot$ are ruled out, as their Hawking radiation lifetime is shorter than the age of the Universe. Black holes between $10^{-7}\,M_\odot-10\,M_\odot$ can be constrained by microlensing \cite{Tisserand:2006zx} searches for MACHO dark matter, those between  $10^{6}\,M_\odot-10^9\,M_\odot$ by  millilensing of  compact radio sources \cite{Wilkinson:2001vv}, and those above $10^5\,M_\odot$ by dynamical constraints on gravitationally bound systems \cite{lacey1985,Carr:1997cn}. Dynamical constraints from wide binary stars in the stellar halo of the Milky Way and the stellar distribution within the Milky Way's satellite population further exclude PBH as a significant contributor to the dark matter budget for PBH masses greater than a few times $10^2\,M_\odot$ \cite{Yoo:2003fr,Quinn:2009zg,Brandt:2016aco,Koushiappas:2017chw}.  Combining all of these constraints (and others) nominally excludes primordial black holes as 100\% of dark matter if the mass spectrum is monochromatic \cite{Carr:2016drx,Carr:2017jsz}. However, astrophysical uncertainties leave three plausible mass ranges where black holes could comprise all of the dark matter: $10^{-17}\,M_\odot-10^{-16}\,M_\odot$, $10^{-13}\,M_\odot -10^{-9}\,M_\odot$,  and $1\,M_\odot-10^3\,M_\odot$.  

As objects with macroscopic masses (though often subatomic radii), it is  somewhat difficult to map primordial black holes into our $\Lambda^{-1}$ parameter. A naive identification of $\Lambda^{-1}$ with the Schwarzschild radius would result in an enormous value for the ``particle physics'' parameter, but this would be misleading, both because no particle physics experiment on Earth can directly access the energy equivalent of even the lightest possible black hole dark matter candidate, and the interaction rate of primordial black holes in the Universe is very low (as their geometric cross section and low number density results in an interaction time which is very small compared to galactic distance scales). As any reasonable definition for $\Lambda^{-1}$ in this case would be much smaller than their gravitational effects, we can safely take
\begin{equation}
\Lambda_{\rm PBH}^{-1} \sim \Lambda_{\rm gravity}^{-1} \sim 10^{-19}~\mbox{GeV}^{-1}.
\end{equation}

\subsubsection*{Baryons}
To put these dark models in context, we consider the subdominant component of the matter distribution most familiar to us: baryons.  To astronomers, ``baryon'' means ``every Standard Model particle but neutrinos and photons," which may be heretical to those with a traditional particle physics upbringing.  Because our Universe is charge-neutral and primarily hydrogen by mass, the dominant species of ``baryon" by mass is the proton, with a mass of 1 GeV.  Determining $\Lambda^{-1}$ is subtle because the baryons can interact through a number of different forces. At astrophysical scales, the most important interactions are electromagnetic. Direct proton-proton scattering in a halo with $v/c \sim 10^{-3}$ via the electromagnetic force resulting in a significant exchange of momentum would have an approximate 
\begin{equation}
\Lambda^{-1} \sim \alpha/m_p v^2 \sim 10^4~\mbox{GeV}^{-1}. 
\end{equation}
Strong-force nucleon-nucleon scattering has 
\begin{equation}
\Lambda^{-1} \sim \mbox{GeV}^{-1},
\end{equation}
and weak-scattering would be 
\begin{equation}
\Lambda^{-1}\sim \alpha/m_W \sim 10^{-4}~\mbox{GeV}^{-1}.
\end{equation}

\vspace{0.4cm}

This representative set of dark matter models gives an idea of the large variation in the figure of merit controlling the ``visibility'' of dark matter to the particle physics experiments. The range of $\Lambda^{-1}$ values for representative dark matter models we have considered are shown in Figure~\ref{fig:fom} (horizontal axis), along with the astrophysical parameter discussed next (vertical axis). Note that this parameter spans $\sim 20$ orders of magnitude --- a wide range, to be sure, but much smaller than the range of either the mass or cross section parameters shown in Figure~\ref{fig:dm_pspace}.  Moreover, the experimentally accessible range is significantly smaller.  The dark matter searches targeting dark matter that was in thermal equilibrium with the Standard Model in the early Universe can cover much of the parameter space above $\sim 10^{-5}$~GeV, though some dark matter models with very small values of $\Lambda^{-1}$ can be detected using specialized detection detection techniques (for example, axions and sterile neutrinos).

\subsection{A Measure of Astrophysical Scales}

\begin{figure}[t]
\begin{center}
\includegraphics[width=0.75\textwidth]{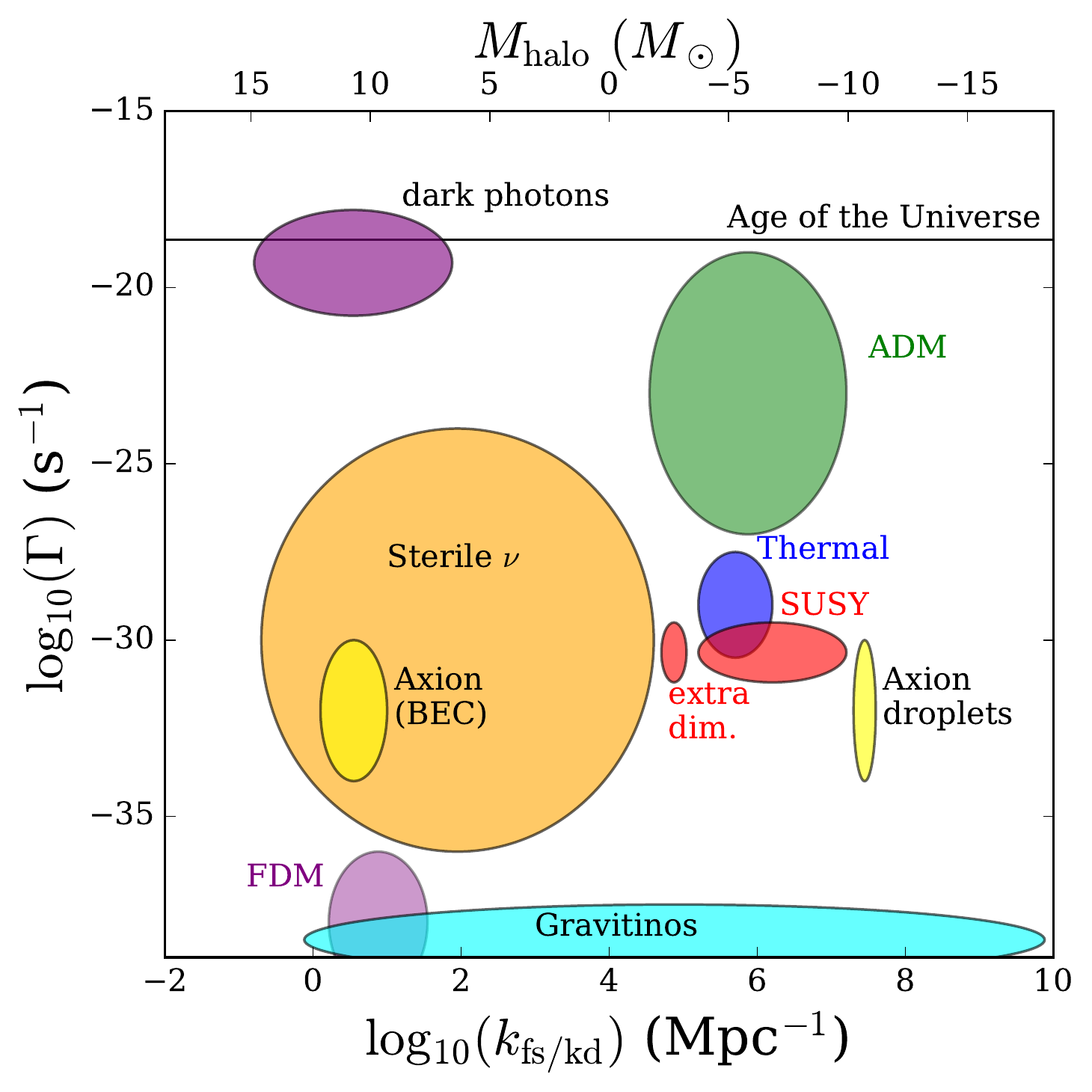}
\end{center}
\caption{\label{fig:dm_astro}Dark matter candidates in an astronomically relevant parameter space.  The horizontal axis is dual-labeled, and represents primordial effects.  The bottom axis shows primordial effects quantified in terms of the characteristic free-streaming wavenumber of the model; the top axis shows how this translates to non-linear scales, quantified in terms of the half-mode halo mass (see text for details).  The vertical axis quantifies evolutionary effects on dark-matter halos and cosmology according to a characteristic interaction or decay rate.}
\end{figure}

The immense range of scales controlling the interaction of dark matter with the visible particles is directly related to the explosion of viable dark matter models that theoretical physicists have devised over the years. This is not to undersell the theoretical motivations underpinning most of these dark matter scenarios. Rather it is an acknowledgment that, without direct experimental evidence, theory alone cannot provide a unique solution to the problem of dark matter. Some of these solutions lend themselves to direct searches in particle physics experiments, but others do not have obvious search strategies given current technology.

However, even for models of dark matter which are invisible to our particle detectors \cite{Bertone:2010at,Bertone:2011kb}, their gravitational imprint on matter and light is unavoidable. This allows for astrophysical probes of dark matter by investigating the structure of the gravitationally bound dark matter \emph{halos}. Indeed, all of our present knowledge of dark matter comes from such investigations: we know that dark matter was non-relativistic at the time of structure formation \cite{Reid:2009xm,irsic2017}, at most has self-interactions below the scale of the strong nuclear force \cite{Kaplinghat:2015aga}, and forms gravitationally bound structures down to a mass of at least $10^8\,M_\odot$ \cite{Viel:2005qj,Seljak:2006bg,Jethwa:2016gra,Kim:2017iwr}.

If dark matter were a purely non-collisional, non-interacting, non-relativistic particle --- a {\em pure CDM} model --- then the power spectrum of the gravitationally-bound dark matter structures (``halos") existing in our Universe today would be set by the primordial density perturbations of the inflaton field (experimentally known to be close to the Harrison-Zel'dovich spectrum \cite{Ade:2015xua}), followed by gravitational evolution as the Universe expanded.  Overdensities in the matter field would grow as modes of the density fluctuations entered the cosmological horizon. Once the perturbations became unstable, matter would decouple from the Hubble flow and collapse in on itself and virializing (converting from bulk to thermal kinetic energy) through a process called ``violent relaxation" to form bound dark matter halos.  Violent relaxation is the process by which the orbits of particles flowing into perturbations get scrambled because of the rapid time evolution of the potential as the perturbations collapse (Sec. 4.10 of Ref. \cite{binney2008}), eventually virializing (e.g., coming to equilibrium) when the potential no longer changes on timescales shorted than the dynamical time.  This would result in hierarchal structures with gravitationally-bound dark matter masses extending from super-clusters down to the mass of individual particles of dark matter. The evolution of this power spectrum to the present day and into the non-linear regime has been well-studied \cite{2005Natur.433..389D,2006Natur.440.1137S,Reid:2009xm}.  This hierarchy of scales is the strongest, the most striking prediction of CDM cosmology.

\subsubsection{Dark Matter Halo Primer \label{sec:primer}}

\begin{figure*}[t]
\includegraphics[width=0.99\textwidth]{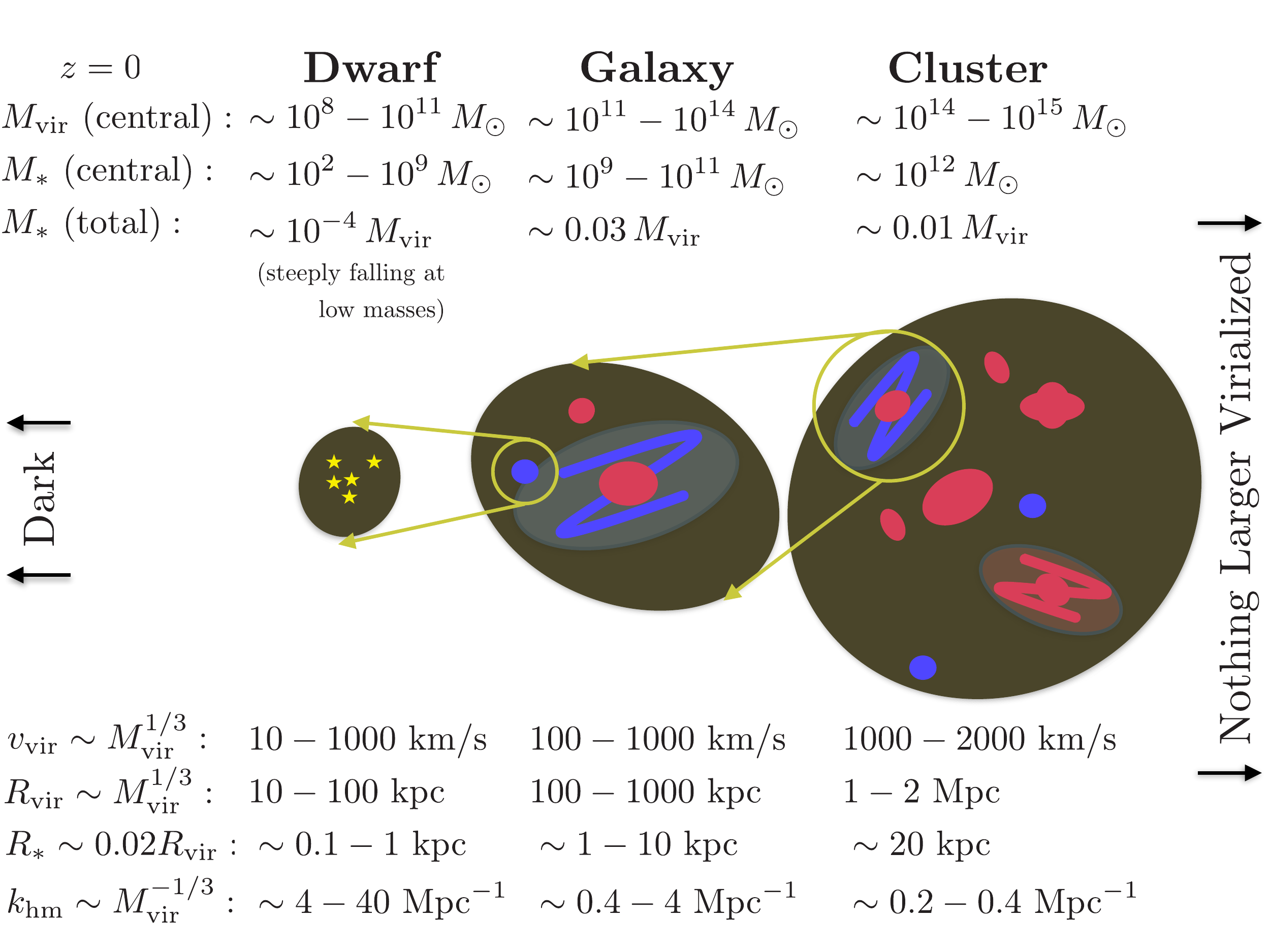}
\caption{\label{fig:halogalaxy}Dark matter halo scales relative to galaxy scales. Cluster-sized halos typically contain thousands of satellite galaxies \cite{Ferrarese:2016kcy}, each in their own subhalo.  Galaxy-sized halos typically have a few hundred (at most) satellites down to ultrafaint scales \cite{Kim:2017iwr}.  Dwarf galaxies are expected to only host a handful of satellites \cite{dooley2016}. For each class of object, we show the typical halo mass if the object is a central or host galaxy (halo), the stellar mass of the central galaxy, and the total stellar mass contained within the halo (including host and satellite contributions).  The scaling of other halo and galaxy properties are given as a function of host halo mass, including the half-mode wavenumber $k_{\mathrm{hm}}$. }
\end{figure*}

Given the importance of dark matter halos have to the astrophysical probes of dark matter (and therefore to this paper), we provide a primer on the basic concepts for the particle physicist, who may be more used to thinking of individual particles of dark matter. The fundamental macroscopic unit of dark matter is the halo --- an overdensity  which decoupled from the Hubble flow to collapse and virialize into a gravitationally bound clump.  Halos are triaxial \cite{Jing:2002np,Bailin:2004wu,Allgood:2005eu}, but we characterize them in terms of spherically-averaged quantities.  Dark matter halos are defined by their {\em virial mass} (which we will typically refer to as \mvir\ in this paper), which is the mass enclosed in a {\em virial radius} $R_\textrm{vir}$, in which the average density is some number $\Delta$ times the critical density $\rho_c$ of the Universe:
\begin{eqnarray}
	M_{\rm vir} = \frac{4\pi}{3} (\Delta \times \rho_c) R_\textrm{vir}^3.
\end{eqnarray}
There is no universal choice for $\Delta$, with different choices having different physical motivation \cite{Diemer:2014xya} though the variation is only a factor of a few.  The most common choices are the virial density $\Delta_v(z)$, which is related to spherical collapse \cite{Bryan:1997dn}; $\Delta_{200c} = 200$ \cite{navarro1997}; and $\Delta_{200b} = 200\Omega_m$.  In galaxy cluster literature, $\Delta_{500c} = 500$ is common. Because structural parameters and substructure counts often depend on the definition of the virial radius, it is important to keep track of which virial radius is meant in a particular paper, and translate to another choice of ``virial" if necessary.  For the order-of-magnitude estimates of interest in this paper, we can treat all these choices of $\Delta$ as approximately the same.

Often we describe halos in terms of velocity instead of mass, because it is typically more robust to uncertainties in the definition of ``virial," also because it represents the characteristic speed or temperature of particles inside.  The {\em virial velocity}
\begin{equation}
v_\textrm{vir} = \sqrt{GM_{\rm vir}/R_\textrm{vir}} \label{eq:vvir}
\end{equation}
is the speed a particle would have on a circular orbit at the virial radius.  It is also common to see the {\em maximum velocity} 
\begin{eqnarray}\label{eq:vmax}
	v_{\mathrm{max}} = \mathrm{max}\left( v_c(r) \right),
\end{eqnarray} 
with the peak of the {\em circular velocity} curve,
\begin{eqnarray}\label{eq:vcirc}
	v_\mathrm{c}(r) = \sqrt{\frac{GM(<r)}{r}}, 
\end{eqnarray}
used as an alternate.  Because $v_\mathrm{c}(r)$ peaks deep inside the halo, $v_\text{max}$ is a good mass proxy when the definition of mass is ambiguous.

Halos grow by ingesting other halos and matter that is not yet subsumed into halos, a process known as hierarchical structure formation.  Halos that have not been wholly dissolved by the {\em host halo} (a.k.a.~the {\em central halo}) are called {\em subhalos}.  These subhalos are typically tidally stripped of much of their dark matter \cite{vandenBosch:2004zs}, with 70-90\% mass stripping being typical.  Halos reach their maximum mass either at $z=0$ (if they are central halos), or just before they fall into bigger systems (if they are to become subhalos) \cite{Reddick:2012qy,Behroozi:2013fqa}.  The maximum mass of a halo throughout cosmic time is called {\em peak mass} $M_\mathrm{peak} = \max( M(z) )$.  For subhalos in particular, where the definition of halo mass is not uniquely defined, it is usually more useful to characterize them in terms of $v_\mathrm{max}$, or in terms of the maximum of $v_{\mathrm{max}}$ throughout cosmic time, $v_\mathrm{peak}$.  If $v_\mathrm{max}/v_\mathrm{peak} < 1$, you can safely assume that the subhalo has been quite significantly stripped.

Dark matter, by its nature, cannot be directly observed, and so the masses and velocities of the halos are not direct astronomical observables. However, halos are the structures in which {\em galaxies} are embedded.  Galaxies are gravitationally bound collections of stars and/or gas with significant spread in metallicity \cite{Willman:2012uj}. The galaxy at the center of the ``host'' or ``central'' halo is called the {\em host} or {\em central galaxy}.  Galaxies inhabiting subhalos are called {\em satellite galaxies}. Galaxies and their halos, including massive galaxies like the Milky Way, may sometimes be embedded in even larger halos of dark matter, forming {\em galaxy clusters}. 

One of the most important developments in galaxy evolution theory and observation in recent decades is the realization that galaxy and halo mass (the {\em stellar-mass--halo-mass} relation, {\em SMHM}) are strongly correlated with each other for halos more massive than $10^{11}M_\odot$. This likely also holds true for smaller halos as well (see Figure~\ref{fig:halogalaxy}), although there are beginning to be hints in simulations \cite{Fitts:2016usl,Garrison-Kimmel:2016szj,Munshi:2017xhq} and observations \cite{errani2018}  that the scatter becomes large.  For small halos, it is possible that there is a floor in galaxy mass on typical star-cluster scales, which could lead to a plateau in the SMHM relation below a fixed halo mass.  There is tentative evidence for such a plateau from the stellar mass function of satellite galaxies of the Milky Way near the Magellanic Clouds found by the Dark Energy Survey \cite{dooley2017}.  Different theoretical frameworks are used to infer the matching from observations, going under the names ``conditional luminosity function'' \cite{yang2003}, ``halo occupation distribution" \cite{berlind2002}, and ``(subhalo) abundance matching" \cite{tasitsiomi2004,vale2004}.  Galaxy stellar mass is tightly correlated to halo mass above Magellanic Cloud ($M_{\rm vir}\sim 10^{11}\,M_\odot$, or $M_* \sim 10^9\,M_\odot$) mass scales.  As we show in the next section, there is significant uncertainty in how smaller galaxies populate halos.  A heuristic mapping between galaxies and halos is shown in Figures~\ref{fig:halogalaxy}.

Galaxies are small compared to halos --- both in volume and in mass.  The typical half-light radius of a galaxy is of order a couple of percent of the virial radius \cite{kravtsov2013,somerville2017,huang2017,Hearin:2017cho}.  The mass in cold gas and stars is never more than a few percent of the total mass of a halo \cite{Behroozi:2013fqa,2013MNRAS.428.3121M}, although large (\mvir\ $\gtrsim 10^{12}\,M_\odot$, Milky Way mass and above) halos retain significant reservoirs of shock-heated gas throughout their volume.  The stellar and gas contents of galaxies are used to trace the dark matter potential wherever the stellar and gas density is high enough to admit an observation.  Spiral galaxies are ``rotationally supported'' --- stars and gas travel in nearly circular, coplanar orbits.  Observations of this motion are used to create {\em rotation curves}, the circular velocity as a function of distance $r$, Eq. (\ref{eq:vcirc}).  Spheroidal and elliptical galaxies are ``dispersion supported'' --- stars and gas exhibit random motions, characterized by a velocity dispersion $\sigma(r)$, and the potential can be recovered under assumptions of virial or hydrostatic equilibrium.

The standard \emph{Hubble Sequence} galaxies, ellipticals and spirals, typically inhabit halos of $M_\text{vir} \sim $ a few $\times 10^{11}\,M_\odot$.  Giant elliptical galaxies --- dispersion-supported systems with stellar mass $M_*=10^{11}-10^{12}\,M_\odot$ --- are hosted by large halos ($M_\text{vir} \gtrsim 10^{13}\,M_\odot$), whereas spiral galaxies are more common at scales of $M_\text{vir}\sim 10^{12} M_\odot$ (this is the approximate mass of the Milky Way).  Dwarf galaxies span an enormous range of stellar mass and origin.  Dwarfs with $M_* \gtrsim 10^5 M_\odot$ are generally forming stars today if they are central galaxies, and have a large gas-to-stellar mass ratio \cite{Geha:2012nq}.  If they are satellites, they are typically devoid of gas and recent star formation \cite{spekkens2014}.  In the Milky Way, such galaxies are called \emph{classical dwarfs}.  Dwarfs with $M_* \lesssim 10^5 M_\odot$ are classified as \emph{ultrafaints}, and are thought to be ancient relics of star formation prior to the reionization of the Universe \cite{Benson:2001at}. Reionization heats the intergalactic medium to the point where it cannot condense into small halos, preventing small halos from forming stars at all but the earliest times \cite{Bullock:2000wn}.  We return to this point in Section~\ref{sec:observations}.

While dark matter halos grow by ingesting smaller halos, much of the galaxy formation occurs in situ.  Satellite galaxies destroyed by the host dissolve into the \emph{stellar halo} (called \emph{intracluster light} on cluster scales).  Because the efficiency of star formation drops rapidly with decreasing halo mass (Figure~\ref{fig:halogalaxy}), the stellar halo of the Milky Way contributes only $\sim 1\%$ of the total stellar mass of the Milky Way \cite{binney2008,mcmillan2017}.  Most of the stellar halo derives from a single or a few massive progenitors \cite{Deason:2016wld}.  The kinematics of halo stars may be used to map out the dark matter distribution of the host.

\subsubsection{Particle Physics and Halo Mass}

In the standard Cold Dark Matter paradigm, the hierarchical structure of large dark matter halos containing smaller halos, which themselves contain smaller halos (i.e., the halo of a galaxy cluster containing many galactic halos, which contain subhalos of varying sizes, and with their own substructure) should continue down to extremely small scales. However, if there exists interactions either between dark matter particles or between dark matter and the Standard Model, then number and structure of these halos can be modified. Depending on the nature of this new interaction, we should expect the modification to only become apparent in halos of some characteristic physical size.

Therefore, deviations from the CDM model can be expressed in terms of the largest physical scale on which deviations from CDM appear noticeable in simulations or analytic theory.  This scale is often expressed in terms of a halo mass, \mhalo, to make contact with astronomical tests on non-linear scales.  This mass scale can also be recast in terms of a distance scale, typically in terms of a comoving wavenumber $k$ (an inverse distance), to make contact with linear theory. This wavenumber might be set by free-streaming $k_{\rm fs}$ or kinetic decoupling $k_{\rm kd}$. We will comment further on exactly how this link is made in the discussion of warm dark matter candidates below.  Figure~\ref{fig:halogalaxy} gives a sense of scale of the kinds of dark-matter halos which are traced by different types of visible structure, and a corresponding wave number; we expand on this figure in Section~\ref{sec:observations}.

Measurements which probe dark matter gravitational structures on a variety of scales can therefore be used to constrain the theoretical model space. Deviations from the CDM model can take two forms, illustrated in Figure~\ref{fig:dm_astro}.  The deviations can be {\it primordial}, when non-trivial physics terminates or modifies the gravitational collapse of dark matter at some comoving wavenumber $k$. It translates to a characteristic mass scale today below which the number of dark matter halos is significantly reduced. Note that if the inflationary power spectrum has non-trivial structure at small scales (rather than the nearly scale-free spectrum measured on large scales) or if the reheating temperature is low, there may be non-trivial structure in the halo mass function even for CDM-like particle dark matter models \cite{Erickcek:2011us,Aslanyan:2015hmi}.

Alternatively, the deviations can be {\it evolutionary}, driven by the interactions of dark matter over the history of the Universe since the original structure formation, with a characteristic rate $\Gamma$. These effects can erase existing structure or change the cosmic density and velocity distributions of dark matter from their non-evolutionary baseline.  While evolutionary deviations are typically characterized by a time scale (e.g., a lifetime for unstable dark matter, or a scattering rate for interacting dark matter), as shown with the vertical axis in Figure~\ref{fig:dm_astro}, the precise choice for this axis has the same type of ambiguity as the $\sigma_{\rm int}$ interaction cross section in Figure~\ref{fig:dm_pspace}.  For example, interaction rates are certainly density-dependent (and likely velocity-dependent as well), and so change from location to location in the Universe.  Instead of classifying the evolutionary deviations of particle models by their late-time interaction or decay rates, we choose to characterize evolutionary processes in terms of a halo mass scale $M_{\rm halo}$ as well. Aside from being pragmatic, this choice has a firm basis in phenomenology.  For example, the kinematics of late-time processes, such as late-time decays or self-interactions, have drastically different impact on halos of different masses \cite{Peter:2010jy,Rocha:2012jg}. While relativistic decays leave their mark on halos of all sizes, non-relativistic decays perturb small halos significantly while leaving large halos untouched \cite{Peter:2010jy}.  Hidden sector models typically yield strongly velocity-dependent self-interaction cross sections, leading to a ``resonant'' effect when the halo virial velocity is well-matched to peak of the self-interaction cross section \cite{Loeb:2010gj,Tulin:2013teo,Kaplinghat:2015aga}.  Thus, models with evolutionary effects can be mapped to \mhalo.

The effects of ``primordial'' and ``evolutionary'' deviations cannot be cleanly divided in any realistic model of dark matter particle physics --- for example, extra interactions which suppress the inner density of dark matter halos also may lead to a suppression in the primordial power spectrum \cite{Buckley:2014hja}. Moreover, any suppression in the matter power spectrum additionally leads to a delay in the formation time of halos relative to a pure $\Lambda$CDM model with a Harrison-Zel'dovich inflationary power spectrum.  Thus, there may be extra \emph{time-dependent} signatures of new dark matter physics even if the origin is purely primordial \cite{Barkana:2001gr,Haiman:2001hc,benson2013,schneider2015}.  Despite this intrinsic mixing it is useful to divide the possible sources of power spectrum deviations in this way for the purposes of this paper, always keeping in mind that applications to any particular model of particle physics necessarily involves a more detailed calculation which includes both the primordial power spectrum {\it and} its evolution. It is convenient to define a single parameter to characterize a scale on which dark-matter microphysics leads to significant departures from CDM predictions, shown for the models described below in Figure~\ref{fig:fom}.

\subsubsection{Primordial Deviations}\label{sec:primordial}

In the standard $\Lambda$CDM model of the Universe, inflation lays down a nearly scale-invariant power spectrum of potential perturbations, described by their wavelength $\lambda$ or wavenumber $k = 2 \pi/\lambda$.  If dark matter is pure CDM, there is no time-evolution in the linear power spectrum other than the standard suppression of growing modes for large $k$ before matter-radiation equality.  The non-linear halo mass function $dn/dM_\text{vir}$ (number density of halos per unit mass) is thus also nearly scale invariant below cluster scales (Figure \ref{fig:halogalaxy}), $dn/dM_\text{vir} \propto M_{\text{vir}}^{-1.9}$ \cite{2001MNRAS.321..372J}.  Non-minimal models of dark matter which influence the power spectrum of dark matter halos during the period of initial Jeans collapse typically lead to cut-off at the comoving wavenumber $k_{\mathrm{cut}}$, although this cut-off is sometimes heavy in features and is rarely a step function. 

However, the effects of a truncation in the matter power spectrum typically become  apparent on much larger scales in the non-linear halo mass function.  The effects of non-minimal dark-matter physics on the linear matter power spectrum are usually expressed in terms of a transfer function 
\begin{equation}
T_{\rm DM}(k) = \left( \frac{P_{\rm DM}(k)}{P_{\rm CDM}(k)} \right)^{1/2},
\end{equation}
where $P_{\rm CDM}(k)$ is the linear CDM power spectrum and $P_{\rm DM}(k)$ is the power spectrum for an arbitrary dark matter model.  Ref.~\cite{schneider2012} define a ``half-mode mass'' $M_{\mathrm{hm}}$ corresponding to the wavenumber $k_{\mathrm{hm}}$ at which $T(k_{\mathrm{hm}}) = 1/2$.  This is also approximately the point at which the halo mass function drops by a factor of order two with respect to CDM when modes become non-linear.  

The half-mode wavenumber can be converted to an equivalent halo mass, below which the number density of halos would deviate significantly from the predictions of hierarchical structure formation in CDM:
\begin{equation}
	M_{\mathrm{hm}} = \frac{4\pi}{3}\bar{\rho}\left( \frac{\pi}{k_{\mathrm{hm}}} \right)^3,
\end{equation}
where $\bar{\rho}$ is the comoving matter density of the Universe \cite{schneider2012,2013MNRAS.433.1573S}.  In other words, this is mass enclosed in an average patch of the Universe with radius $\pi/k_{\mathrm{hm}} = \lambda_{\mathrm{hm}}/2$. For primordial effects, we choose 
\begin{equation}
	M_{\mathrm{halo}} = M_{\mathrm{hm}}.
\end{equation}

For the most frequently encountered (thermal relic warm dark matter) transfer functions, $k_{\mathrm{cut}} \sim 10\, k_{\mathrm{hm}}$, or $M_{\mathrm{hm}} \sim (\hbox{a few})\times 10^3 \,M_{\mathrm{cut}}$, where $M_{\mathrm{cut}}$ is the halo mass corresponding to $k_\mathrm{cut}$ \cite{schneider2012}.  We caution that when encountering a new model, it is always best to calculate the half-mode mass directly from the power spectrum. 

There are two primary physical pathways leading to an early-time truncation or suppression in the matter power spectrum.  First, for particles may free-stream out of small density perturbations if they  ``born hot'', i.e., have a (semi-) relativistic momentum distribution in the early universe.  This typically leads to a smooth truncation of the matter power spectrum.  Second, interactions with Standard Model or other hidden-sector particles can lead to truncations that have acoustic-type oscillations structure.  Often, both types of primordial features are present in the power spectrum.

\subsubsection*{Thermal relics}

Assuming dark matter is a thermal relic, the power spectrum is exponentially suppressed once the dark matter kinetically decouples from the Standard Model particle bath. This occurs at a temperature $T_{\rm kd}$, with the approximate relation \cite{Hofmann:2001bi,Green:2003un,Profumo:2006bv,Bringmann:2009vf}
\begin{equation}
M_{\rm halo} \sim \left(10^{-1}\,M_\odot\right) \left(\frac{T_{\rm kd} g_{\rm eff}^{1/4}}{50~\mbox{MeV}}\right)^{-3}
\end{equation}
where $g_{\rm eff}$ is the number of effective degrees of freedom in the thermal bath at $T_{\rm kd}$. For an arbitrary thermal relic model with mass $m_\chi$ and scattering via the $\ell^{\rm th}$ partial wave is given by \cite{Green:2003un}
\begin{equation}
\frac{m_\chi}{T_{\rm kd}} \sim \left[ \frac{\xi(3)}{\pi^2} \left( \frac{90 g_{\rm eff}}{8\pi^3}\right)^{1/2} m_{\rm Pl} m_\chi \sigma_0^{\rm el}(m) \right]^{1/(3+\ell)}.
\end{equation} 
Here $\sigma^{\rm el}_0$ is the elastic scattering cross section. For 100~GeV dark matter, if $\sigma^{\rm el}_0$ is the thermally averaged cross section ($3\times 10^{-26}$~cm$^3$/s) then
\begin{equation}
T_{\rm kd} \sim 17(154)~\mbox{MeV}
\end{equation}
for $\ell = 0(1)$. Note that the assumption that $\sigma^{\rm el}_0$ is exactly the thermal cross section is not typically realized in a specific model of thermally produced dark matter. That said, in this simplified model, our assumptions result in a minimum halo mass of 
\begin{equation}
\left(M_{\rm halo}\right)^{\rm thermal} \sim 10^{-5}\,M_\odot-10^{-2}\,M_\odot.
\end{equation}

Free streaming can further damp the matter power spectrum, with \cite{green2005}
\begin{equation}
k_{\rm fs} \approx (10^6~\mbox{Mpc}^{-1})\frac{\sqrt{(m_\chi/100~\mbox{GeV})(T_{\rm kd}/30~\mbox{MeV})}}{1+0.05\ln(T_{\rm kd}/30~\mbox{MeV})}
\end{equation}
The relative importance of free-streaming and kinetic decoupling depends on the specifics of the model. For the thermally-coupled models considered here, free-streaming is less important than  kinetic decoupling, and so is ignored.

Note that our numbers for \mhalo\, may appear large to experienced practitioners.  As described above, this is because we are considering \mhalo\, to be the scale on which deviations from CDM begin to become important, rather than (as is usually done) calculating the minimum possible halo mass.  The difference between the two halo masses can be orders of magnitude in scale.  This prediction can be refined in specific models of thermal relic dark matter, where the elastic scattering cross section can be calculated from fundamental parameters and related to the annihilation cross section.

\subsubsection*{WIMPs}

For supersymmetric WIMP dark matter, scanning over the parameter space of the minimal supersymmetric Standard Model, $T_{\rm kd}$ can vary from $15-1500$~MeV \cite{CollierCameron:2009ja}, resulting in
\begin{equation}
\left(M_{\rm halo}\right)^{\rm SUSY} \sim 10^{-8}\,M_\odot-10^{-2}\,M_\odot.
\end{equation}
A similar scan over the parameters of Kaluza-Klein dark matter from models with extra spatial dimensions found
\begin{equation}
\left(M_{\rm halo}\right)^{\rm extra-dim} \sim 10^{-2}\,M_\odot-1\,M_\odot.
\end{equation}

\subsubsection*{Dark Photons}

Dark matter charged under a new force with a massless or nearly massless gauge boson can induce dark acoustic oscillations in the early Universe. This will suppress the number of small halos \cite{Cyr-Racine:2013fsa}, as well as decrease the density profile of smaller dark matter halos \cite{Buckley:2014hja}. While it is in principle possible to construct dark photon models that suppress structure on almost any scale (by appropriate selection of masses and couplings), we take the parameters of Ref.~\cite{Buckley:2014hja} as an upper limit, and estimate
\begin{equation}
\left(M_{\rm halo}\right)^{\rm dark~photon} \lesssim 10^{12}\,M_\odot.
\end{equation}

\subsubsection*{Asymmetric dark matter}

Asymmetric dark matter models could stay in kinetic equilibrium for much longer than thermal dark matter, due to the larger interaction cross section required to suppress the thermal component of dark matter. Thus, considering only the interactions required to annihilate away the thermal component, we can estimate
\begin{equation}
\left(M_{\rm halo}\right)^{\rm ADM} \lesssim 10^{-1}\,M_\odot,
\end{equation}
with the upper limit in a particular model set either by the decoupling of the dark matter from the baryons, or by the collapse of the baryon sound speed at the QCD phase transition, depending on the details of the model. If additional dark matter physics is added to an asymmetric model (for example, dark photons), then \mhalo\, could be much larger.

\subsubsection*{Axions, QCD and otherwise}

It has been suggested that axion dark matter formed from field misalignment could form small-scale Bose-Einstein condensates (BEC), taking the form of high-density ``droplets'' of axions with a mass \cite{RindlerDaller:2009er,Chavanis:2011zi,Chavanis:2011zm,RindlerDaller:2011kx,Liebling:2012fv,RindlerDaller:2012vj,Guth:2014hsa,Davidson:2016uok}
\begin{equation}
\left(M_{\rm halo}\right)^{\rm axion-droplet} \sim 10^{-10}\,M_\odot-10^{-13}\,M_\odot,
\end{equation}
with the droplet mass proportional to $f_a^2$. 
These droplets then cluster into larger cosmological structures, as in other forms of dark matter. This primordial deviation from the cold dark matter power spectrum is distinct from the possibility of large-scale axion BECs, which would effect the structure of galaxies, and will be discussed in the next section.

The axion-like Fuzzy Dark Matter models also can alter the structure of dark matter halos, in this case by erasing structure below the de Broglie wavelength of the dark matter. By focusing on ultra-light $10^{-21}-10^{-22}$~eV particles, this wavelength can be made larger than dwarf galaxies, suppressing the number of structures below \cite{Hui:2016ltb}:
\begin{equation}
\left(M_{\rm halo}\right)^{\rm FDM-pr.} \sim 10^{10}\,M_\odot\left(\frac{m_\chi}{10^{-22}~\mbox{eV}} \right)^{-4/3}.
\end{equation}
These models have evolutionary effects in addition to this cut-off in the primordial power spectrum (see also Ref.~\cite{johnson2008}). Note that the choice of dark matter mass is driven purely by astrophysical phenomenology. 

\subsubsection*{Sterile neutrinos}
Sterile neutrinos are a warm dark matter candidate \cite{Shi:1998km,Dodelson:1993je,Abazajian:2001nj,Abazajian:2005gj}, with a mass range of $\sim 0.4-10^5$~keV. The lower end of this mass range is set by astrophysics: for lighter sterile neutrinos, the structure of dark matter halos would deviate too much from observations of the Lyman-$\alpha$ forest or from Milky Way satellite counts \cite{irsic2017,Jethwa:2016gra,Kim:2017iwr}. The mixing angle to the active neutrinos $\sin^2\theta$ is proportional to $m_\chi^{-5}$ \cite{Boyarsky:2009ix}. 
The free-streaming length depends strongly on the sterile neutrino production mechanism \cite{Venumadhav:2015pla}. Contrary to how limits are almost always presented in the literature, \emph{there is no general one-to-one mapping between sterile neutrino mass and a free-streaming scale}.  However, for the specific choice of the neutrino minimal standard model ($\nu$MSM) \cite{Boyarsky:2009ix}, and for some non-resonant production models (requiring a primordial lepton asymmetry) \cite{Shi:1998km,Abazajian:2001nj,Abazajian:2005gj}, the free-streaming length is approximately 
\begin{equation}
k_{\rm fs} \sim \left(0.5~\mbox{Mpc}^{-1}\right)\left(\frac{m_\chi}{\mbox{keV}}\right)
\end{equation}
This results in a range of $M_{\rm halo}$ for sterile neutrinos of
\begin{equation}
\left(M_{\rm halo}\right)^{\rm sterile-\nu} = 10^{-6}-10^{11}\,M_\odot.
\end{equation}
The presence of small-scale structure has long been used to constrain neutrino dark matter \cite{Bond:1980ha}. 

This astrophysical figure of merit is therefore correlated to the particle physics parameter $\Lambda^{-1}$, which is proportional to $\sin\theta \propto m_\chi^{-5/2}$. As a result, $\log \Lambda^{-1} \propto \tfrac{5}{6}\log M_{\rm halo}$. As mentioned previously, the exact production mechanism for the sterile neutrinos can modify this relationship, introducing additional dependence of $M_{\rm halo}$ on $\sin^2\theta$, but for our purposes, this rough estimate suffices.

\subsubsection*{Gravitinos}
Similarly, gravitino dark matter has a dark matter mass set by the supersymmetry breaking scale $m_F$
\begin{equation}
m_\chi \sim \frac{m_F^2}{m_{\rm Pl}}.
\end{equation}
With $m_F$ in the range of $10^6-10^{11}$~GeV, the gravitino masses can range from $\sim 100$~eV up to 100~TeV. The resulting free-streaming length sets a minimum halo mass range of
\begin{equation}
\left(M_{\rm halo}\right)^{\rm gravitino} = 10^{-17}\,M_\odot-10^{13}\,M_\odot.
\end{equation}
As with the sterile neutrinos, we expect the particle physics and astrophysics figures of merit to be correlated, with $\log \Lambda^{-1} \propto -\tfrac{1}{3}\log M_{\rm halo}$.

\subsubsection*{Self-interacting dark matter (SIDM)}

In recent years, there has been a great deal of interest in models of SIDM \cite{Spergel:1999mh,Tulin:2017ara}, driven in large part by the possibility of evolutionarily altering the structure of dark matter halos relative to CDM (see our discussion of evolutionary deviations below). Motivations for these models \cite{Weinberg:2013aya} 
will be discussed in more detail in Section~\ref{sec:observations}. Here, we consider a subset of SIDM models which also cause a primordial deviation from the cold dark matter power spectrum. 

Such deviations can occur if the SIDM model contains a relatively light force carrier, such as the dark photon discussed in the previous section \cite{Ackerman:mha,Feng:2009mn,Kaplan:2009de,CyrRacine:2012fz} or dark pion from a confining sector \cite{Buckley:2012ky}. Though massless states in the dark sector face severe constraints from CMB measurements \cite{Feng:2008mu,Cyr-Racine:2013fsa}, a relatively low-mass force mediator can keep the dark matter in kinetic equilibrium with itself long after decoupling from the Standard Model. This can suppress and alter primordial structure both through collisional damping and ``dark'' acoustic oscillations (analogous to baryon acoustic oscillations) \cite{Cyr-Racine:2013fsa}. 

The resulting deviations are more complicated than a straightforward exponential suppression of the power spectrum, and can include a reduction of the central densities of collapsed dark matter halos, caused by a delay in the formation time. A full calculation of the effects of a particular model of SIDM on the primordial power spectrum requires specialized $N$-body simulation \cite{Buckley:2014hja}, but it has been demonstrated that SIDM models with light mediators can affect the primordial power spectrum on scales corresponding to
\begin{equation}
\left(M_{\rm halo}\right)^{\rm SIDM-pr.} \lesssim 10^{11}\,M_\odot. 
\end{equation}
There are numerous mechanisms for the generation of SIDM dark matter, so we cannot assign a single particle physics parameter $\Lambda^{-1}$ to all of these models. 

\subsubsection*{Primordial Black Holes}

As discussed above, it is possible for primordial black holes to be 100\% of the dark matter in the Universe if their mass spectrum falls in one of three windows:
\begin{equation}
(M_{\rm halo})^{\rm PBH} \approx 10^{-(17-16)}\,M_\odot, 10^{-(14-9)}\,M_\odot,10^{(0-3)}\,M_\odot,
\end{equation}
due to a combination of constraints from MACHO microlensing searches, disruption of gravitationally bound binary objects, and searches for Hawking radiation from the evaporation of light black holes \cite{Carr:2017jsz,Carr:2017jsz}.  We treat these ranges of PBH masses as \mhalo~for this model.

\subsubsection*{Gravity-Only Interactions}

If dark matter interacts only via gravity, the mechanism for dark matter production is completely unspecified. As a result, no general statement can be made about the deviation of gravity-mediated dark matter structure from the predictions of pure cold dark matter.

\subsubsection*{Baryons}
As in Section~\ref{sec:standardmodel}, we set the primordial effects of baryon physics in contrast to these dark-matter models.  Because baryons are tightly coupled to photons in the early Universe, sound waves propagate up until the point of recombination.  Not only does this leave an imprint in the cosmic microwave background, but leads to the set of ``wiggles'' in the matter power spectrum, the BAO that we discussed in Section~\ref{sec:darkscience}.  In contrast to the dark photon model discussed above, the BAO do not lead to a cut-off in the matter power spectrum because baryons are a subdominant form of matter.  In a baryon + CDM model, the scale on which these wiggles occur is $\sim 100$ Mpc, well in the linear regime today \cite{boss2013}.  These result in features in the two-point correlation function of galaxies rather than in galaxy or halo number counts.

A second primordial effect imprints itself onto the matter power spectrum in different patches of the Universe --- in other words, there is a large scale-dependence of this small-scale effect.  It arises from the relative velocity between dark and baryonic perturbations at the epoch of recombination \cite{Tseliakhovich:2010bj}.  The relative velocity between baryons and dark matter suddenly becomes supersonic for baryons as photons break free from the previously tightly coupled baryon-photon fluid.  This bulk flow of baryons with respect to dark matter suppresses the growth of small halos at early times \cite{Tseliakhovich:2010yw,Bovy:2012af}, on scales of \mvir\ $\sim 10^8 M_\odot$, but the suppression is modulated on $\sim 100$ Mpc scales.

\subsubsection{Evolutionary Deviations}\label{sec:evolution}

The structure of dark matter halos may also be altered by late-time effects. These effects are distinct from the suppression of power in the primordial spectrum, though many models of dark matter may cause both primordial and evolutionary deviations.  Primordial deviations may lead to time-dependent changes in the growth of structure relative to CDM, notably that structure forms later when there exists a cut-off in the matter power spectrum \cite{Barkana:2001gr,Governato:2014gja,Schive:2017biq}.  But we focus this section on evolutionary deviations that have their origin in ongoing interactions or decays, rather than time-dependent but primordial-in-origin effects.

\subsubsection*{WIMPs and thermal relics}
Dark matter-dark matter interactions become important in reshaping halos when the per-particle interaction rate $\Gamma \sim \langle \sigma v\rangle \rho/m_\chi$ is of order the inverse of the local dynamical time $t_d \sim r/v\sim 1-100$ Myr, or higher.  The central regions of halos have densities of order $\sim 0.1\,M_\odot/\hbox{pc}^3 \sim 10^{-23}\hbox{ g/cm}^3$ or less \cite{strigari2008}.  If we assume that any dark matter self-interaction cross section is of order the thermal relic cross section $\langle \sigma v \rangle \sim 10^{-26}\hbox{ cm}^3/\hbox{s}$, and that the WIMP is $\sim 100$ GeV in mass, then $\Gamma \sim 10^{-27} \hbox{s}^{-1} \sim 10^{-11}\hbox{ Gyr}^{-1}$.  Thus, the relative evolution of WIMP-like thermal relic dark matter halos vs.~pure CDM halos is completely negligible.

\subsubsection*{Axions: QCD and otherwise}

Axion models may form high-density ``droplets'' from BEC effects in the early Universe, of order $M_{\mathrm{halo}} \sim 10^{-10}\,M_\odot$ in size (see our previous discussion) \cite{Guth:2014hsa}.  It has also been argued that Milky Way-sized dark matter halos composed of axions would form caustics from long-range BEC effects \cite{Sikivie:2009qn,Erken:2011dz,Banik:2013rxa,Dumas:2015wba}. These deviations do not come in the form of the complete suppression of structure on any mass scale, but rather alterations of the density and velocity distributions on the physical scale of $\sim 10$~kpc, which is to say a mass scale of
\begin{equation}
\left( M_{\rm halo}\right)^{\rm axion-BEC} \sim 10^{12}\,M_\odot, 
\end{equation}
Recent studies of axion physics call the existence of the long-range BEC into question \cite{Guth:2014hsa}, finding that the attractive interactions that are typical between axions cannot lead to long-range correlations  \cite{fan2016}.  Ref.~\cite{Guth:2014hsa} argues that repulsive interactions may allow long-range forces and large-scale BEC to exist, but Ref.~\cite{fan2016} shows that it is difficult to construct a light scalar theory with repulsive, not attractive, interactions.  We include the possibility of large-scale BEC here for completeness even though the current conventional wisdom for QCD axions comes down against it (see also Ref.~\cite{hertzberg2016}).

Fuzzy dark matter, in addition to erasing structure on scales below the de Broglie wavelength, also acts as a quantum fluid, rather than a collection of discrete particles. This can result in a ``solitonic core'' \cite{Schive:2014hza,Hui:2016ltb}, which affects the inner kpc of the galaxies that form above the primordial $M_{\rm halo}$ cut-off. Dark-matter-only $N$-body simulation indicates that this softens the dark matter cusp predicted by CDM into a core \cite{Zhang:2016uiy}, resulting in deviations from the predictions of CDM on the scale of
\begin{equation}
\left(M_{\rm halo}\right)^{\rm FDM-ev.} \sim 10^{8}\,M_\odot-10^{10}\,M_\odot.
\end{equation}
In addition, quantum interference outside the core can lead to clumpy structure throughout the rest of the halo \cite{Suarez:2013iw,Schive:2014dra,Schive:2014hza,Schive:2015kza,Schive:2017biq,Mocz:2017wlg}.

\subsubsection*{Sterile neutrinos}
In Section~\ref{sec:standardmodel}, we found that the sterile neutrino lifetime $\tau_{\rm sterile-\nu} \sim \left(10^{22}~\mbox{s}\right)\sin^{-2}\theta\left( \frac{m_\chi}{{\rm keV}}\right)^{-5}.$  For typical values of $m_\chi$ and $\sin^2\theta$, this decay time is many orders of magnitude larger than the age of the Universe.  Only a tiny fraction of sterile neutrinos decay, so sterile neutrinos have negligible evolutionary deviations from CDM.  Their deviations are essentially purely primordial.

\subsubsection*{SIDM}

Interactions lead to the redistribution of energy and angular momentum of particles in individual dark matter halos, typically on scales where the per-particle interaction rate $\Gamma \sim \rho \sigma v/m_\chi$ is of order $1/t_d$, where $t_d$ is the dynamical time and $\sigma$ is the (velocity-dependent) self-interaction cross section \cite{Spergel:1999mh,Rocha:2012jg,Peter:2012jh}.  Note that the cross section to which we refer here is the momentum-transfer or viscosity cross section, depending on the specific SIDM model \cite{Tulin:2013teo,Boddy:2016bbu}.  The redistribution of energy and angular momentum leads to the formation of flat, spherically symmetric density cores. For velocity-independent cross sections, these cores become significant at approximately the same fraction of the virial radius across a wide range of halo mass (up to cluster scales) \cite{Lin:2015fza}.  For velocity-dependent cross sections, cores can show up for only a narrow range of halo mass, leaving halos outside this mass range relatively unperturbed with respect to CDM \cite{Buckley:2009in,Feng:2009hw,Loeb:2010gj,Zavala:2012us,Kaplinghat:2015aga,Tulin:2017ara}.   

Under some circumstances, interactions trigger central densities that may be \emph{higher} than predicted in CDM theory, at least in a subset of halos.  First, if the scattering cross section (even purely elastic) is sufficiently high, it is possible to trigger a ``gravothermal catastrophe,'' in which case the halo center becomes increasingly dense with time \cite{Kochanek:2000pi,Balberg:2001qg,Balberg:2002ue,Pollack:2014rja}.  Cross sections typically need to be $\gtrsim 10\hbox{ cm}^2/\hbox{g}$ for this process to begin \cite{Elbert:2014bma}.  Second, if some fraction of dark matter has significant self-interactions, including massless mediators allowing for radiative cooling, then more exotic deviations are possible \cite{Boddy:2016bbu}, for example the formation of dark disks in spiral galaxies \cite{Fan:2013yva}. 

Interactions may become important when halos accrete smaller halos, in the process of hierarchical structure growth.  Originally, a major motivation for SIDM models was to eliminate subhalos by the ejection of particles from, and eventual evaporation of, subhalos through particle interactions with the host halo \cite{Spergel:1999mh}.  Although this mechanism is largely inefficient for currently allowed cross sections \cite{Gnedin:2000ea,Vogelsberger:2012ku,Dooley:2016ajo}, which are of order the scale of the strong nuclear force, the idea that mergers can probe interactions is alive on cluster scales ($M_{\mathrm{halo}}\sim 10^{14}\,M_\odot - 10^{15}\,M_\odot$)  \cite{Harvey:2013tfa,Kahlhoefer:2013dca,Kahlhoefer:2015vua,Massey:2015dkw,Robertson:2016qef,Robertson:2016xjh,Kim:2016ujt}.  We will discuss this more in Section~\ref{sec:future}.

Thus, SIDM models can affect dark matter structure on the scales
\begin{equation}
\left( M_{\rm halo}\right)^{\rm SIDM-ev.} \lesssim 10^{15}\,M_\odot. 
\end{equation}

\subsubsection*{Baryons}
Baryons are special for two reasons: their interaction rates are high, and they can dissipate energy in these interactions.  Thus, after halos virialize, baryons can continue to cool, as we showed in Section~\ref{sec:darkscience}.  Baryons condense in the centers of halos.  As we discuss in Section~\ref{sec:observations}, baryons can alter the total matter (baryon + dark matter) density profile on scales 
\begin{equation}
M_{\mathrm{vir}}\gtrsim 10^{8}\,M_\odot.  
\end{equation}
Below this scale, baryons cool inefficiently, which we see by the very high mass-to-light ratios in dwarf galaxies.  The dynamical effects of baryons on the matter profile for small halos are limited in scope.  Above this scale, baryons do affect halo structure, by an amount still under debate.  Thus,
\begin{equation}
	10^8 M_\odot < M_\mathrm{halo} < 10^{15} M_\odot.
\end{equation}

For completeness we note that if baryons scattered significantly but elastically, it would also alter the matter density profile, but not dramatically unless the scattering were high enough to trigger a gravothermal catastrophe \cite{binney2008}.

\vspace{0.5cm}

To conclude this section, we note that the scale on which departures are expected can span a range of halo mass below \mhalo. To discover dark matter, the key is to see consistency in departures from CDM across a range of scales.

\section{Open Problems: Hints and Known Systematic Uncertainties}\label{sec:observations}

While for particle theorists, it is natural to consider a wide range of particle models for dark matter, for astronomers, one paradigm reigns: CDM.  All observations on small scales are couched in terms of how well they are or are not perceived to be in line with the two main predictions of the CDM paradigm.  In CDM,
\begin{enumerate}
\item There exists a \emph{nearly scale-free hierarchy} of bound dark matter structures, called halos, down to astronomically irrelevant scales.
\item Dark matter halos are stratified in density, sharply \emph{cusped} in their centers.
\end{enumerate}
Any deviation from this paradigm may suggest non-minimal extensions to the CDM paradigm, of the type we described in Section~\ref{sec:fom}.  In recent years, there have been hints from data implying that there are deviations from the predictions of CDM, affecting the structure and evolution of astrophysical systems.  These hints are the topic of a vigorous debate in the astronomical community, largely over the interplay between dark matter and baryonic physics.  For particle theorists, the hints have influenced the development of specific types of dark matter models.  This is reflected in the number of dark matter models in Figure~\ref{fig:fom} which have $M_{\rm halo}$ on the range $10^8\,M_\odot-10^{10}\,M_\odot$. In both fields, we must fully understand both baryonic effects on halos if we are to continue exploring dark matter interactions via astrophysics, even if the current set of hints are the result of the interplay between CDM and baryonic physics.  

\begin{figure*}
    \includegraphics[width=0.49\textwidth]{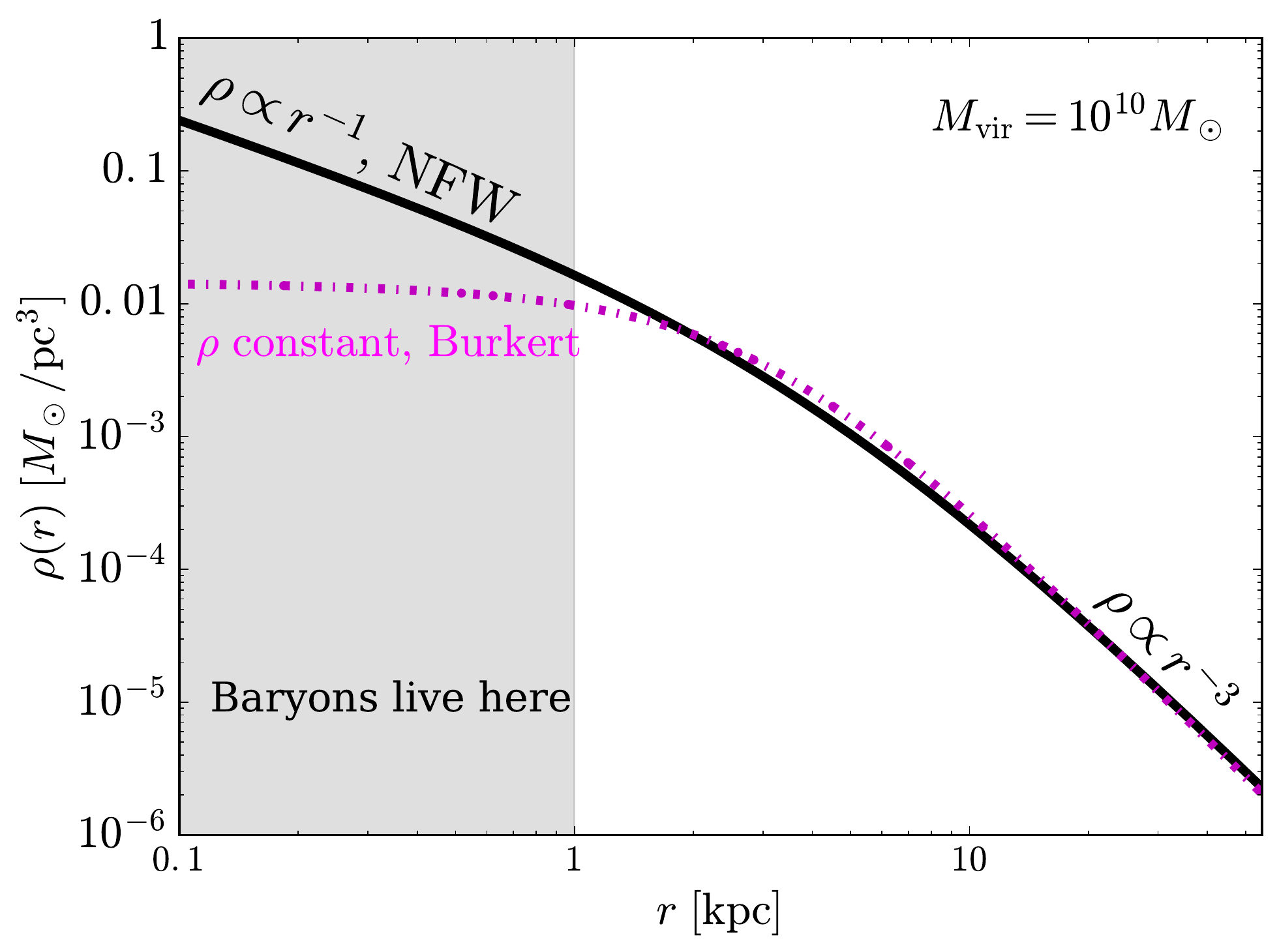} \includegraphics[width=0.48\textwidth]{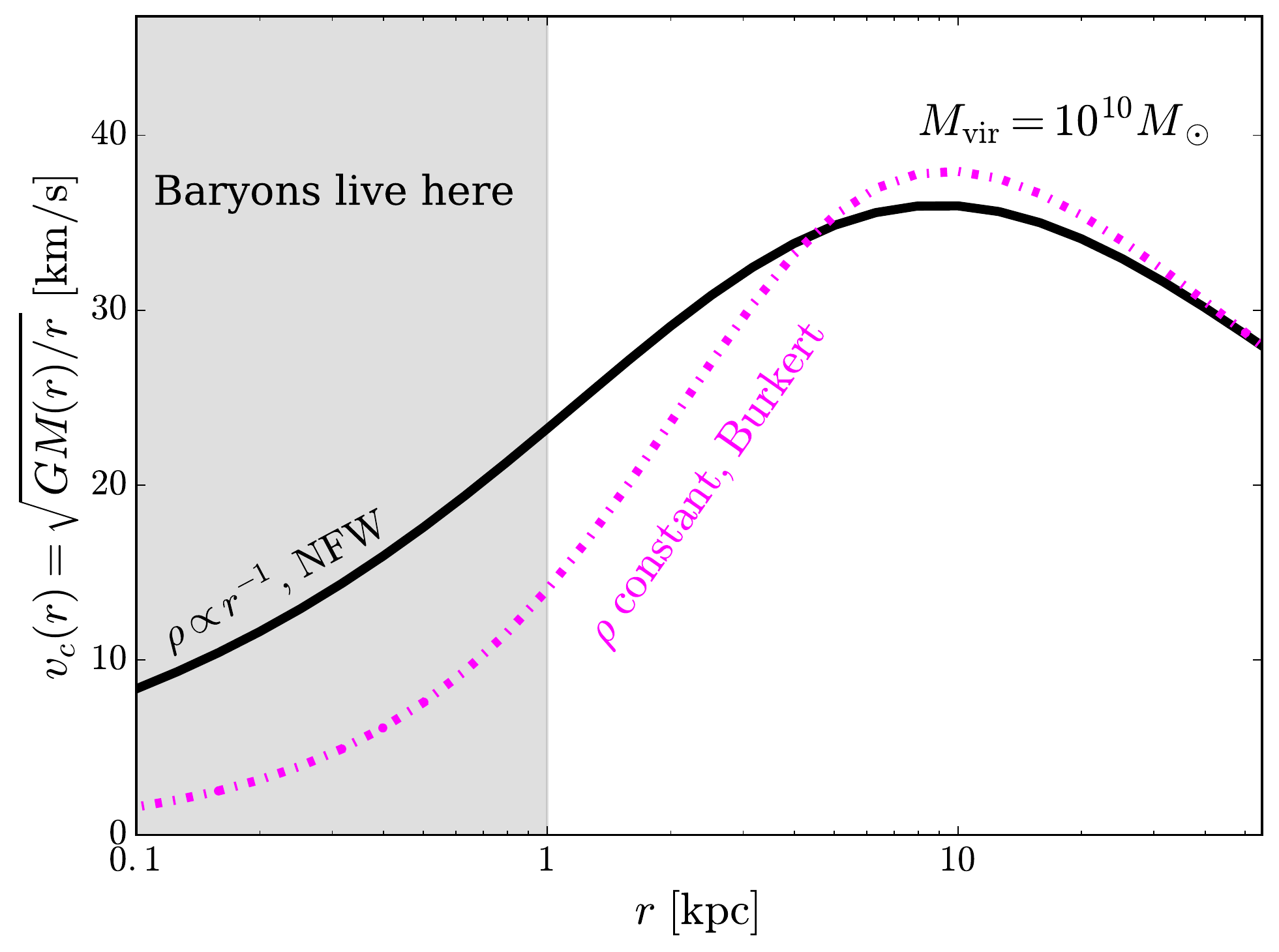}
	\caption{\label{fig:cuspcore} Density profiles of $10^{10}\,M_\odot$ halos (\emph{left}) and the resulting integrated mass profile plotted as $v_c$ (\emph{right}), comparing ``cored'' and ``cusped'' density profiles, out to the virial radius (55 kpc).  Any density profile with $\rho \propto r^{-\alpha}$ toward the center with $\alpha \gtrsim 1$ is considered a ``cusped" profile.  The case of $\alpha = 1$ is specific to the NFW profile \cite{navarro1996}.  By ``cored", people \emph{either} mean the specific case of $\alpha = 0$, \emph{or} more generally $\alpha < 1$.  Here, we compare a cusped NFW profile to a cored profile (in the strict, $\alpha = 0$ sense): a Burkert profile with a scale radius of $0.7r_s$, which is a good fit to both dwarf rotation curves and SIDM halos \cite{burkert1995,Rocha:2012jg}.  The mass profiles are often plotted as the circular velocity $v_c$, which is the orbit a particle on a circular orbit in a spherical potential would have.  The maximum value of $v_c$ is called $v_{\mathrm{max}}$.  The largest value of $v_{\mathrm{max}}$ a halo ever achieves in its history (typically either $z=0$ for isolated halos, or just before a satellite halo falls into a larger system) is called $v_{\mathrm{peak}}$ (Section~ \ref{sec:primer}).}
\end{figure*}

In this section, we discuss the present state of observations, which have been interpreted as hints of physics beyond CDM; and the major theoretical uncertainty complication in transforming hints to measurements, the dynamical interplay between galaxy formation and dark matter physics. We describe the current state of these possible hints of new physics from astrophysical observations, namely (see also Ref.~\cite{Bullock:2017xww}):
\begin{enumerate}
\item The cusp/core problem (see Figure~\ref{fig:cuspcore} for a summary),

\item The Missing Satellites problem (Figure~\ref{fig:cdmproblems_msp}),

\item ``Too Big To Fail'' (Figure~\ref{fig:cdmproblems_tbtf}),

\item The baryonic Tully-Fisher relation (Figure~\ref{fig:cdmproblems_shm}) and ``too big to fail in the field" (Figure~\ref{fig:cdmproblems_msp}).
\end{enumerate}
We will end the section with a summary on the state of simulations that include galaxy evolution physics, and show that it is not currently possible to unambiguously determine whether these hints are evidence of new physics or the result of the gravitational interplay between dark matter and baryons. We summarize our main conclusions about these hints of possible deviations from CDM in Figure~\ref{fig:summary4}. A key insight in this section is that a major observational uncertainty behind the lack of resolution to these problems is the not-well-quantified relationship between galaxy mass and halo mass on dwarf-galaxy scales.  In  Section~\ref{sec:future}, we describe a path forward to break through present-day barriers to progress, summarized in Figure~\ref{fig:future_probes}.  We recommend that readers less familiar with galaxies and halos consult Section~\ref{sec:primer} before continuing.

\subsection{Hints}\label{sec:hints}

There are a number of identified ``problems" on small scales.  We discuss each in turn, and show where they land in astrophysical parameter $M_{\rm halo}$. A summary of these problems can be found in Figure~\ref{fig:summary4}. For an expanded discussion of these problems, see Ref.~\cite{Bullock:2017xww}. 

\subsubsection*{Cusp/Core Problem}

From the time of the earliest CDM simulations and calculations, it was found that dark matter halos should achieve extremely high densities at their centers (e.g., Ref.~\cite{fillmore1984}).  The primordial fine-grain phase-space density, which sets the upper bound on the central phase-space density of the halo, is high compared to any material with non-zero velocity dispersions or interactions \cite{tremaine1979}.   Thus, both the central phase-space density and the configuration-space mass density (the integral of the phase-space density over velocity space) are high compared to the hot baryonic halo gas permeating large dark-matter halos.  By the mid-1990's, it was clear from dark-matter-only simulations that CDM halos did not follow the scale-free $\rho \propto r^{-2}$ isothermal profile,\footnote{Named because the velocity dispersion, a proxy for temperature, is constant as a function of $r$ for a self-gravitating halo \cite{binney2008}.} but instead had another nearly universal profile, a broken Navarro-Frenk-White (NFW) power-law profile \cite{navarro1996,navarro1997}
\begin{eqnarray}
	\rho(r) = \frac{\rho_s}{(r/r_s)\left( 1 + r/r_s\right)^2}.
\end{eqnarray}
Since then, there have been refinements to the inner slope of the density profile \cite{navarro2004,navarro2010}, but in general the $\rho \propto r^{-1}$ (at small $r$) cusped profile fits CDM-only simulations well, on average.

Evidence that this simulation-derived profile did not match observed mass profiles appeared around the same time as the original NFW papers (Refs.~\cite{navarro1996,navarro1997}).  The first hint came from low surface-brightness galaxies (LSBs): spiral galaxies with unusually diffuse disks \cite{1997MNRAS.290..533D}.  Because these disks have low baryon density, the rotation curves of these galaxies should be dark matter dominated.  It was found that these galaxies have rotation curves which rise (almost) linearly as a function of radius \cite{2000AJ....119.1579V,2001ApJ...552L..23D,simon2003,simon2005,kuzio2006,kuzio2008}, which are characteristic of halos having $\rho \propto r^{-\alpha}$ with $\alpha \approx 0$, that is, a core (Figure~\ref{fig:cuspcore}).  There are actually two discrepancies: in the \emph{shape} of the density profile, where this can be measured; and the mean density within some radius, which is more robustly measured for most galaxies.  This discrepancy in the density profile between CDM simulations and observations at small radii has become known as the ``cusp/core problem."  

Since then, deviations from NFW predictions have been found in all systems for which dynamical estimates of the density profile are possible. We step through the evidence from small-halo to large-halo scales.  Notably, there is as yet no observation of the density profile of the smallest known galaxies: ultrafaint dwarfs (stellar mass $<10^5\,M_\odot$, \mvir\, likely of order $\sim 10^8\,M_\odot - 10^9\,M_\odot$). This is on account of their paucity of stars (but see Refs.~\cite{penarrubia2016,inoue2017} for ideas on photometric estimates of density profiles). For these systems, we are in the near-term limited to estimates of the mean density within the half-light radius of each system.  The density is remarkably constant as a function of stellar mass, a fact which is actually consistent with CDM interpretations \cite{strigari2008,walker2009}.  

For somewhat larger objects, like the classical dwarf satellites of the Milky Way (with stellar mass $M_* \gtrsim 10^5\,M_\odot$, many of which were recently star-forming) or the late-type (spiral) 
dwarf galaxies in the \textsc{Little Things} surveys \cite{Hunter:2012un,Oh:2015xoa} (\mhalo$\sim 10^{10}\,M_\odot-10^{11}\,M_\odot$), it is clear that their central densities are lower than expected in CDM-only simulations \cite{mateo1998,walker2007}.  The existence of cores appear likely outside the Milky Way \cite{Oh:2010ea,Governato:2012fa}.

For the Milky Way satellites, there is still a vigorous debate as to whether they have cusped or cored density profiles (e.g., Refs.~\cite{strigari2006,gilmore2007,Walker:2011zu,2012MNRAS.419..184A,2014ApJ...791L...3B,2016arXiv160706479F,2017ApJ...838..123S}).  This ambiguity arises because these systems are dispersion-supported, and there are well-known degeneracies between the orbit anisotropies, 3D stellar density, and dark matter density profile that are difficult to disentangle with the observational tools available \cite{binney2008,2017ApJ...838..123S}.  As we describe in Section~\ref{sec:future}, proper motion measurements in the 2020's with the WFIRST satellite may break this degeneracy.

One major surprise in this area is the recently discovered Milky Way companion Crater 2 \cite{Torrealba16b}.  It is bright like the classical satellites, but its surface brightness is so low that it was only recently discovered.  It has an unusually low central density (although see Ref. \cite{errani2018}), even for a dwarf galaxy, and it lies far from the center of the Milky Way, so tidal effects should  be small. When better understood, this object may help clarify the cusp/core problem. 

The story becomes more complicated at larger scales.  For LSBs and larger spiral galaxies ($M_{\rm halo} \sim 10^{12}\,M_\odot$), halos also appear less dense than expected from CDM-only simulations (in addition to the references above, see Refs.~\cite{deblok2008,dutton2011}).  Interestingly, while the total mass profile of cluster-scale halos follows the NFW prediction well \cite{2012ApJ...755...56U,newman2013a,newman2013b}, the most massive galaxy clusters also appear to have underdense halos ($M_{\rm halo}\sim 10^{15}\,M_\odot$). Typically, the center of a cluster is occupied by a massive galaxy: the brightest cluster galaxy (BCG).  Because BCGs are compact objects relative to dark-matter halos, and dominate the gravitational potential where they sit, this implies that the density profile of the dark matter \emph{outside} the BCG is shallower than NFW. 

The one outlier to this trend of low-density halos is found in of elliptical galaxies in group-scale halos ($M_{\rm halo} \sim 10^{13}\,M_\odot-10^{14}\,M_\odot$).  Strong gravitational lensing by these systems consistently shows that the total mass density profile is consistent with $\rho \propto r^{-2}$ for a large part of the halo \cite{2003ApJ...583..606K,2012MNRAS.423.1073B}.  The dark matter profiles are typically found to be either consistent with or steeper than NFW  \cite{2010MNRAS.408.1463S,dutton2011,2015ApJ...814...26N}.  

To summarize, in the parameter $M_{\rm halo}$, the dark-matter solution to the cusp/core problem would have to show visible deviations from pure CDM on scales of $M_{\rm halo} \sim 10^{9}\,M_\odot - 10^{15}\,M_\odot$. 

Originally, the cusp/core model motivated investigations into WDM models, because of their much-lower-than CDM primordial phase-space densities \cite{bode2001,Dalcanton:2000hn}.  This would imply that a dark matter solution is primordial in origin, and would thus occupy the moderate $k_{\rm fs}$, low $\Gamma_{\rm int}$ region of parameter space in Figure~\ref{fig:dm_astro}. However, it has since been recognized that the cores produced in WDM cosmologies in the absence of baryons are far too small to be consistent with observations \cite{Strigari:2006ue,deNaray:2009xj}.  To obtain the relatively large core sizes needed to resolve the cusp/core problem, the number densities of those same halos would also be massively suppressed. This is the ``Catch 22 problem" \cite{2001ApJ...559..516A,2012MNRAS.424.1105M,Colin:2014sga}.  Thus, most dark-matter interpretations of the core/cusp problem now focus on ultralight dark matter \cite{Hui:2016ltb} (another primordial solution) or self-interacting dark matter (large evolutionary effect) \cite{Kaplinghat:2015aga,Valli:2017ktb}).   Notice that astrophysics is already informing the parameter space for viable particle physics models of dark matter.

\subsubsection*{``Missing Satellites" problem, and other substructure problems}

\begin{figure*}[t]
	\includegraphics[width=0.50\textwidth]{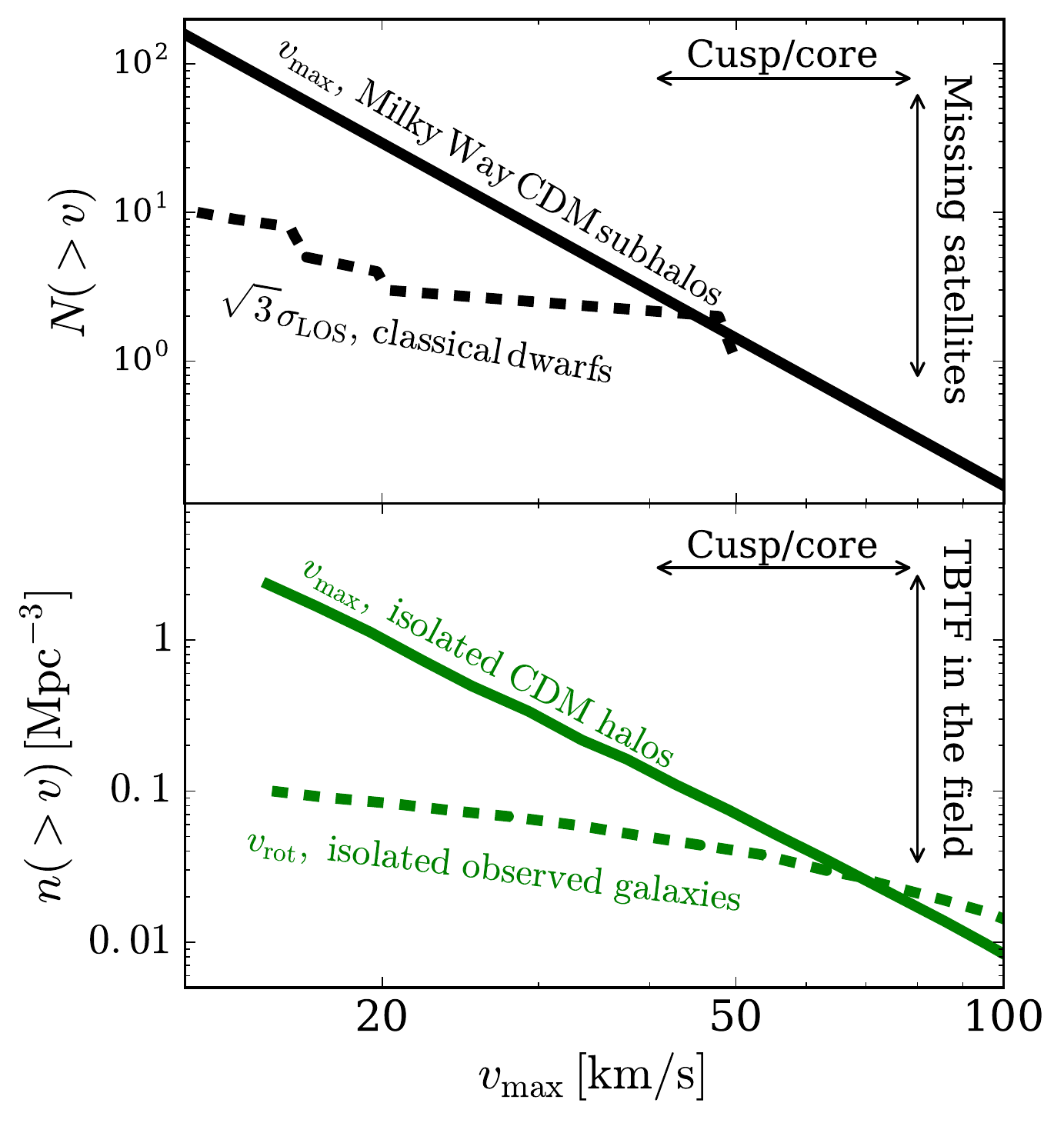} \includegraphics[width=0.46\textwidth]{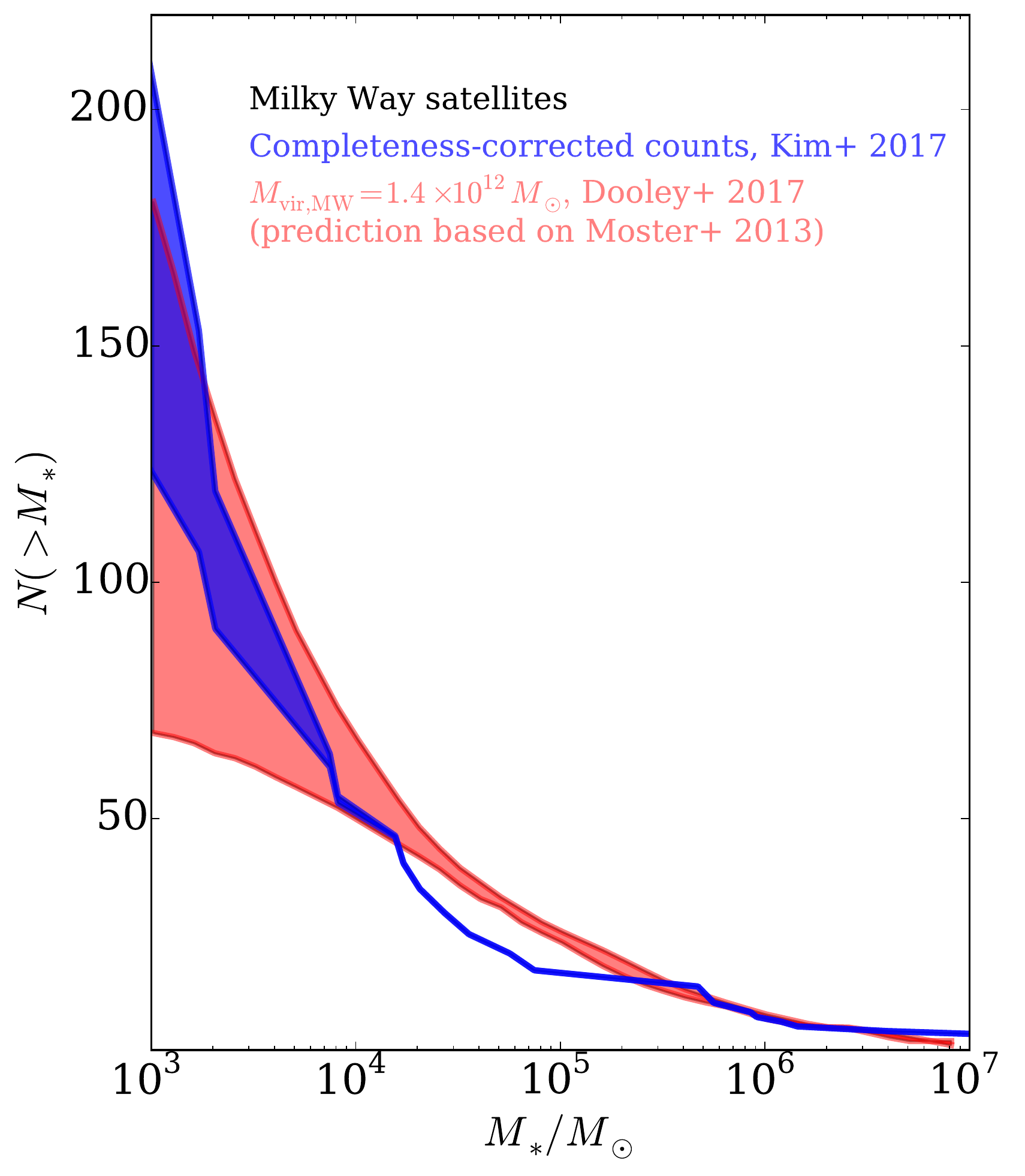}
    \caption{\label{fig:cdmproblems_msp} \emph{Left:} The two counting problems in CDM: the missing satellites problem in the Milky Way \cite{Klypin:1999uc,Moore:1999nt} (top panel) and ``too big to fail in the field'' \cite{2015A&A...574A.113P} (bottom panel) in their original incarnations in velocity space. The predicted curves of CDM (solid lines) are $v_\mathrm{max}$ functions from dark-matter-only CDM simulations.  The missing satellites problem is typically expressed in number counts above a threshold, the ``too big to fail in the field'' problem in terms of a number density of objects above a velocity threshold.  For observations, the corrected velocity dispersions for Milky Way classical dwarfs come from Ref.~\cite{mcconnachie2012} following the correction by Ref.~\cite{Moore:1999nt}, and the peak of the gas rotation curves for field dwarf spiral galaxies come from Ref.~\cite{2015A&A...574A.113P}.  \emph{Right:} Resolution to the missing satellites problem.  The blue shaded region shows the completeness-corrected number of satellites orbiting the Milky Way, assuming an isotropic angular distribution of satellites in the halo \cite{Kim:2017iwr}.  The width of the shaded region reflects ambiguity in the radial distribution of satellites but not does include Poisson uncertainty in satellite counts.  The red region shows Dooley {\em et al.}~(2017)'s \cite{dooley2016} predictions for the satellite counts using the Moster {\em et al.}~(2013) \cite{2013MNRAS.428.3121M} SMHM relation applied to Monte Carlo realizations of a \mhalo$=1.4\times 10^{12} M_\odot$ halo's subhalo population. The width of the prediction curve comes from including early reionization suppression (bottom of region) versus no reionization suppression (top of region). The gap between the completeness-corrected satellite counts and the analytic prescription near $M_* = 10^5 M_\odot$ may result from the tidal disruption of cored satellite galaxies by the Milky Way disk \cite{Brooks:2012ah}.}

\end{figure*}

While the cusp/core problem revolves around the smooth distribution of matter within individual halos, the missing satellites problem and other substructure problems arise from the non-smooth components.  In a hierarchical Universe, large objects grow by the accretion of smaller objects.  However, the accretion of these smaller objects is incomplete in many cases, as is demonstrated by the existence of satellite galaxies and unmixed stellar streams.  Until the late 1990's, simulations did not have enough resolution to reliably identify halo substructure in Milky Way-mass systems.  When the simulations did achieve sufficient resolution,\footnote{However, see Ref.~\cite{vandenBosch:2016hjf} for a serious warning about numerical overmerging even today.} it was discovered that Milky Way-like systems have as much substructure as galaxy clusters do, if normalized to host properties \cite{Klypin:1999uc,Moore:1999nt}. In the context of substructure, because of the ambiguity in the definition of subhalo mass (Section~\ref{sec:primer}) abundances are typically presented in terms of cumulative number of subhalos of a given $v_{\rm max}$ per unit host halo, $N(> v_{\mathrm{max}})$ (Figure~\ref{fig:cdmproblems_msp}), often normalized by the host $v_{\mathrm{max}}^{\mathrm{host}}$, $N(>v_{\mathrm{max}}/v_{\mathrm{max}}^{\mathrm{host}})$.

In Refs.~\cite{Klypin:1999uc,Moore:1999nt}, this substructure velocity function was compared against the velocity function of known Milky Way satellites, which numbered a dozen at the time (see Ref.~\cite{mcconnachie2012} for a compilation of known satellites and their 
properties\footnote{\url{http://www.astro.uvic.ca/~alan/Nearby_Dwarf_Database.html}}). 
Converting the observed line-of-sight stellar velocity dispersion $\sigma_{LOS}$ to $v_{\mathrm{max}}$ is non-trivial, but for many years a conversion of $v_{\mathrm{\max}}$ (or $v_{\mathrm{vir}}$, the virial speed) equal to $\sqrt{3}\sigma_{LOS}$ was made on the assumption that the stellar velocity dispersion tensor was isotropic.  As demonstrated in Figure~\ref{fig:cdmproblems_msp}, there was a significant mismatch in the velocity functions.  If one interprets the mismatch in the velocity functions ``vertically'' (i.e., assuming our match between observational and simulated velocities is perfect, thus indicating that the discrepancy is in the vertical axis of the plot), it suggests that there are far fewer large satellites than expected in CDM.  This was dubbed the ``missing satellites problem."  It inspired the theoretical development of WDM and SIDM models \cite{Spergel:1999mh}, and is a significant driver of the prominence these two classes of theory have in the particle and astronomical worlds. 

Since this time, there have been a number of important developments and restatements of this problem.  Importantly, dozens of new satellites of the Milky Way and Andromeda galaxies have been discovered in the past fifteen years, notably in the SDSS (e.g., Ref.~\cite{Willman:2004kk}), \textsc{PAndAS} \cite{martin2016}, DES (e.g., Ref.~\cite{Drlica-Wagner:2015ufc}), \textsc{Pan-STARRS} (e.g., Ref.~\cite{Laevens:2015kla}), \textsc{MagLiteS} \cite{Drlica-Wagner:2016hwk}, HSC (Hyper Suprime-Cam; \cite{homma2017}), and \textsc{VST ATLAS} surveys \cite{Torrealba2016a}.  Most of these new satellites have properties distinct from the those of the classical (pre-SDSS) satellites: they are tiny ($M_* \ll 10^5\,M_\odot$), ancient (consistent with being reionization fossils \cite{bovill2009,2014ApJ...796...91B}), and extremely low in surface brightness.  All of these properties allowed them to remain hidden before the advent of large, deep surveys.  Taking into account survey selection functions, it is now thought that there could be hundreds (if not thousands) of satellites of the Milky Way (Figure \ref{fig:cdmproblems_msp}) \cite{koposov2008,tollerud2008,walsh2009,bullock2010,hargis2014,Jethwa:2016gra,Kim:2017iwr}.  Simultaneously, the connection between galaxies and the halos they live in has been much better quantified (e.g., \cite{berlind2002,vale2004,2013MNRAS.428.3121M,2013ApJ...770...57B}), as has the effect of reionization to prevent star formation in small halos \cite{Gnedin:2000uj,Bullock:2000wn,Benson:2001at,somerville2002}. While the velocity function of these satellites remains unknown, it is the case that if the mismatch of velocity functions in Refs.~\cite{Klypin:1999uc,Moore:1999nt} is interpreted as missing satellites, then the new discoveries and estimated completeness corrections ameliorate this counting problem essentially entirely (Ref.~\cite{Kim:2017iwr}; Section \ref{sec:baryonproblems}).  

This puts significant pressure on any dark-matter model with primordial effects.  
If the missing satellite problem's initial incarnation in 2000 had a dark matter solution, the particle model would have lead to either primordial or evolutionary deviations on the scale of $10^7\,M_\odot \lesssim M_{\rm halo} \lesssim 10^{10}\,M_\odot$ (though with at least an order of magnitude uncertainty on both ends).  As seen in Figure~\ref{fig:fom}, a number of dark-matter models result in deviations in this regime.  This is perhaps not surprising, because the missing satellites problem motivated many of these models (e.g., Ref.~\cite{Spergel:1999mh}).  However, the reevaluation of this problem in light of the new dwarf galaxy surveys suggests that primordial deviations from CDM can appear on scales likely no greater than \mhalo$\sim 10^8\, M_\odot$ (evolutionary deviations up to cluster scales are still allowed for some models) \cite{Jethwa:2016gra,Kim:2017iwr}.  Refs.~\cite{Jethwa:2016gra} and \cite{Kim:2017iwr} show that WDM models equivalent to a thermal relic mass $m \lesssim 4-8$ keV are under severe pressure. 

The remaining uncertainty in the solution to the missing satellites problem revolves around how and why, precisely, galaxies inhabit halos; and at what scale halos remain totally dark (i.e., devoid of luminous gas and stars). This problem can only be probed by gravity, that is to say, only by astrophysics, rather than particle physics.  We discuss some possibilities in Section~\ref{sec:baryonproblems}.  

One way around the baryonic ambiguities is to find a way to count subhalos regardless of their baryonic content.  This is the approach of strong gravitational lensing \cite{2006glsw.conf...91K,2010ARA&A..48...87T}.  (We will discuss other probes of dark subhalos, including stream gaps, in Section~\ref{sec:future}.) Strong gravitational lensing occurs when the light from a source is significantly distorted by a lens along the line of sight to the observer.  Most dramatically, this leads to ``Einstein ring" or ``Einstein cross" systems, in which the source appears as a ring around the lens or where it is split into multiple images.  It is sensitive to substructure in three different ways, probing three different derivatives of the gravitational potential and thus different properties of the substructure \cite{2009arXiv0908.3001K}:
\begin{enumerate}
\item Substructures can change the positions of images, either in the ring or multiple-image cases  \cite{2009arXiv0908.3001K}; 
\item Substructure can perturb the relative brightnesses of images \cite{1998MNRAS.295..587M}; 
\item Substructure can induce time-delay anomalies \cite{keeton2009}, on account of the extra bending of the light travel path relative to the smooth lens case.
\end{enumerate}
Studies of Einstein ring systems are sometimes referred to as ``gravitational imaging"; tests of substructure using the fluxes of Einstein crosses use ``flux-ratio anomalies" among the images in the crosses.  Currently, the best constraints on the substructure mass function come from flux ratio anomalies and gravitational imaging studies \cite{dalal2002,vegetti2010,2016ApJ...823...37H,nierenberg2017,2017JCAP...05..037B}.  Interestingly, these studies prefer an amount of structure \emph{consistent with or in excess of CDM predictions}, albeit with large theoretical and observational uncertainties \cite{dalal2002}.  In Section~\ref{sec:future}, we discuss near-term prospects to reduce the uncertainties and home in on dark matter physics with strong lensing and new ideas for probes of tiny subhalos in the Milky Way halo using both the phase-mixed stellar halo and individual stellar streams.

The missing satellites and substructure lensing problems are both nearing twenty years old, but have evolved rapidly in just the past few years.  We discuss prospects for continued progress in Section~\ref{sec:future}.  The classic and excellent Refs.~\cite{2010arXiv1009.4505B,2010AdAst2010E...8K} provide more in-depth discussion of the missing satellites problem, Ref.~\cite{Kim:2017iwr} discusses the likely resolution; Refs.~\cite{2006glsw.conf...91K,2009arXiv0908.3001K,2010ARA&A..48...87T} provide illuminating descriptions of strong and substructure lensing; and Ref. \cite{johnston2016} for an introduction to stellar streams. 

\subsubsection*{``Too Big to Fail" (TBTF)}

\begin{figure}[t]
\begin{center}
\includegraphics[width=0.70\textwidth]{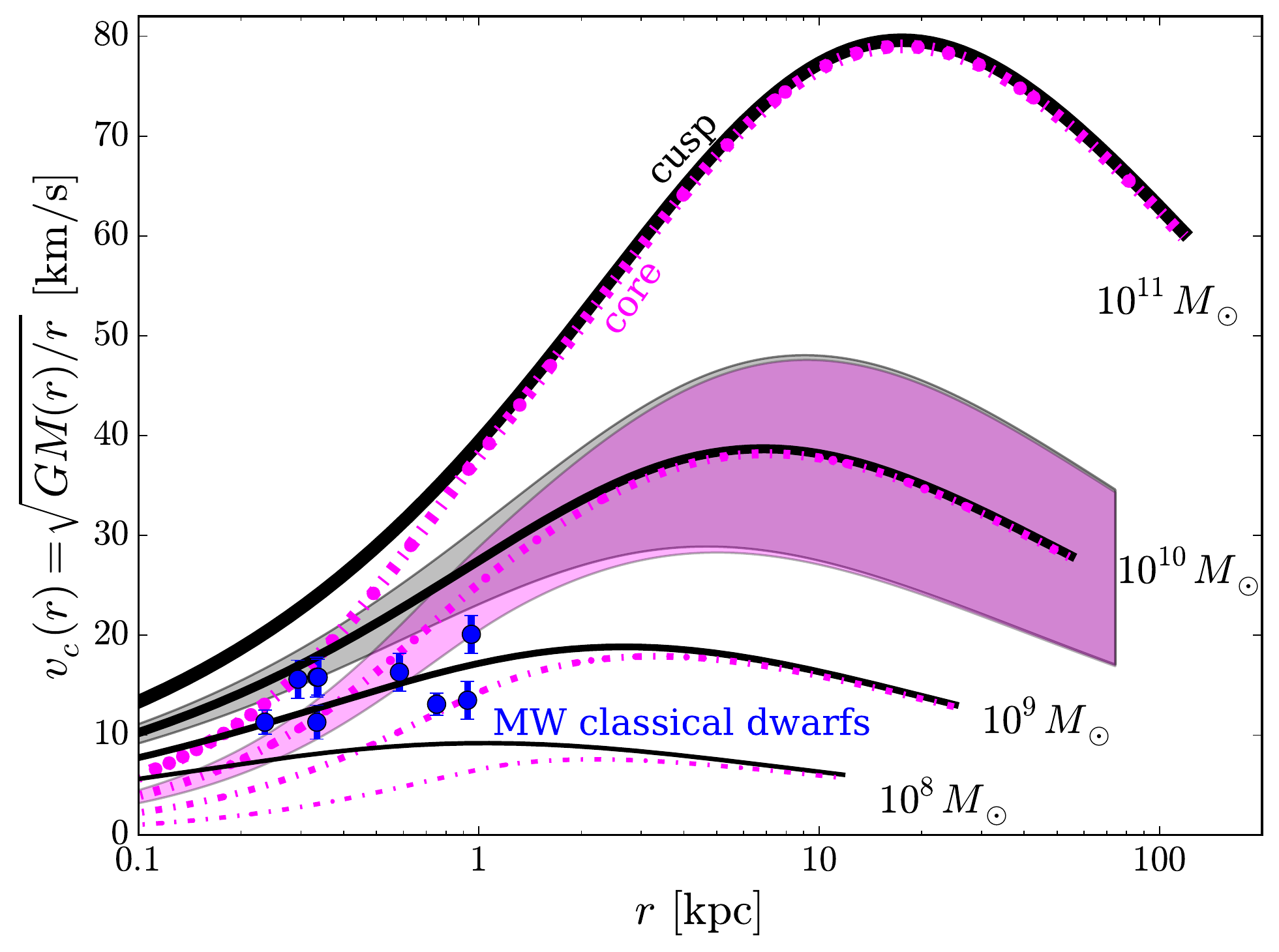} 
\end{center}
    \caption{\label{fig:cdmproblems_tbtf} Illustration of the TBTF problem.  We show enclosed mass curves for dark matter halos, quantified in terms of $v_{c}(r)$; and the Milky Way classical satellite dark masses represented as $v_{1/2}$ vs.~$r_{1/2}$ ($v_{1/2}$ values taken from Ref.~\cite{2016arXiv160706479F} and represented with blue points).  Black curves show predictions for unstripped CDM halos with NFW profiles following the mean mass-concentration relation for subhalos \cite{rodriguez-puebla2016,Okoli2017}.  The dashed magenta lines show the $v_c(r)$ curves of halos with identical mass and NFW scale radius as the CDM halos, but where all halos are able to form 1 kpc cores.  The grey shaded region shows the likely $v_c(r)$ curves for the Milky Way classical satellites according to the SMHM relation of Moster {\em et al.}~\cite{2013MNRAS.428.3121M} (Figure~\ref{fig:cdmproblems_shm}) assuming CDM NFW profiles, and the magenta shaded region is $v_c(r)$ assuming cored profiles.  The TBTF problem is essentially that the grey shaded region does not overlap with the observed $v_c$ values of the classical dwarfs.  We do not include tidal stripping in our estimates, although the Milky Way classical dwarfs are satellites prone to tidal stripping by our galaxy; dark-matter-only simulations show that this is insufficient to reduce the CDM halo central densities of the most massive subhalos to match the observed dark masses.}
 
\end{figure}

An alternative way to view the missing satellites problem is to compare the central densities of satellites derived from observations to those from simulations.  This can be more unambiguously measured and related in simulations and in observations than a relation between $\sigma_{LOS}$ and $v_{\mathrm{max}}$ \cite{strigari2007missingsat,strigari2008,madau2008,madau2008b,walker2009,wolf2010}.  A mismatch similar to the formulation of the missing satellites problem in terms of $v_{\mathrm{max}}$ persists, with CDM predicting many more dense subhalos than the observed number of dense satellites.  This formulation of the missing satellites problem was the direct progenitor of the ``Too Big to Fail'' (TBTF) problem \cite{BoylanKolchin:2011de,BoylanKolchin:2011dk}.  Specifically, TBTF was the observation that the largest subhalos in CDM dark-matter-only simulations had circular velocities larger than those measured in the large Milky Way classical dwarfs (see Figure~\ref{fig:cdmproblems_tbtf}). That is, the most massive subhalos in simulations, which should be ``too big to fail'' to produce stars and therefore should host visible satellites, did not seem to be present in observations.

The TBTF problem is a statement about the central densities of the classical dwarf satellites of the Milky Way ($M_* > 10^5\,M_\odot$, i.e., the classical dwarfs known prior to SDSS) as measured and in simulations.  Although there is a degeneracy between the velocity anisotropy of stellar orbits, the mass profile of dark matter, and the stellar density profile, the degeneracies can be approximately broken at one point in a dwarf galaxy: the half-light radius \cite{walker2009,wolf2010,errani2018}. 

This occurs because (so far) we can only measure line-of-sight velocities of stars, and because there are so few stars, especially going toward and into the ultrafaint dwarf regime.  Let us consider the first point.  Line-of-sight velocities are mostly dominated by radial orbits at the center, and tangential orbits near the outskirts of the galaxy.  At the half-light radius, we sample comparable proportions of stars on radial or tangential orbits.  Because the strongest degeneracy is between mass and the relative distributions of tangential and radial orbits, which are totally degenerate for a spherically symmetric equilibrium system, we can only break the degeneracy if we have the other two components of the velocity vector (discussed in Section~\ref{sec:future}) or if we measure at the half-light radius. However, even with proper motion measurement of individual stars in dwarfs to get the other components of the velocity vector, there is still considerable uncertainty on the density profile because of the low number of stars to trace the gravitational potential \cite{strigari2007,read2017b}.  Fortunately, the degeneracy breaks at one point in the dwarf, and so one may estimate the mass enclosed within approximately the half-light radius \cite{walker2009,wolf2010,errani2018}.  For multiple distinct stellar populations, we can find the mass within the half-light radius for each population, and infer a slope for the mass profile as well \cite{Walker:2011zu}.  

The mass within the half-light radius, $M_{1/2} = M(<r_{1/2})$, can also be reinterpreted in terms of the circular velocity: the velocity of a particle on a circular orbit at the half-light radius ($v_{1/2}$; Eq.~\eqref{eq:vcirc}).  Ref.~\cite{BoylanKolchin:2011de} plotted $v_{1/2}$ for the classical satellites of the Milky Way, as well as the $v_c$ (a proxy for the enclosed mass) curves of the most massive subhalos of Milky Way analogs in several CDM-only simulations.  In principle, one may match an observed satellite to a simulated subhalo if $v_{1/2}$ lies on top of one of the $v_c$ lines from a simulations.  

Based on our expectation from galaxy evolution theory, the biggest satellites (in terms of number of stars) should live in the biggest subhalos.  Successful models of galaxy formation --- at least for large galaxies where they can be empirically tested --- suggest that the stellar mass of a galaxy is highly correlated with halo mass, the SMHM relation (see Section~\ref{sec:primer}).  There is some evidence that the scatter in the SMHM relation increases with decreasing halo mass, through a combination of feedback and reionization regulation of gas flow and star formation \cite{Fitts:2016usl,Garrison-Kimmel:2016szj}, but the overall correlation is observed in simulations down to ultrafaint galaxy scales \cite{Hopkins:2013vha}.  We show the range of (sub)halo masses expected to host the classical Milky Way satellites according to the SMHM relation of Moster {\it et al.}~\cite{2013MNRAS.428.3121M} in Figure~\ref{fig:cdmproblems_tbtf}.  (Although this is empirically tested and constrained for galaxies more massive than the Magellanic Clouds, anything below a stellar mass of $M_* \sim 5 \times 10^7\,M_\odot$ requires extrapolation and assumptions about the surface brightness distribution function of galaxies around that mass scale \cite{2013ApJ...770...57B,2013MNRAS.428.3121M,wright2017}.) 

If we have a strong prior that CDM is the correct phenomenological model for dark matter, that dark-matter density profiles are unaffected by baryonic physics, and that dwarf galaxies follow the extrapolated SMHM relation, we expect the $v_{1/2}$ points of the classical galaxies to lie on the NFW $v_c(c)$ curves of the most massive simulated Milky Way subhalos. However, on the small scales of Milky Way satellites, Ref.~\cite{BoylanKolchin:2011de} found that the largest subhalos in CDM dark-matter-only simulations had circular velocity curves far above the $v_{1/2}$ points for known satellites.  In fact, it appeared that only lower-mass simulated subhalos had circular velocity curves that were more consistent with the observed $v_{1/2}$ points, but even then it was a challenge to match some surprisingly low-density satellites (e.g., Fornax, which is the brightest Milky Way satellite other than the Magellanic Clouds and the disrupting Sagittarius dwarf).  We illustrate this point in Figure~\ref{fig:cdmproblems_tbtf}, showing $v_{1/2}$ for the classical satellites and the  $v_c$ profiles of SMHM-matched subhalos with typical density profiles in CDM.

This observation can be interpreted in several ways.  The TBTF interpretation is that the big subhalos ($\gtrsim 10^{10}M_\odot$) 
are devoid of baryons. Thus, these massive objects are not seen in astronomical surveys, leaving only the smaller subhalos ($\lesssim 10^{10}M_\odot$) 
with stars in the sample of galaxies to be matched with simulation.  However, this idea is at odds with our intuition regarding galaxy evolution.  While we do expect star formation to become stochastic (that is, difficult to predict from halo mass alone) on small scales (\mvir$\lesssim 10^8 M_\odot - 10^9 M_\odot$) \cite{2015ApJ...807L..12O}, we expect that scale to be much smaller than the scale on which TBTF would require that halos be dark \cite{Jethwa:2016gra,Kim:2017iwr,Munshi:2017xhq,2015ApJ...807L..12O}.  Again, this is the source of the moniker ``Too Big to Fail,'' as such massive subhalos are expected to be ``too big'' to fail to produce stars. 

There are three commonly invoked alternative explanations to explain the mismatch between observed $v_{1/2}$ and the CDM-only circular velocity curves.  First, the Milky Way may have a smaller dark matter halo than assumed by Ref.~\cite{BoylanKolchin:2011de}.  The mass of the most-massive subhalos depends on the parent halo's mass, and so reducing the mass of the Milky Way would reduce the expected mass of the largest subhalos orbiting it.  Several have suggested that a Milky Way halo mass of $\sim 8\times 10^{11}\,M_\odot$ has no TBTF problem if CDM-only simulations are to be believed \cite{2012MNRAS.424.2715W,2012JCAP...12..007P,2013MNRAS.428.1696V,jiang2015}.  While such a low halo mass is consistent with some Milky Way mass estimates \cite{2017MNRAS.468.2359W}, it is inconsistent with others \cite{BoylanKolchin:2012xy}.  A low mass also makes the presence of the Magellanic Clouds much more problematic \cite{2011ApJ...738..102T,2011ApJ...743..117B}, and is inconsistent with expectations from the SMHM relation \cite{2013ApJ...770...57B,2013MNRAS.428.3121M}.  

Second, the Milky Way could simply be an outlier among Milky-Way-mass systems \cite{2012JCAP...12..007P,jiang2015}. There is significant halo-to-halo scatter in satellite and subhalo populations \cite{lu2016,geha2017}, so perhaps the Milky Way lies some ways off the mean satellite population.  

The third explanation, one closely tied to the cusp/core problem, is that the mass profile of CDM-only simulated halos may just be a poor match to reality.  Connecting the enclosed mass measured at the half-light radius to a total halo mass requires an assumption about the density profile of the dark matter, since the virial radius is one to two orders of magnitude larger than the half-light radius (in Figure \ref{fig:cdmproblems_tbtf} we plot $v_c$ curves out to the virial radius) \cite{kravtsov2013,somerville2017,huang2017,Hearin:2017cho}. 
 If the density profiles are modified by physics which is not correctly modeled by the CDM-only simulations, the matching between $v_{1/2}$ and $M_\mathrm{vir}$ changes significantly \cite{Vogelsberger:2012ku,Elbert:2014bma,2016arXiv160706479F,vandenBosch:2016hjf}.  Such a modification can be achieved either with new dark-matter physics or via baryonic supernova feedback (see below), but due to the computational cost of high-resolution simulations of galaxies with either baryonic physics or non-trivial dark matter physics, the exact goodness of fit of any solution remains to be determined.  As shown in Figure \ref{fig:cdmproblems_tbtf}, if either dark-matter physics or baryons can produce cores in dark-matter halos, the $v_c$ curves of the halos expected to host the Milky Way classical dwarfs (according to the SMHM relation) pass through the classical dwarf $v_{1/2}$ points. 

Interpreted in the context of dark matter in Figure~\ref{fig:fom}, the TBTF problems suggests we should look for new dark-matter physics in halos of $\sim 10^{10}\,M_\odot$.  New physics could show up on smaller scales than this, but a solution to TBTF specifically must address this scale.  The solution to this problem alone can either be primordial (suppression of moderate-sized halos) or evolutionary (modification of density profiles).

\subsubsection*{The Baryonic Tully-Fisher (BTF) relation and field dwarf counts}

One of the longest standing distance measures for galaxies is the Tully-Fisher (TF) relation \cite{tully1977}.  The distance measure exploits the tight power-law correlation between the intrinsic luminosity of spiral galaxies and their spectroscopically determined rotation speed --- that is, a tight correlation between the luminosity of a galaxy and the depth of its potential well.  The TF relation, as derived from stellar luminosity, has much higher scatter for small galaxies (roughly at the scale of the Large Magellanic Cloud, $M_{\rm halo} \approx 10^{11}\,M_\odot$), because such galaxies are so gas-dominated. Big galaxies have comparatively little cold gas, but low-mass isolated field dwarf galaxies are, to first order, clouds of gas lightly sprinkled with stars \cite{2006ApJ...653..240G,2012ApJ...759..138P,2015ApJ...809..146B}.  When cool gas and stars are summed, the tight power-law correlation between the rotation velocity of galaxies and baryonic mass reappears \cite{2014ApJ...782..115M,2016ApJ...832...11B}, called the baryonic Tully-Fisher (BTF) relation \cite{2000ApJ...533L..99M}.

\begin{figure}[t]
\begin{center}
	\includegraphics[width=0.70\textwidth]{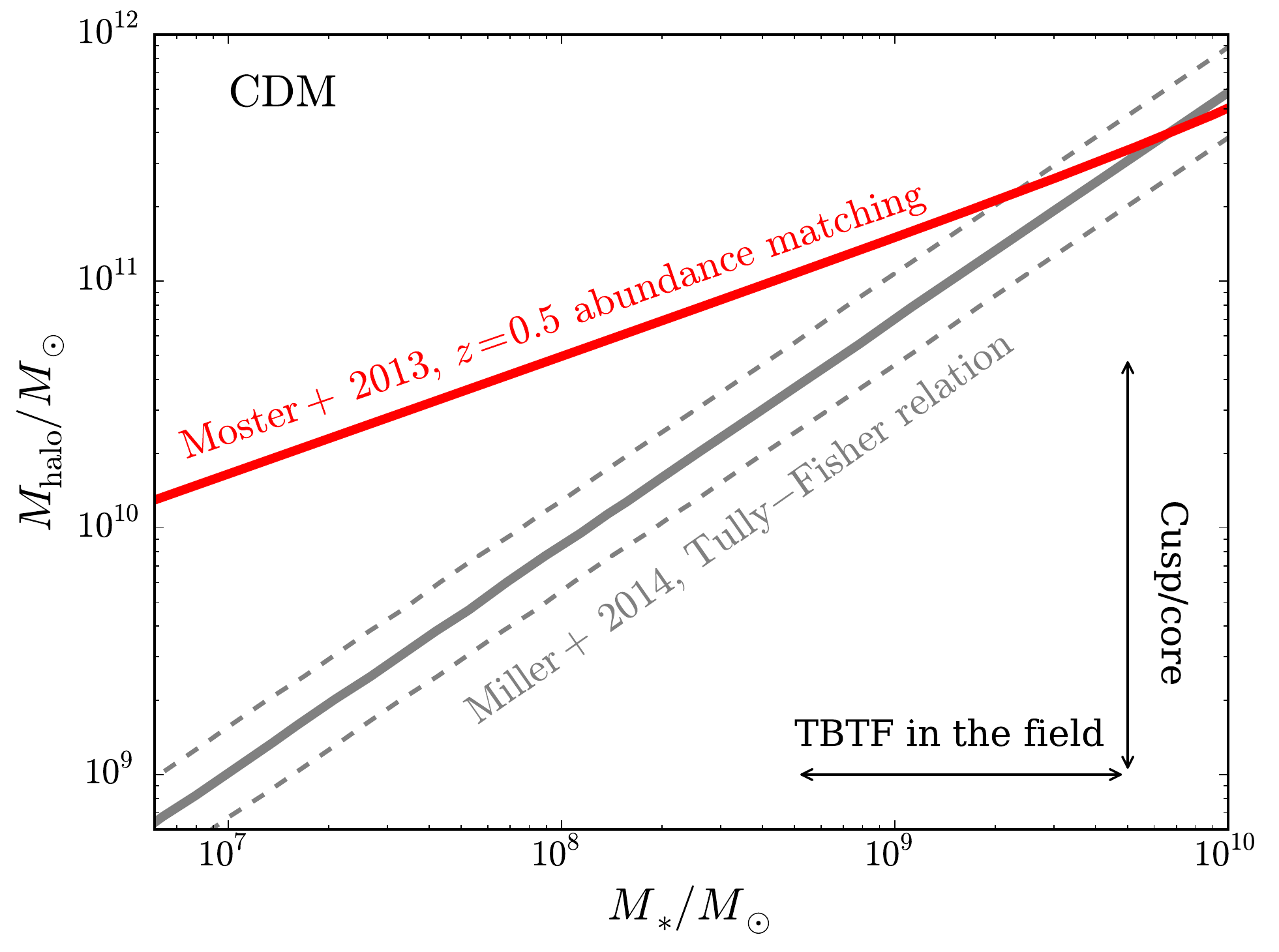}
	\end{center}
    \caption{\label{fig:cdmproblems_shm} The stellar-mass--halo-mass (SMHM) relation inferred using the star-formation-rate matching method of Moster {\em et al.}~(2013) \cite{2013MNRAS.428.3121M} relative to the BTF measurement of Miller {\it et al.}~(2014) \cite{2014ApJ...782..115M} for a similar redshift range.  The velocity rotation curves derived by Miller {\it et al.}~for their moderate redshift sample are mapped onto virial masses according to the empirical optical-to-virial velocity relation, as determined for higher-mass galaxies, by Ref.~\cite{2012MNRAS.425.2610R}.  We show the cusp/core and TBTF in the field interpretations of the mismatch in the SMHM relations depicted here.}
\end{figure}

However, the TF relation (shown in Figure~\ref{fig:cdmproblems_shm}, mapping from rotation curves to \mhalo) is in tension with other measurements of the relation between the halo mass and baryonic mass of galaxies.  If the furthest measured point of the rotation curve is a good tracer of the halo potential (allowing a mapping between velocity and halo mass), it appears that dwarf galaxies live in systematically smaller halos than expected from the extrapolated SMHM relation \cite{2012ApJ...759..138P,2012MNRAS.425.2817F,2014ApJ...782..115M,2016A&A...593A..39P,2017MNRAS.464.2419S}.  This again suggests that if CDM is the correct description of dark matter, there must be a ``too big to fail'' problem for small galaxies in the field (that, is far from a large Milky Way-type galaxy) not just for satellite galaxies of the Milky Way.  As there are many more small halos than big halos in CDM, if the TF relation is to hold, there must be many halos which are dark but relatively massive. 

This counting problem shows up in field in other ways.  Using the \textsc{Alfalfa} radio data (measuring 21 cm radiation from atomic hydrogen, and used to quantify the BTF relation \cite{2016A&A...593A..39P}), Ref.~\cite{2015A&A...574A.113P} turned their galaxy counts into a velocity function for these isolated galaxies, just as done for the missing satellites work.  They found a flattening of the velocity function below 80 km/s relative to the CDM $v_{\mathrm{max}}$ function (Figure~\ref{fig:cdmproblems_msp}, bottom panel).  A similar discrepancy was found on a similar scale by Ref.~\cite{2015MNRAS.454.1798K}, which focused on optically selected galaxies within 10 Mpc from the Milky Way rather than the radio-selected sample of Ref.~\cite{2015A&A...574A.113P}.  These measurements again suggest a ``Too Big to Fail'' problem in the field. 

One of the major problems with this interpretation is the difficulty in relating the observed velocity data to halo masses.  There is a close relation between the characteristic speed of the rotation curve of large galaxies and the virial velocity (Eq.~\eqref{eq:vvir}) 
of their hosts \cite{2012MNRAS.425.2610R}.  However, we know this because we have a number of different tools available to measure halo masses for large galaxies, notably galaxy-galaxy weak gravitational lensing \cite{Mandelbaum:2005nx}.  For small halos, fewer tools are available, and the mapping between observed rotation speed and halo mass is more ambiguous.  Often, the rotation curves are still rising when the gaseous or stellar tracers disappear, well within the radius at which $v_{\mathrm{max}}$ occurs \cite{2017MNRAS.464.2419S}.  Typically, even if the rotation curves are not well-matched to NFW density profiles, the last point on the rotation curve is used to assign galaxies to a halo mass under the assumption that the NFW profile is a reasonable description outside that radius \cite{2012MNRAS.425.2817F,2015A&A...574A.113P}, which may not be the case.  If the halos are cored rather than cusped, we break the neat relation between the measured rotation curve and the virial mass, and in fact a low-amplitude rotation curve can be mapped to a halo mass that is substantially higher than inferred if one assumes that the halos have NFW density profiles (Figure \ref{fig:cdmproblems_tbtf}).  Thus, the mismatch between the predicted and observed velocity functions, or the TF-derived SMHM relation and the abundance-matching prediction, may be hints of a cusp/core problem instead.  As we will see below, the interpretation of the problem, additional systematic uncertainties, and the necessity of either new baryonic or dark-matter-physics solutions are hotly debated \cite{read2016,2016arXiv161009335T,2016arXiv161109362S,2016arXiv161102716K,2017MNRAS.468.2283C,2017arXiv170107835B}.  

Interpreted in the context of our astrophysical parameter $M_{\rm halo}$, solutions to the BTF problem in the form of new dark matter physics are needed for the $M_{\rm halo} \sim 10^9 \,M_\odot - 10^{11}\,M_\odot$ mass range.  

\subsubsection*{Are all these problems aspects of one central problem?}\label{sec:alloneproblem}

\begin{figure*}
\begin{center}
\includegraphics[width=0.95\textwidth]{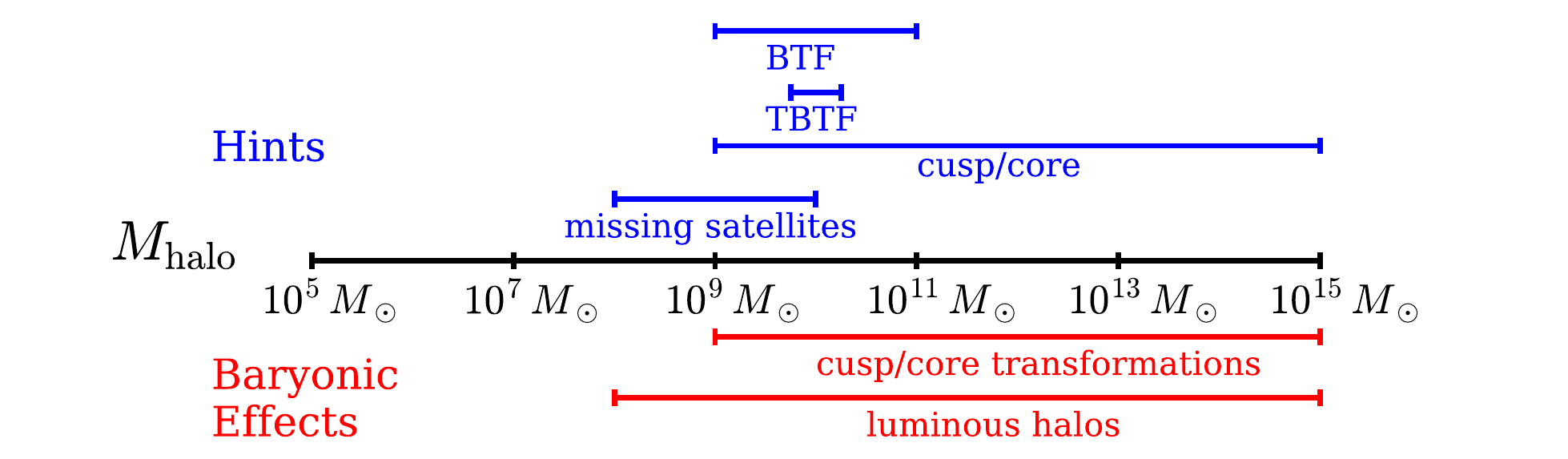}
\end{center}
\caption{\label{fig:summary4} Summary of the halo mass $M_{\rm halo}$ where hints of deviations from $\Lambda$CDM have been claimed, as well as the range of $M_{\rm halo}$ where baryons are expected to influence the structure of halos. ``TBTF'' is ``Too big to fail'' and ``BTF'' is the baryonic Tully-Fisher relation.}
\end{figure*}

We described the four most commonly cited problems with CDM on small scales, summarized in Figure~\ref{fig:summary4}.  However, we argue here that these problems are different views of one central problem, which is that the relationship between the mass of galaxies and that of their dark matter halos is uncertain on small scales.

Why is this the central problem, and how will a reduced uncertainty help illuminate dark matter microphysics?  Let us consider first the missing satellites problem, the TBTF problem, and the BTF relation.  The existence of these problems as they are stated depends on an accurate mapping from the observed kinematics of baryons to a halo mass.  If we assume that the currently proposed mappings are correct, then the horizontal axes in Figure~\ref{fig:cdmproblems_msp} and the vertical axis in Figure~\ref{fig:cdmproblems_shm} are correct and the discrepancy between the velocity functions in the Milky Way or the field can be interpreted by their difference in the vertical (Figure~\ref{fig:cdmproblems_msp}) or horizontal (Figure~\ref{fig:cdmproblems_shm}).  That is, there is a mismatch in simulated CDM halo and observed galaxy abundances; the problem is a counting problem.  

However, if the mapping between the internal kinematics of galaxies and their host halo is incorrect, either because of systematics or because the dark matter mass profile of halos is fundamentally different from CDM-only simulations, then the problem is instead in the horizontal axis of Figure~\ref{fig:cdmproblems_msp} and vertical axis of Figure~\ref{fig:cdmproblems_shm}.  It turns the problem into a cusp/core problem.  If interpreted as such, the important implication is that by assuming CDM-only density profiles, we underestimated the halo mass for fixed baryonic mass or for fixed baryonic kinematics (Figure~\ref{fig:cdmproblems_tbtf}).  Cored halos generically have higher ratio of virial velocity (or virial mass) to the velocity of the baryonic tracer in the inner part of the halo than do cusped halos.  Thus, if we understood exactly what types of halos are inhabited by different types of galaxies, there would be no ambiguity in interpreting galaxy abundances in the context of dark matter theories---we would know if we had a serious counting problem on our hands, or mostly a cusp/core problem.

Likewise, if we knew how to match galaxies with halos, we could assess the origin of the cusp/core problem.  Currently, baryonic solutions to this problem in the context of baryonic feedback (discussed below) depends on how much energy is available in the baryonic sector to push the dark matter around in halos \cite{penarrubia2012,read2016}.  The amount of energy required to push dark matter out of halo centers depends on the depth of the potential well, which depends on the halo mass.  If we knew how to match galaxies to halos, we could determine if the pattern of central dark matter densities as a function of galaxy mass were consistent with a CDM + baryon physics interpretation, or not.

Thus, pushing direct measurements of halo mass down to $M_{\rm vir}\lesssim 10^{11}\,M_\odot$ with, for example, galaxy-galaxy lensing \cite{zu2015} (see Section~\ref{sec:future}) should be a \emph{high} priority for the field.  Because several of the small-scale ``problems" are Milky Way-specific, we also need better constraints on the Milky Way halo mass, which may be obtained in the future with distant satellite orbits and with stellar streams \cite{BoylanKolchin:2012xy,bonaca2014,price-whelan2014,johnston2016,sanderson2017a}.

\subsection{The Known Unknowns of Theoretical Predictions for Astrophysical Dark Matter: It's All About the Baryons}\label{sec:baryonproblems}

These four discrepancies between the dark matter predictions and observations have been immense interest to both astrophysicists and particles physicists, driving the creation of a large number of dark matter models which modify dark matter halos on the scale of $M_{\rm vir} \sim 10^8\,M_\odot - 10^{15}\,M_\odot$. However, the biggest known problem for dark matter structure predictions is that most predictions are made without baryons.  Although baryons are relatively minor contributors in the energy budget of the Universe, they are dynamically important precisely in the parts of the Universe where  dark-matter-induced deviations from CDM might appear.  Thus, a true mapping between dark matter models and \mhalo\, requires an accurate treatment of baryons.  Here, we consider the impact of baryons on the ``hints'' described above, and highlight current uncertainties.  We focus largely on hydrodynamic simulations with CDM, as hydrodynamic simulations with other types of dark matter are still in their infancy \cite{Vogelsberger:2014pda,Fry:2015rta,2017MNRAS.468.2836L}.  We turn to non-CDM hydrodynamic simulations at the end of the section.

An alternative way to cast this discussion is in the context of Figure~\ref{fig:fom}, where baryons are placed at $M_{\rm halo} \sim 10^8\,M_\odot - 10^{15}\,M_\odot$ --- that is, we expect the existence of baryons, and in particular the non-trivial interactions that baryonic physics allows (scattering, cooling, stellar formation, supernovae, {\em etc.}) to modify the structure of dark matter halos in this mass range.  Baryons are a form of matter, with a unique microphysical description.  If we want to understand the microphysics of the particles making up astrophysical systems (which include baryons) from the observations of the structure of these macroscopic systems, we need a dictionary to translate between the particle and galactic scales.  Simulations serve as this dictionary, and so we we discuss the current state of the art.

\subsubsection*{Cusp/Core Problem}
When hydrodynamic simulations were first performed, it appeared that baryons exacerbated the cusp/core problem.  Baryons cooled efficiently in halos, and dragged dark matter in with them as they deepened the gravitational potential wells of their systems.  Thus, halos appeared denser in hydrodynamic simulations than in dark-matter-only simulations \cite{Blumenthal:1985qy,Gnedin:2004cx}.  There were additional problems with the simulations at the time, namely that they were \emph{too} efficient at forming stars (this is still somewhat of a problem on Milky Way scales), and that the disks of galaxies were too dense (the ``angular momentum catastrophe") \cite{governato2004}. 

Ref.~\cite{Governato:2009bg} showed that a key reason for these problems was insufficient spatial resolution.  Star formation in simulations is implemented using semi-analytic prescriptions rather than \emph{ab initio} with all relevant physics self-consistently implemented (star formation remains a very active area of research \cite{krumholz2015}).  Star formation is inherently clumpy, taking place in individual molecular clouds ($\sim$ pc scale).  Due to its spatial concentration, the radiation and kinetic energy flux from the death throes of stellar evolution have a (relatively) easy time ``punching'' through the interstellar medium and out of the galaxy, removing significant quantities of gas in the process.  However, the simulations of the early 2000's were averaging star formation over very large scales relative to the scales on which stars actually form. As a result, in these simulations, energy and gas had a hard time getting out of the interstellar medium, and stars formed very efficiently.  When the simulations of Ref.~\cite{Governato:2009bg} and others began resolving star formation at more realistic scales, star formation became clumpy, energy and gas escaped from the interstellar medium, and dark matter halos responded by becoming \emph{less dense} than the CDM-only simulations predicted \citep{Pontzen:2011ty}.

Despite differences in code and the implementation of semi-analytic star formation prescriptions, there are now largely consistent results about the sign of the effect of baryons on halos on dwarf scales \citep{Governato:2012fa,Brooks:2012ah,Brooks:2012vi,Zolotov:2012xd,Teyssier:2012ie,Arraki:2012bu,dicintio2014,schaller2015,tollet2016,dutton2016,read2016,2016arXiv160706479F}.  However, there are disagreements in the details.  In some simulations, halos show true cores, while in others they appear to have low-density cusps (Figure~\ref{fig:cuspcore}).  Moreover, below some mass scale, there are never enough baryons at the centers of halos to be dynamically important.  On cluster scales, simulations show anything from steeper-than-NFW cusps to mild cores \cite{martizzi2013,schaller2015,dutton2016}.  Determining why there are these differences among simulations, and which (if any) is closest to reality, is the subject of vigorous ongoing activity. 

Thus, there are indications that including baryons in simulations goes at least part of the way to resolve the cusp/core problem. It is as yet unclear if baryons are solely responsible.  The halo mass scale on which baryons can affect the central density of dark matter is around $M_{\rm halo} \sim 10^{9}\,M_\odot - 10^{15}\,M_\odot$, where the bottom of the range is uncertain by about an order of magnitude because of the uncertainty on the scale on which baryons significantly affect halo structure \cite{penarrubia2012,amorisco2014,read2016}.

\subsubsection*{``Missing Satellites" problem, and other substructure problems}
Baryonic solutions to the missing satellites problem involve three separate physical effects: 
\begin{enumerate}
\item subhalo mass functions and survival, 
\item the mapping between the line-of-sight velocities of individual stars in galaxies and the $v_{\mathrm{max}}$ of dark-matter halos, and
\item the probability that some subhalos have no visible baryons in them at all.
\end{enumerate}
Together, these solutions give strong evidence that the missing satellites problem is resolved in the context of CDM, but also that new methods are required to test the halo mass function below \mvir\ $\sim  10^8 M_\odot$.  Galaxy evolution theorists consider the missing satellites problem solved.

Why?  The missing satellites problem was initially formulated in the context of dark-matter-only simulations.  However, baryons can significantly reduce the abundance of dark matter halos at fixed $v_{\mathrm{max}}$ in several ways.  First, baryonic outflows from halos can prevent halos from growing by accretion as fast as they would in the absence of baryons. This is a $\sim 20\%$ effect on the mass function (although it varies between simulation codes) \cite{Sawala:2012cn,Despali:2016meh}.
Second, once halos fall into larger halos, the small halos are more easily destroyed in hydrodynamic simulations as compared to their dark-matter-only analogues.  This is largely driven by the extra tidal field of the central galaxy in the host \cite{Garrison-Kimmel:2017zes,Brooks:2012ah,Brooks:2012vi}, but also because the lower central density of subhalos in hydrodynamic simulations makes satellite galaxies more vulnerable to tidal heating or disruption \cite{Penarrubia:2010jk,Brooks:2012vi,Dooley:2016ajo,errani2018,donghia2010}.  This leads to a ${\cal O}(10\%)$ reduction in the subhalo mass function.  However, neither of these effects is the dominant driver of the solution to the missing satellites problem.  

Next, we consider the mapping of the observed properties of galaxies to those found in simulations.  In Section~\ref{sec:hints}, we saw that the missing satellites problem was conceptualized as a velocity problem, and that we needed to relate line-of-sight stellar velocity dispersion $\sigma_{LOS}$ to the halo $v_{\mathrm{max}}$ to compare an observed property of a galaxy to a halo.  It is now possible to bypass the $v_{\mathrm{max}} \to \sigma_{LOS}$ conversion in hydrodynamic simulations, and instead compared the simulated stellar $\sigma_{LOS}$ to  observed values.  Recent high-resolution simulations now show good agreement between the $\sigma_{LOS}$ distribution functions of simulated and observed satellites \cite{Sawala:2015cdf,Wetzel:2016wro,2016arXiv160706479F}.  If the missing satellites problem is cast in terms of a luminosity or stellar mass function, there is a similar level of agreement on classical dwarf scales \citep{Zolotov:2012xd,Sawala:2015cdf}.  However, these simulations can only resolve classical satellites, not yet the ultrafaint galaxies that are expected to dominate the Milky Way satellite luminosity function below a stellar mass $M_* \lesssim 10^5\,M_\odot$  \cite{tollerud2008,koposov2008,hargis2014,Jethwa:2016gra,Kim:2017iwr}.  Thus, the velocity functions are now a good match to classical satellites, but there are still many more subhalos in simulations than ultrafaint dwarfs orbiting the Milky Way.

The dominant solution to the missing satellites problem comes when we consider which halos are actually occupied by visible baryons, which is especially important for ultrafaint dwarfs.  It is an interesting and instructive historical note that the existence of ultrafaint dwarfs was predicted before their discovery in SDSS \cite{Bullock:2000wn,Gnedin:2000uj,Benson:2001at}.  The origin of this prediction resulted from the following arguments.  Galaxies form from gas enriched by the first generation of stars, and cannot form in arbitrarily small halos \cite{bromm2009,bromm2011}.  Even if molecular hydrogen and metals are only minute constituents of the baryon budget before reionization ($z \gtrsim 6$), it can allow gas to collapse and cool in halos of \mvir\ $(z)\,\sim \,10^6 \,M_\odot - 10^7\,M_\odot$ ($v_{\mathrm{max}} \sim  \hbox{ a few km/s}$) \cite{Thoul:1994ir,Tegmark:1996yt,2015ApJ...807L..12O}.  There are thousands of subhalos larger than those in the Milky Way today.  

However, and very importantly, the reionization of the Universe sets an important limit on how many of such small galaxies may form.  When the Universe reionizes, the intergalactic medium is heated and pressurized; it can no longer collapse into small halos, and what gas exists in small halos at reionization may be boiled out \cite{Barkana:1999apa}.  This has two consequences.  First, few halos reach the halo mass threshold for cooling before reionization, so only a minority of halos host ancient stellar populations.  Second, galaxies forming after reionization can only form in much larger halos ($v_{\mathrm{max}} \gtrsim 20\hbox{ km/s}-50\hbox{ km/s}$) than before \cite{Gnedin:2000uj}.  Depending on the exact prescription for star formation and the redshift of reionization, the Milky Way is estimated to have a few hundred to a few thousand reionization fossils in its halo in a CDM cosmology \cite{Peter:2010sz,lunnan2012,starkenburg2013,dooley2016,Kim:2017iwr}.  In fact, the number of completeness-corrected dwarf satellites projected to inhabit the Milky Way halo based on present-day observations matches almost exactly the number predicted by the SMHM relation and simulation-tested reionization models \cite{Kim:2017iwr}, shown in Figure~\ref{fig:cdmproblems_msp}. 

Where does this leave us?  The current state of theory and simulations suggest that baryon physics solves the missing satellites problem.  It should be noted, though, that there do not exist fully self-consistent simulations of the Milky Way satellite population down to ultrafaint scales (see Ref.~\cite{jeon2017} for a novel variant of standard ``zoom-in" simulations).  

The new frontier for tests of primordial deviations from CDM is on the \mvir\ $\sim 10^8\,M_\odot$ scale and below, where halos are expected to be sans baryons in galaxy evolution theory.  Methods that do not rely on luminous tracers, like substructure lensing or stellar stream perturbations, are the future.  However, the highest resolution hydrodynamic CDM simulations focus almost exclusively on Milky Way analogs or isolated dwarfs.  The host halos of strong lenses tend to be significantly more massive than the Milky Way, and require resolution down to $M_{\rm vir} \lesssim 10^6\,M_\odot$ scales. We discuss this point in greater detail in Section~\ref{sec:future}.  Note that evolutionary deviations are possible on scales above $10^8 M_\odot$.   

\subsubsection*{Too Big to Fail}
As with the missing satellites problem and the core/cusp problem, hydrodynamic CDM simulations indicate that baryons play an important role in shaping the central densities of Milky Way satellites to solve TBTF \cite{Wetzel:2016wro,2016arXiv160706479F}.  This is not surprising, given the core/cusp interpretation of TBTF; if baryons can push dark matter out of the cores of halos, then the central densities of halos can be low even if the total halo mass is high.  However, the surprisingly low densities of Fornax (a \emph{big} satellite), Canis Venatici I, and Sextans remain difficult to reconcile with simulations of CDM + baryons \cite{Simon:2007dq,walker2009b,Sawala:2015cdf,Wetzel:2016wro}.  The conclusion is that baryons can address TBTF on halo mass scales \mhalo$\,\sim 10^{10}\,M_\odot$, but are still not bringing about perfect agreement.  Fornax, in particular, remains an intriguing problem \cite{Wang:2015fia}.

\subsubsection*{Baryonic Tully-Fisher Relation}
This arguably is the hardest problem to solve in the context of baryons.  Let us consider problems with BTF separately from the problem found in the number of dwarf galaxies in the field.  Despite the relatively good agreement between observations and the \textsc{Apostles} simulations for the TBTF problem \cite{2016arXiv160706479F}, the field galaxies in this simulation suite consistently show high rotation curves for fixed stellar mass relative to observations, below $M_* \sim 10^8\,M_\odot$ ($v_{\mathrm{max}} \lesssim 40 \hbox{ km/s}$) \cite{2017MNRAS.464.2419S}.  This discrepancy persists even when the authors ``measure" (i.e., use the rotation speed derived from the simulated mass distribution) the simulated rotation curves in the same regions of the galaxies within which rotation curves are measured for real galaxies.  The problem can be stated one of two ways.  If one believes the simulated stellar masses are accurate, the problem is that the simulated dark-matter halos are too dense.  On the other hand, if we believe the velocity numbers instead, it suggests that star formation remains too efficient for fixed halo mass.   We return to this point after we discuss the counting problem.

The counting problem begins for even higher halo masses, $v_{\mathrm{max}} \lesssim 80\hbox{ km/s}$, which in the context of the Milky Way corresponds to an analog of the Magellanic Clouds (Figure~\ref{fig:cdmproblems_msp}).  It is curious that the missing satellites problem can be solved on those scales, while the problem in the field begins on higher mass scales.  While other attempts to reconcile the counting problem in the field below this scale in the context of CDM have not worked well \cite{2016A&A...591A..58P,2016arXiv161009335T}, a new study by Ref.~\cite{2017arXiv170107835B} indicates that the problem may be the way in which simulators compare their results to observations.  Many groups create rotation curves using the distribution of simulation gas, star, and dark matter particles or fields.  Instead, Ref.~\cite{2017arXiv170107835B} ``observed" the atomic hydrogen in their simulated galaxies, analyzing their simulated observations the same way as real observations.  The agreement between theory and observation was greatly improved, especially for the lowest-atomic-gas-mass objects.  The origin of the agreement is that much of the highest-velocity material in small galaxies is in gas whose column density is too low to produce flux above the background level of the experiment, and thus the rotation curves for these dwarfs is systematically observationally underestimated.  For the smallest gas-containing galaxies, all of the material is too low in density to be picked up with the current 21-cm surveys.  

The idea that there are systematics remaining in comparing the velocity profiles of real and simulated galaxies is appealing, in no small part because other simulation methods of matching galaxies to halos work remarkably well. In particular, the relationship between stellar and dark matter mass (the SMHM relation; Section~\ref{sec:primer}) found in hydrodynamic + CDM simulations \cite{Hopkins:2013vha,Munshi:2017xhq,read2017} shows good agreement with extrapolations from empirical measurements for larger masses \cite{2013MNRAS.428.3121M} and inferences based on Local Group satellite populations \cite{dooley2017,Kim:2017iwr}. It is attractive therefore to assume that the problem lies not with the efficiency of star formation in simulation, but rather in the interpretation of rotation curves. These lines of evidence point to the problem with BTF lies in the assignment of galaxies to halos based on rotation curves, both on the simulation and observational sides.   

\subsubsection*{Baryons and non-CDM models}

Our figure of merit $M_{\rm halo}$ encodes the characteristic size of dark matter haloes where non-gravitational physics enters and causes a departure of the power spectrum or internal properties of the haloes from the predictions of CDM. The interpretation of the hints in Section~\ref{sec:hints} depends at minimum on accurate predictions for CDM \emph{with} baryons, to set the benchmark for CDM cosmology.  Tests of specific dark-matter models other than CDM must also incorporate the physics of baryons in order to define \mhalo, the halo mass scale(s) at which deviations from CDM arise.  

At present, there are many more non-CDM, dark-matter-only simulations than simulations with baryons (e.g., Refs.~\cite{Vogelsberger:2012ku,Rocha:2012jg,Peter:2012jh,Buckley:2012ky,RindlerDaller:2012vj,Suarez:2013iw,Lovell:2013ola,Bozek:2015bdo,Horiuchi:2015qri,Vogelsberger:2015gpr,Bose:2016irl,Mocz:2017wlg,Robertson:2017mgj}).  Most of the deviations from CDM, quantified by \mhalo, shown in Figure~\ref{fig:fom} and Section~\ref{sec:fom} were made in the context of dark-matter-only simulations.

As with CDM, the field is transitioning from making accurate predictions for the wrong problem (increasingly high-resolution studies of structure formation without baryons) to increasingly precise and more robust predictions for the right problem (simulations of dark matter {\em and} baryons).  In the context of dark matter physics, the two classes of astronomical systems simulated most often are isolated dwarf galaxies \cite{Governato:2009bg,Teyssier:2012ie,Madau:2014ija,onorbe2015,wheeler2015,Munshi:2017xhq,Vogelsberger:2014pda,Fry:2015rta,Governato:2014gja,Chau:2016jzi,robles2017,Lovell:2017eec} and Milky Way analogs \cite{Zolotov:2012xd,Hopkins:2013vha,Wetzel:2016wro,2016arXiv160706479F,Lovell:2016fec,2017MNRAS.468.2836L,DiCintio:2017zdz} (although some are simulated with simplified models of the baryonic component \cite{elbert2016,2017MNRAS.468.2283C}).  The former is a popular choice for simulations because one may achieve much higher spatial resolution for fixed computation time than for more massive systems.  Hydrodynamic simulations are much more computationally costly to run than dark-matter-only simulations ($\sim$ an order of magnitude more, even without potentially important physics like radiative transfer), so this is a practical strategy.  The latter is popular because of the Milky Way-centric missing satellites and TBTF problems.   

How are non-CDM hydrodynamic simulation results different than those without galaxy evolution physics?  The most important finding so far is that baryon physics brings CDM and non-CDM predictions closer together, at least when it comes to halo density profiles and bright galaxy counts.  As pointed out in Ref.~\cite{Kaplinghat:2013xca}, dark matter with self interactions behaves as a fluid, and is in hydrostatic equilibrium unless the system is undergoing a major merger.  Thus, in baryon-dominated central potentials, the self-interacting dark matter nearly forms a cusp unlike the low-density cores found in dark-matter-only simulations \cite{Vogelsberger:2014pda,elbert2016,2016arXiv161102716K,2017MNRAS.468.2283C,DiCintio:2017zdz,Robertson:2017mgj}. For non-baryon-dominated systems, there remain differences in the density profiles between CDM and SIDM halos, but they are not as extreme as the dark-matter-only simulations suggest \cite{robles2017}.  Galaxies in cosmologies with a truncation in the matter power spectrum also show convergent evolution if residing in halo masses above the cutoff \cite{Lovell:2016fec,2017MNRAS.468.2836L,robles2017}, even if there are some differences in the star-formation histories at early times \cite{Governato:2014gja,Chau:2016jzi,Lovell:2017eec}.  However, any cutoff in the power spectrum does lead to distinct differences in galaxy and halo counts regardless of baryon physics \cite{Kim:2017iwr}.

\vskip 0.3cm

In summary, we described a number of small-scale hints that CDM is an inadequate description of dark matter on scales of $10^9\,M_\odot\lesssim M_{\rm vir} \lesssim 10^{15}\,M_\odot$. These hints are shown in Figure~\ref{fig:summary4}.  They highlights a region in Figure \ref{fig:fom} for which dark matter candidates with novel physics may warrant significantly more study.  However, we also know of another form of matter, baryons, that also affect halos on these scales (again, see Figure~\ref{fig:summary4}).  Baryons are important for $10^8\,M_\odot \lesssim M_{\rm vir} \lesssim 10^{15}\,M_\odot$ and above.  It is not clear if baryons can solve all small-scale ``problems'' with CDM, but the physics goes in the direction of better, rather than worse, agreement.  Baryons also affect non-CDM predictions significantly, especially with respect to halo density profiles.  Critically, we argued that a central observational problem preventing further progress in determining the dark or baryonic origin of these hints is the fact that the mapping between galaxies and the halos they inhabit is still highly uncertain on the dwarf galaxy scale.

Ultimately, Figure~\ref{fig:fom} must be made in the context of dark matter + baryon predictions.  Regardless of what the true nature of dark matter is, its cosmic distribution is governed by the gravitational effects of baryons in addition to dark physics, and so the baryons must be well understood before non-gravitational interactions in the dark sector can be identified or ruled out with any confidence. We now turn to discussing what the future prospects are for this effort.

\section{Future Directions \label{sec:future}}

In this section, we lay out a plan for dark matter constraints from astronomy.  As the previous sections have demonstrated, the space of dark matter theory is enormous, and encompasses particles far removed from the canonical CDM or WIMP dark matter picture. Critically, an entire axis characterizing the effects of dark sector physics on astrophysical 
structures (\mhalo\, in Figure~\ref{fig:fom}), is orthogonal to the particle interactions which have been the primary interest of many dark matter physicists (characterized by $\Lambda^{-1}$ in Figure~\ref{fig:fom}). The study of dark matter structures has to-date been the source of all positive statements about the properties of dark matter, and has informed many of the constraints on the possible nature of this mysterious material.  Additionally, we showed in Section~\ref{sec:observations} that there are tantalizing hints for non-minimal dark matter on a wide variety of halo mass scales, but we also highlighted some of the present-day systematic uncertainties in turning these hints into measurements of dark matter microphysics.

Here, we advocate a specific plan to turn the promise of measurements of dark matter from astronomy into reality.  We focus on two specific paths.  First, we outline a theory program to make the mapping between the specific types of dark matter models described in Section~\ref{sec:fom} and \mhalo\,(and, more generally, what observables are relevant on those scales) much more precise.  Importantly, we will show why this program is relevant for particle physicists even if dark matter turns out to be a vanilla WIMP.  Second, we describe some ideas of how to leverage the next decade's impressive increase in the quantity and quality of large astronomical survey data for dark matter science, to tightly constrain deviations from CDM as a function of \mvir.  We highlight the parts of the \mhalo\, space which are especially in need of new ideas for observations.  Our main message is that the capabilities to make dramatic improvements in astrophysical dark matter constraints are here now, both in theory and observation.  

\subsection{Theory}

The ability of astronomical observations to measure dark-matter physics is predicated on an accurate mapping between particle theory space and astronomical observable, which is tied to a specific \mvir\, scale or scales.  Building on Section~\ref{sec:baryonproblems}, we advocate for a specific theory program, highlighting how this approach leads to improvements in traditional astroparticle dark-matter searches as well as for measurements of dark-matter microphysics with astronomy.  The key is accurate theoretical predictions for structure formation for different classes of dark matter models that include the effects of galaxy evolution physics on the dark matter distribution, and that these predictions are tuned to observationally relevant scales.

What are the necessary requirements to this program?  Let us start with the big picture first: theoretical predictions for \mhalo\, must be made for the types of astronomical systems most amenable to study, and the theory program must be flexible enough to respond to new (and unexpected) signatures of dark matter physics and influence observational programs.  

The former is easier to plan for than the latter, but still requires a potentially new approach. Here we give an example of the issues the community faces to make these predictions, and show what kinds of new approaches may be required.  As we described in Section~\ref{sec:baryonproblems}, most high-resolution hydrodynamic simulations (especially those focused on cusp/core problem or missing satellites problems) focus on isolated dwarfs or Milky Way analogs, out of a combination of technical feasibility issues and taste in problems.  However, as we describe below in Section~\ref{sec:obsfuture}, substructure lensing of group-scale dark matter halos is a promising observational method to measure the halo mass function down to low masses ($\sim 10^6\,M_\odot - 10^7\,M_\odot$, depending on a precise definition of halo mass).  Accurate predictions for these systems as a function of dark matter physics, after including galaxy evolution systematics, 
are essential to realizing the potential of these systems for dark matter physics.

The challenges are:
\begin{itemize}
\item The large dynamic range of the problem. The host halo masses are $\sim 10^{13}\,M_\odot - 10^{14}\,M_\odot$, or 10 to 100 times more massive than the Milky Way, yet we need to accurately resolve substructure that is $10^7-10^8$ times less massive in size. This does not even include line-of-sight effects on up to Gpc scales, which may be as or more important than structure within individual halos \cite{Xu:2011ru,Birrer:2016xku,Despali:2017ksx}. 
\item Each system has unique properties and uncertainties in those properties --- halo mass and density profile, central galaxy matter distribution, large-scale environment that may add to the lensing signal, and specific realization of substructure --- and not all lensing configurations are equally useful for dark matter constraints \cite{2006glsw.conf...91K}. 
\item We have ever-increasing ensemble sizes of these lens systems, and that the necessary physics (dark and baryonic) to understand these systems to the level required for dark matter constraints is either missing or not well quantified.  
\end{itemize}
Even in the absence of baryons, predictions for these systems are challenging in detail, in part because of line-of-sight effects and the difficulty in following small subhalos as they evolve in their hosts \cite{Xu:2011ru,xu2015,vandenBosch:2016hjf,Despali:2017ksx}.  Uncertainties in galaxy evolution physics, detailed in Section~\ref{sec:baryonproblems}, add significant computing time and uncertainties even for a fixed dark matter model \cite{Despali:2016meh,Graus:2017rrr}. 

How do we make predictions for observations of ensembles of objects as a function of dark matter physics, given the enormous computational cost to model even a single system for fixed dark matter physics?  The solution is almost certainly a hybrid program of simulations of various types, and semi-analytic and analytic modeling \cite{benson2010,bensonreview2010,galacticus2010,somerville2015}.  Simulations are powerful, but are so costly that they are best suited to case studies.  They are also incredibly important to isolate specific physical effects, which may then be modeled analytically.  Different types of simulations --- large volume but low-resolution cosmological simulations, cosmological zoomed simulations, and high-resolution idealized simulations --- are necessary for different parts of the problem.  But it will be a challenge to model substructure lensing observations with simulations alone.  Semi-analytic models (SAMs) layer parameterized models for galaxy formation physics on halo merger trees.  They are fast and flexible tools, and particularly well suited to creating realizations of large ensembles of observational systems.  It is straightforward to quantify and parametrize uncertain physics, and to connect the model to likelihood functions to constrain dark-matter physics.  However, they must be calibrated with simulations, which may be idealized in order to isolate specific physical processes (e.g., dynamical friction).  Efforts to incorporate new dark-matter physics into semi-analytic models are underway \cite{benson2013,pullen2014,schneider2015,lovell2016,menci2017}. 

In addition to tailoring an astrophysics theory program to model known classes of interesting systems, it is also important to pay attention to novel signatures of dark matter physics. Due to the small-scale structure problems (described in Section~\ref{sec:hints}), a great deal of theoretical thought has been devoted to dark matter physics scenarios that modify CDM predictions on the scale of $10^9\,M_\odot - 10^{10}\,M_\odot$. However, this is not the unique scale that can be affected by new physics in the dark sector (merely the scale at which hints currently appear), and other ranges of \mvir\, have not received as much theoretical consideration. While theory should always take cues from experiment and observation, it can also proactively suggest new targets and benchmarks against which observation can be tested. As we show below, new ideas are desperately needed for large swaths of \mvir, possibly leading to better use of astronomical objects already identified as interesting. For example: the implications for two-body gravitational relaxation in ultrafaint dwarf galaxies \cite{Brandt:2016aco,penarrubia2016,inoue2017}

We now turn to more technical issues, namely the mapping between dark-matter microphysics and the macroscopic implementation of that physics in simulations.  Importantly, the algorithms to model dark matter properties and their accuracy used to simulate dark matter and galaxies together must be well-matched to the observed system under consideration.  Leaving aside improvements in the hydrodynamic modeling of gas and stellar evolution in galaxies, for which a vigorous program is underway in the galaxy simulation community (e.g., Refs.~\cite{springel2010,read2012,scannapieco2012,Kim:2013jpa,hopkins2015,chang2017,wadsley2017}), there are other technical challenges to achieve an accurate calibration for the \mhalo\, figure of merit.  Fundamentally, the problem is again a matter of scales; in simulations, we represent enormous patches of phase space with single simulation particles.  In the case of self-interacting dark matter, significant progress has been made to map microphysical parameters onto cosmological phenomenological parameters (e.g., Refs.~\cite{Cyr-Racine:2015ihg,Vogelsberger:2015gpr}).  However, the interactions between particles that drive the exchange of energy and momentum at late times pose problems, as the dark matter cannot be modeled in either of the ``simple" limits of collisionless or maximally collisional (i.e., fluid), for which there is a vast body of literature.  To model the macroscopic effects of these moderately frequent microscopic collisions, various authors recommend using different forms of the cross section (viscosity, transfer) in different limits (e.g., a Debye-like plasma limit) to model interactions of simulation particles \cite{Feng:2009hw,Tulin:2013teo,Boddy:2016bbu,Tulin:2017ara}, but it is not clear which is on the firmest theoretical ground.  Velocity-dependent cross sections are especially problematic if the cross section is highly peaked in velocity space in a region of phase space poorly sampled by simulation particles.  Even in the case of non-interacting sterile neutrinos and light thermal WDM models, the non-cold momentum distribution is difficult to map to non-linear scales.  There have been recent technical advances in calculating transfer functions and removing numerical artifacts from simulations \cite{Angulo:2013sza,Lovell:2013ola,Venumadhav:2015pla}.  More work is needed on this subject.

In summary, the key points of the astrophysics theory program are:
\begin{enumerate}
    \item The highest priority for predictions are for the types of astronomical objects for which observations currently or will soon exist (e.g., the exponentially growing ensemble of substructure lenses), where predictions should include simulations of observations (a la Ref. \cite{2017arXiv170107835B}).  However, the theory program must be flexible enough to go on ``fishing expeditions'' for novel signatures of dark matter microphysics.
    \item Predictions for the effects of dark matter on structure formation must be made with the baryonic physics of galaxy evolution included.  The uncertainties in galaxy evolution physics should be marginalized over in predictions.
    \item It is important that the algorithms used to model non-minimal dark matter physics in simulations be on firm theoretical footing.  This is an example of an issue where close collaboration between particle physicists and simulators is \emph{essential}.
    \item Predictions for observations as a function of dark matter microphysics will require a hybrid approach of simulations and semi-analytic modeling.  Simulations have high spatial resolution but are extremely costly.  Semi-analytic models must be informed by simulations, but are fast and well-suited to statistical studies of astronomical objects and for parameter constraints from observations.
    \item As much as possible, phenomenological models of dark matter for cosmic structure evolution that can encompass many microphysical models (e.g., Refs.~\cite{Cyr-Racine:2015jwa,Vogelsberger:2015gpr,Murgia:2017lwo}) should be used in order to minimize the computational overhead for theory predictions.
\end{enumerate}

\subsubsection*{Applications to Direct and Indirect Detection}

We are proposing a program to improve our understanding of astrophysical structures in order to aid the discovery of new physics in the dark sector. However, imagine that --- ten years from now --- a new WIMP-like particle is found at the LHC, and that astronomical observations are completely consistent with a minimal inflationary $\Lambda$CDM Universe.  How will the theory program described above, meant to carefully calibrate \mhalo\, and identify specific signatures of new dark-matter physics in observations, be helpful in this scenario?  Would all the work done to search for departures from the predictions of CDM across multiple scales of \mhalo\, be in vain, if a particle candidate for cold dark matter was found? 

Importantly, the theory program outlined in this paper quantifies the baseline CDM+baryons model from which \mhalo\, (the scale of {\em deviations} from CDM) is defined. If a CDM-like particle is discovered, we can use the baseline CDM+baryon astronomical model to greatly improve the ``traditional'' astroparticle experiments searching for WIMPs: direct and indirect detection. Any particle discovered at the LHC which is ``dark matter-like'' requires consistent positive signals from one or both of these classes of experiments to unambiguously identify the new particle discovered at colliders as the same particle that makes up $\sim 25\%$ of the Universe's energy budget. 

The program we describe will lead to much better constraints on the density and velocity distribution of dark matter in halos, and thus will reduce astrophysical uncertainties and improve the experimental reach of direct and indirect detection experiments.  In other words, with better astrophysical modeling of dark matter, we sharpen measurements of $\Lambda^{-1}$ with direct and indirect detection methods.

Indirect detection searches for dark matter annihilation or decay in the Universe today can improve signal over background by targeting high-density regions --- be it the Galactic Center \cite{Goodenough:2009gk,Hooper:2010mq,Boyarsky:2010dr,Hooper:2011ti,Abazajian:2012pn,Hooper:2012sr,Hooper:2013rwa,Gordon:2013vta,Huang:2013pda,Abazajian:2014fta,Daylan:2014rsa,Calore:2014xka,Zhou:2014lva,TheFermi-LAT:2015kwa,deBoer:2016esu,TheFermi-LAT:2017vmf}, satellite galaxies \cite{GeringerSameth:2011iw,Ackermann:2011wa,Ackermann:2013yva,Geringer-Sameth:2014qqa,Buckley:2015doa,Caputo:2016ryl}, or galaxy clusters \cite{Ackermann:2010rg,Dugger:2010ys,Lisanti:2017qlb,Lisanti:2017qoz}. Improving our understanding of these systems will both provide more targets for indirect detection \cite{Geringer-Sameth:2015lua,Drlica-Wagner:2015xua}, and remove sources of astrophysical uncertainty (e.g., the density profile of halos) which limit our knowledge of the underlying particle physics parameters \cite{Boddy:2017vpe}. 

Similarly, direct detection of dark matter relies on high-velocity dark matter scattering on target nuclei (in the case of WIMP-like dark matter) \cite{Goodman:1984dc}, or on resonant conversion of dark matter into visible particles (as in the case of the ADMX search for axions \cite{Asztalos:2009yp}). Currently, constraints on the particle properties of dark matter from direct and indirect detection are completely degenerate with the poorly quantified dark matter density and velocity structure of halos.  In all cases, knowledge of the local density and velocity distribution of dark matter can improve the sensitivity of the searches, reduce the astrophysical uncertainties, and sharpen particle physics particle constraints \cite{Bozorgnia:2016ogo,Kelso:2016qqj,Sloane:2016kyi}. Therefore, even in cases where there is no physics in the dark sector beyond that consistent with CDM (i.e., $M_{\rm halo}$ immeasurable small), the results of astrophysical studies that probe this parameter will be of practical use in the search for dark matter.

\subsection{Observations}\label{sec:obsfuture}
Fundamentally, the theory and observational programs are intertwined.  Theory predictions are required to interpret observations, as well as suggest new types of observations that have the potential to be good probes of dark matter physics.  On the other hand, as illustrated in Section~\ref{sec:hints}, observations can reveal unexpected phenomenology that demands a theoretical explanation.  In this section, we will highlight near-term prospects for observations relevant to different \mhalo\, scales, and show where new ideas and new types of observation are needed. 

A major theme in this section is that current and next-generation wide-field astronomical surveys designed for dark-energy and time-domain science should yield major results for dark matter, at a small marginal cost.  There are many opportunities for motivated dark-matter enthusiasts and cosmologists interested using their dark-energy-finding skills on the other dark side.  Moreover, with the next astronomy decadal survey just around the corner, now is the time to start identifying new types of observational projects that will extend the reach of dark matter astrophysics in the future beyond the projects currently under construction. 

Throughout this section, we will refer to a number of astronomical and cosmological surveys and facilities.  We provide a summary of these surveys in Table~\ref{tab:surveys}.

\begin{table}[t]

\begin{tabular}{|c|c|}
\hline \multirow{3}{2.5cm}{Astrometry (space based)} & Gaia\footnote{\url{http://sci.esa.int/gaia/}} \cite{gaia2016} \\ 
 & JWST\footnote{\url{https://www.jwst.nasa.gov/}} \cite{jwst2009} \\ 
& WFIRST \cite{Spergel:2015sza} \\ \hline
\multirow{6}{*}{CMB} & ACT\footnote{\url{https://act.princeton.edu/}} \\ 
& CMB-S4 \cite{cmbs42016,cmbs42017} \\ 
& \textsc{COrE} \cite{Bouchet:2015arn} \\
& Keck Array\footnote{\url{http://bicepkeck.org/}}  \cite{Grayson:2016smb} \\ 
& PIXIE  \cite{Kogut:2011xw} \\
& Simons Observatory\footnote{\url{https://simonsobservatory.org/}} \cite{Stebor:2016hgt} \\ \hline
\end{tabular}
\begin{tabular}{|c|c|}
\hline
\multirow{9}{2.5cm}{Galaxy surveys (optical/IR)} & DES\footnote{\url{https://www.darkenergysurvey.org/}} (photometry)  \cite{des2016} \\ 
& DESI (spectroscopy)  \cite{desi2016} \\
& \textsc{Euclid}\footnote{\url{http://www.euclid-ec.org/}} (photometry, \\ 
&  slitless spectroscopy)  \cite{euclid2011,Tereno:2015hja} \\
& HSC\footnote{\url{https://hsc-release.mtk.nao.ac.jp/doc/index.php/survey/}} (photometry)  \cite{aihara2017} \\ 
& LSST\footnote{\url{https://www.lsst.org/}} (photometry)  \cite{Abate:2012za} \\
& PFS (spectroscopy) \cite{Tamura:2016wsg} \\ & WFIRST (photometry, \\
 & spectroscopy) \cite{Spergel:2015sza} \\ \hline
 \end{tabular}
\begin{tabular}{|c|c|}
\hline
 \multirow{6}{2.5cm}{Intensity mapping \cite{kovetz2017} \& radio galaxy surveys \cite{giovanelli2015}} & CHIME \cite{chime2014} \\
& \textsc{SPHEREx}\footnote{\url{http://spherex.caltech.edu/}}  \cite{Dore:2016tfs} \\
& SKA and its pathfinders\footnote{\url{https://skatelescope.org/} and \url{https://skatelescope.org/precursors-pathfinders-design-studies/}}  \\
& FIGGS  \cite{nath2016} \\
& GALFA-HI  \cite{peek2011} \\
& SHIELD \cite{Cannon:2011fe} \\ \hline
\end{tabular}
\hskip 0.74cm
\begin{tabular}{|c|c|}
\hline
 \multirow{3}{2.5cm}{Giant optical telescopes} & Extremely Large Telescope \\ & (ELT)\footnote{\url{https://www.eso.org/sci/facilities/eelt/}} \\
& Giant Magellan Telescope \\ & (GMT) \footnote{\url{https://www.gmto.org/}} \\
& Thirty Meter Telescope \\ &(TMT) \footnote{\url{https://www.tmt.org/}} \\

\hline
\end{tabular} 

\caption{Summary of present and near-term astronomical surveys or instruments whose capability to probe aspects of dark matter physics is discussed in the text. The surveys are all multi-purpose --- for example, although listed under astrometry, JWST will also dramatically enable substructure lensing measurements and distance measurements to Local Volume ultrafaint and ultradiffuse galaxies. \label{tab:surveys}}
\end{table}

\begin{figure*}[t]
\begin{center}
\includegraphics[width=0.95\textwidth]{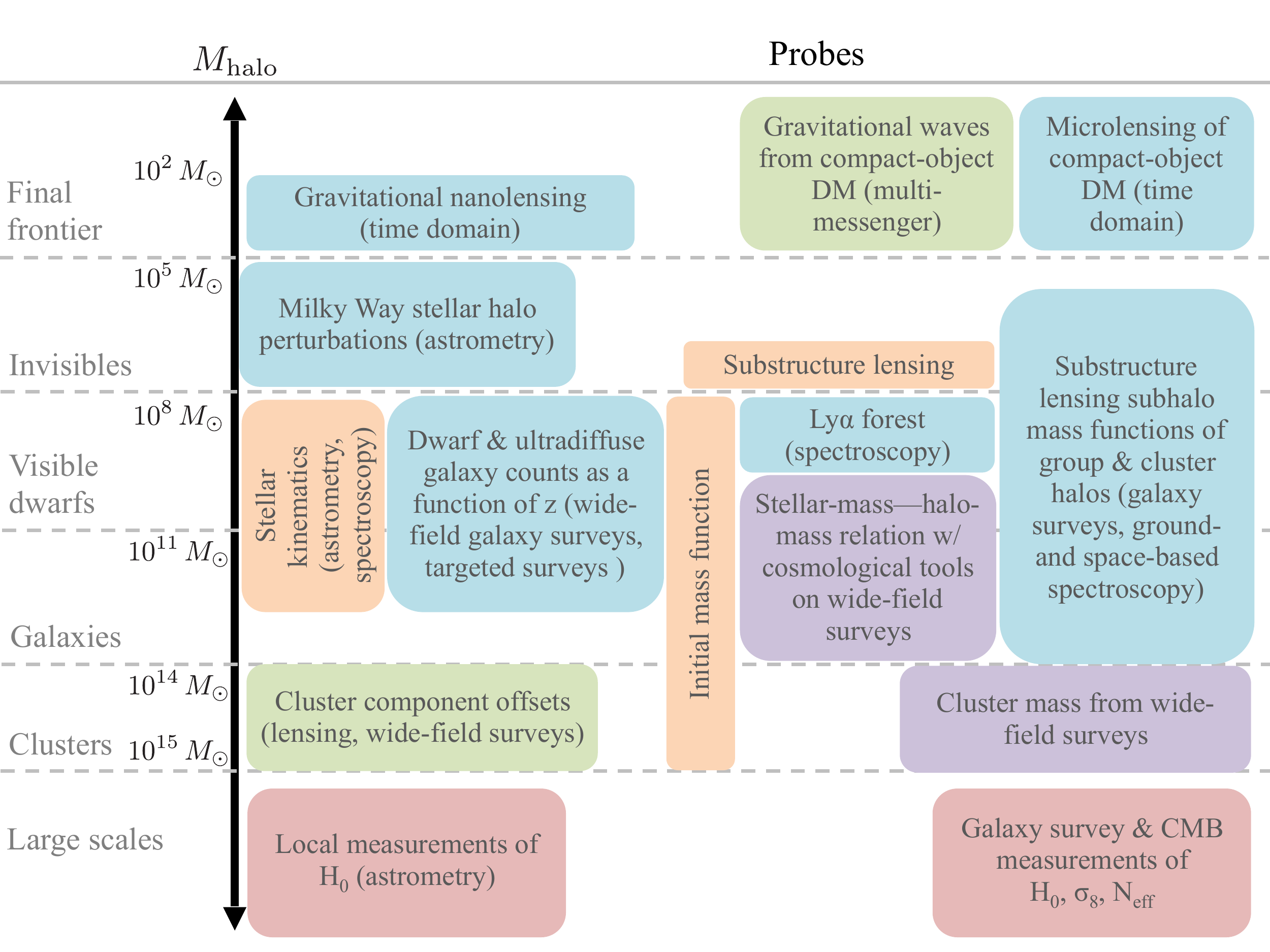}
\end{center}
\caption{\label{fig:future_probes}\mhalo\, scales accessible with future probes, color-coded by what aspect of dark matter astrophysics they will address.  Observations marked in blue will weigh in on the (sub)halo mass function, those marked in purple will enable halo mass estimates, while those in orange will facilitate measurements of the mass distribution within halos.  Red marks cosmic-scale measurements, while those probes marked in green measure aspects of dark matter astrophysics not captured by the other probes. This figure summarizes Section~\ref{sec:obsfuture}, and the halo mass ranges map onto our section titles.  Please see each relevant subsection for more details.}
\end{figure*}

We organize this discussion by \mvir\, scale, working from large to small scales.  This section is summarized in Figure~\ref{fig:future_probes}.

\subsubsection*{Cosmological scales (\mvir$\gg 10^{15}\,M_\odot$; homogeneous and linear-regime Universe)}
Measurements from the cosmic microwave background and galaxy clustering at low redshift tell a consistent story about the contents of the Universe, about the abundance of dark matter, dark energy, and baryons.  And yet in this age of precision cosmology, a few points of tension remain between the Universe at high- and low-redshift.  We briefly discuss these, their implications for dark matter, and upcoming projects that will be able to address these tensions head-on.

Measurement of the Hubble constant $H_0$ has been an active area of research for nearly a century.  At the present, measurements from the nearby cosmic distance ladder \cite{Riess:2016jrr,Beaton:2016nsw,Freedman:2017yms} and from the CMB and large-scale structure \cite{Aubourg:2014yra,Ross:2014qpa,Ade:2015xua,Abbott:2017wau,Abbott:2017smn} achieve a precision of a few percent, but the two types of measurement are in tension with each other at the $2\sigma-4\sigma$ level (central values ranging from $H_0 = 67$~km/s/Mpc to 73~km/s/Mpc).  Measurements in the local Universe consistently find larger values of $H_0$ than those from cosmological scales (but see Ref.~\cite{Abbott:2017smn}).  It is at present difficult to reconcile these measurements in the standard cosmological model \cite{Bernal:2016gxb,Wu:2017fpr,Addison:2017fdm}, although it should be noted that the level of discrepancy is insufficient to definitively point to new physics.

In addition, there exists tension between high- and low-redshift estimates of the amplitude of matter fluctuations, quantified in terms of $\sigma_8$, although the tensions are not as severe as for $H_0$ \cite{Troxel:2017xyo}.  In this case, early-Universe estimates of $\sigma_8$ are larger than those from low-redshift cosmological probes.

In both cases, non-minimal dark matter solutions have been invoked to reconcile the measurements of these fundamental cosmological parameters.  Decaying dark matter can reduce the amplitude of matter fluctuations with time, as well as change the expansion history of the Universe compared to the standard $\Lambda$CDM case \cite{Zentner:2001zr,cen2001,Peter:2010au,Lesgourgues:2015wza,Chudaykin:2016xhx}.  Hidden sector dark matter with a dark photon can also ameliorate these tensions, by introducing light degrees of freedom (either weakly or strongly coupled) before recombination.  This changes the expansion history of the Universe at early times (notably changing the redshift of matter-radiation equality), and thus affects acoustic horizon \cite{Riess:2016jrr,Chacko:2016kgg,Ko:2016uft,Agrawal:2017rvu,Baumann:2015rya,Baumann:2017lmt}.  It also affects the growth of small-scale structure, through a combination of free streaming at late times and the shift in matter-radiation equality (before which time the growth of structure is suppressed). 

The current and next generation of telescopes will be critical for determining if these anomalies are real, or statistical fluctuations.  On the most local scale, the Hubble Space Telescope (HST) and the Gaia mission will enable 1\% measurements of Cepheid stars, the bottom of the cosmic distance ladder \cite{casertano2017,Dalal:2017lpy}.  Hence, low-$z$ measurements of the Hubble constant should approach percent-level precision in just a few years.  Stage-IV CMB \cite{Kogut:2011xw,cmbs42017,Bouchet:2015arn} and galaxy \cite{Abate:2012za,Tereno:2015hja,Jain:2015cpa,Spergel:2015sza,Tamura:2016wsg,desi2016,Dore:2016tfs} surveys (see Table~\ref{tab:surveys}) should shrink error bars on the number of light degrees of freedom, N$_\text{eff}$, and $\sigma_8$, by factors of several to factors of nearly ten \cite{Baumann:2015rya,cmbs42016,Chacko:2016kgg,Brust:2017nmv,Baumann:2017lmt}. We also note that LIGO observations of ``standard sirens'' can provide an orthogonal measurement of $H_0$ with competitive error bars in the near future \cite{Schutz:1986gp,Abbott:2017xzu}. It is important to continue to consider novel effects from dark matter physics that may affect observables in these experiments, especially as it pertains to $H_0$, N$_\text{eff}$, and $\sigma_8$ \cite{Baumann:2015rya,Amendola:2016saw,Baumann:2017lmt}.

\subsubsection*{Clusters: $10^{14}\,M_\odot \lesssim$ \mvir\ $\lesssim 10^{15}\,M_\odot$}
Galaxy clusters, the largest gravitationally bound objects in the Universe, are excellent laboratories for dark matter physics on account of their deep potential wells and abundance of substructure.  In Section~\ref{sec:hints}, we already saw some evidence pointing to galaxy clusters (\mvir\ $\sim 10^{14}\,M_\odot-10^{15}\,M_\odot$) having density profiles shallower than NFW, an intriguing hint for dark matter science.  Currently, the density profiles of cluster-scale halos are well fit with self-interaction cross sections of order $\sigma/m \sim 0.01 - 0.1$~cm$^2$/g (notably lower than suggested for dwarf systems) \cite{Kaplinghat:2015aga,Robertson:2017mgj}.  Two other classes of observation shed light on other aspects of dark matter physics on this scale.

First, enabled by the Hubble Frontier Field initiative \cite{lotz2017}, strong lensing maps of a handful of clusters reveal an exquisite amount of substructure down to $M_{\rm vir}\sim 10^{9.5}\,M_\odot$ (with the caveat that this is the tidally stripped --- not infall --- subhalo mass) \cite{jauzac2016,mohammed2016,natarajan2017}.  The amount of substructure is consistent with CDM predictions, although large subhalos are perhaps \emph{too} abundant in the merging cluster Abell 2744 \cite{jauzac2016}.  

Future surveys will be hard-pressed to improve on the data from the Frontier Fields for individual systems: the excellent spatial resolution combined with the extreme depth of the field (measured in 100's of kiloseconds in exposure) which are needed to find many faint strongly lensed background galaxies will be hard to beat without a dedicated campaign.  However, constraints may be improved by stacking clusters, and with dedicated campaigns.  Expanding the number of well-studied clusters --- especially relaxed morphologies which make for more straightforward comparisons with theory --- would be possible with WFIRST.  The large footprint of the WFIRST camera makes it possible to achieve similar depth to the Frontier Fields with far fewer pointings than with HST.

Second, strong lensing maps enable a study of the distribution of dark matter, gas, and galaxies in clusters, making them sensitive to offsets between these components.  During the process of cluster growth, either by the minor merger of individual galaxies or small groups, or through major mergers (with a mass ratio between primary components of 10:1 or smaller), it is expected that galaxies trace the dark matter component.  This is because galaxies (as compact objects on the scale of clusters and comprising only $\sim 2\%$ of the cluster mass budget), effectively behave as collisionless particles during cluster mergers, just like collisionless dark matter.  Since gas strongly interacts, there may be separation between the gas and the galaxies/dark matter \cite{markevitch2004}.  Gas is easily ram-pressure stripped during the merger process.  Galaxies can be traced by their light, gas by its X-ray emission, and dark matter by gravitational lensing.

In observations of cluster-cluster mergers, or of galaxies falling onto clusters, there are frequently observations of separations between the galaxies and the dark matter, in addition to the expected separation between galaxies/dark matter and gas (e.g., \cite{williams2011,dawson2012,Massey:2015dkw}).  It has been speculated that this may be a sign of SIDM, where the self-interactions (characterized by a cross section per mass $\sigma/m$) induce a drag-like force on halos \cite{markevitch2004,Randall:2007ph,Harvey:2013tfa,Kahlhoefer:2013dca,Kim:2016ujt}.  Because the column density of dark matter in clusters is much higher than in smaller halos, the interaction probability of dark matter particles is likewise higher in clusters \cite{Kim:2016ujt}.  However, recent works shows that the separation between dark matter and galaxies is expected to be small in SIDM during the merger, in some cases much smaller than the observed separations, even for large ($\sigma/m \sim 10$~cm$^2$/g) cross sections \cite{Kahlhoefer:2013dca,Kahlhoefer:2015vua,Kim:2016ujt,Robertson:2016qef,Robertson:2016xjh}.  The smallness arises because the galaxies are gravitationally bound to the halo, and thus respond gravitationally to the halos' motion and evolution.  The recent work casts doubt on the Bullet Cluster limit of Ref.~\cite{Randall:2007ph}.  The observed offsets between galaxies and halos remains a mystery, although some of it is almost certainly a result of sparse galaxy sampling effects on centroiding the galaxy distribution \cite{ng2017}, and lensing systematics \cite{Wittman:2017gxn}. 

However, although offsets predicted by SIDM are smaller than originally anticipated, they are not negligible.  Moreover, with an ever-increasing sample of merging clusters, powered by wide-field radio surveys \cite{vanweeren2011,dawson2015,Golovich:2017tke}, continuing into the SKA era, we will be able to follow those systems that are most likely to be clean tests of SIDM (Table~\ref{tab:surveys}).  The wide-field strong lensing enabled by WFIRST will be particularly valuable here, assuming that the systematics underlying the anomalously large offsets described above can be brought under control.

Perhaps more excitingly, Ref.~\cite{Kim:2016ujt} showed that, for SIDM models, separations between galaxies and the center of the dark matter halo are expected in the relaxed merger remnant, and that these offsets may be much more significant, both in magnitude and duration, than the transient offsets found during the merger process.  This is because SIDM halos are cored, and dynamical friction is ineffective in cored (i.e., harmonic oscillator) potentials \cite{Read:2006fq,petts2016}.  As such, galaxies at the center of the halo ``slosh'' \cite{Kim:2016ujt} or ``wobble'' \cite{Harvey:2017afv} about the center of the halo.  Departures of $\sim 100$ kpc are expected for self-interaction cross sections of order $\sim 1$~cm$^2$/g,
rather than the $\sim 10$ kpc for offsets during the merger for a similar cross section.  However, new hydrodynamics simulations \cite{Robertson:2017mgj} hint that the amplitude of the wobble may be somewhat smaller than found in the dissipationless simulations of Ref.~\cite{Kim:2016ujt}. Clearly more simulation work is needed on this front. Intriguingly, there is already an abundance of relaxed cluster data that could be used for a sloshing study \cite{skibba2011,george2012,lauer2014,hikage2017}.  Sloshing is a smoking gun for a shallow density profile toward the center of halos, and investigation thereof should be a high priority for the field.  Finding the ``true" centers of halos is important for dark energy science, which has motivated nearly all of the work performed so far on galaxy cluster centers; the synergy with dark matter science makes this measurement even more compelling. 

These measurements can also be used to measure the shapes of cluster-scale halos.  Dark matter self-interactions can alter the shapes of halos, making them rounder than expected from CDM cosmologies \cite{Peter:2012jh}.  While cross section constraints are presently at the 1 cm$^2$/g level --- not competitive with these other measurements --- we expect improvements in constraints as more hydrodynamic simulations of clusters are made \cite{Robertson:2017mgj}.  The extraction of SIDM constraints from cluster scales is currently theory-limited rather than limited by observation, a problem which will be exacerbated without serious theory effort as the observational data sets are expected to mushroom in size in the LSST and WFIRST eras.  

Finally, we close this section with a discussion of halo masses.  Because of the exponential drop-off of the cluster mass function and its exquisite sensitivity to the cosmological growth function, clusters are an important dark energy probe.  Characterizing the mass function correctly depends on a faithful mapping between cluster masses measured in simulation and observation.  There is enormous effort in this direction in the dark energy community (e.g., Refs.~\cite{vonderLinden:2012kh,Old:2014jza,Mantz:2015vua}); the dark matter community will benefit from this work, as the cluster mass function is also sensitive to late-decaying dark matter models \cite{cen2001,Peter:2010au}.  Better mappings of the baryon and dark content of clusters will also be beneficial for more accurate estimates of self-interaction cross sections from observations of offsets and density profiles.

\subsubsection*{Galaxies and galaxy groups ($10^{11}\,M_\odot \lesssim$\mvir$\lesssim 10^{14}\,M_\odot$)}
On this scale, the connection between dark-matter halos and the galaxies that inhabit them is well-measured, on average, with the plethora of available tools (including lensing \cite{Mandelbaum:2005nx}, two-point statistics \cite{zehavi2011}, abundance matching \cite{Reddick:2012qy}).  The stellar mass of the central galaxy (Section~\ref{sec:primer}) is estimated to between $M_*\sim 10^9\,M_\odot$ to $\sim 10^{11}\,M_\odot$ for this range of halo masses.  The consistency with vanilla $\Lambda$CDM is generally good for objects in this mass range, but with a new class of object (ultradiffuse galaxies) beginning to perhaps muddy the waters again.  The outstanding ``hint" (Section~\ref{sec:hints}) is the cusp/core problem, although the puzzle over ultradiffuse galaxy origins is starting to hint at new additions to field galaxy (and halo) counts. In this section, we explore this new ultradiffuse direction and the big question for this halo mass range: What is the central dark matter density profile, and does that profile make sense in the context of baryonic physics for a $\Lambda$CDM cosmology?  Addressing these questions is essential to assess the (baryonic or dark) origin of the observed cusp/core problem, and to properly counting dark matter halos (Section~\ref{sec:hints}).

Answering both of these questions requires understanding two properties of galaxies at these scales: the link between galaxies and their halos (the SMHM relation) as a function of galaxy properties, and the initial mass function (IMF) of stellar populations.  The former sets the energy budget for baryons or dark matter physics to alter the dark matter profile shape from the CDM prediction of NFW, and the latter determines how well we can infer the density profile of the central regions of dark-matter halos from dynamical mass measurements.

While the link between galaxy and halo mass in the galaxy-group range is well-established for typical high surface brightness galaxies (e.g., Refs.~\cite{Mandelbaum:2005nx,vanuitert2016}), there is increasingly strong evidence that the average halo mass varies significantly as a function of galaxy properties for fixed stellar mass \cite{vandenBosch:2002zn,zu2015}.  In fact, Refs.~\cite{Zu:2015xwa,Mandelbaum:2015zwa} find that red (non-star-forming) galaxies live in halos at least twice as massive as those galaxies that are blue (and star-forming).  

In recent years, a major surprise (and new mystery) on these mass scales is the discovery of extreme outliers to the SMHM relation.  A new class of ultradiffuse (extremely low projected luminosity density) galaxies has been discovered, largely in galaxy clusters (because that is where people looked first \cite{merritt2014,mihos2015,Koda:2015gwa,munoz2015,vanDokkum:2015pla,Ferrarese:2016kcy,bennet2017}), but also in the field \cite{leisman2017,greco2017,papastergis2017}.  Many, if not most, of these galaxies are ``conventional'' dwarf galaxies  in low-mass halos (stellar masses $M_* \ll 10^9\,M_\odot$ in halos of \mvir$\lesssim 10^{11} M_\odot$), which on average have lower luminosity densities than large galaxies \cite{munoz2015}. However, others appear to be dwarf-luminosity galaxies living in LMC- to Milky-Way-mass dark-matter halos \cite{vanDokkum:2016uwg,beasley2016,amorisco2017,jones2017}.  The most famous example is Dragonfly 44 in the Coma cluster, which is a galaxy with a stellar mass only 0.5\% that of the Milky Way's, but residing in a halo approximately as massive \cite{vanDokkum:2016uwg}. Understanding these systems is important both for galaxy evolution and dark-matter studies, and suggests that the current census of even high-mass halos is incomplete.  Thus, ultradiffuse galaxies hint at a new counting problem on galaxy-halo-mass-scales, but also further complicate the cusp/core problem. 

Because hydrodynamic simulations predict that the halo profile is a strong function of star-formation history and halo mass (see the discussion in Section~\ref{sec:baryonproblems}), an accurate observational mapping between galaxies and their halos is essential to assessing the proposed baryonic solutions to small-scale problems.  We strongly endorse efforts to further tease apart the connection between galaxies and their halos as a function of galaxy properties other than mass, down to much lower stellar masses ($M_* \lesssim 10^{10}\,M_\odot$), which will be enabled by clever new methods (e.g., Refs.~\cite{zu2015,amorisco2017}) applied to the next generation of deep wide-field surveys (Table~\ref{tab:surveys} optical survey list)  \cite{Abate:2012za,Tereno:2015hja,Jain:2015cpa,Spergel:2015sza,Tamura:2016wsg,aihara2017}.  The spectroscopic capability of thirty-meter-class giant optical telescopes (Table~\ref{tab:surveys}) is essential to characterizing the halo masses of ultradiffuse galaxies---their low surface brightness currently makes follow-up spectroscopy infeasible for samples greater than $\mathcal{O}(10^2)$ objects for ten-meter-class telescopes \cite{vanDokkum:2016uwg,kadowaki2017}. 

While the total matter profile of galaxy- and group-scale systems are well-measured observationally, interpreting measurements in the context of dark-matter halo profiles is complicated by low-mass stars \cite{kroupa2001,kuzio2008,lee2009,deblok2008,dutton2011,2012MNRAS.423.1073B,2015ApJ...814...26N}.  Low-mass ($< M_\odot$) stars dominate the number count and mass of stars in galaxies, yet contribute little to the total luminosity.  Thus, determining the stellar mass function at birth (the IMF) is critical for accurately inferring the mass of stars in a galaxy from the galaxy light. Only then can one subtract the baryons from the total mass profile to determine the dark-matter distribution in the inner parts of dark-matter halos.  Recently, Refs.~\cite{conroy2012a,vandokkum2012a} used stellar absorption features of elliptical galaxies to demonstrate that elliptical galaxies tend to have relatively fewer low-mass stars than the Milky Way.  Since then, evidence has suggested that the initial mass function varies from galaxy to galaxy, and also within galaxies \cite{conroy2017,vandokkum2017,newman2017}.  This is an active area of research, and has significant implications for the dynamical modeling of galaxies.  It should be noted that the mass distribution of halos outside the luminous region can be measured with gravitational lensing and are consistent with NFW profiles, although it is not expected that baryonic physics affects halos so far from the centrally-concentrated stellar and cold gas components \cite{Mandelbaum:2006pw}.

Thus, we expect significant progress in assessing solutions to the cusp/core problem on the scale of large-ish galaxies, and to a more accurate accounting of large halos with the discovery of more ultradiffuse galaxies.  Next, we consider the visible and invisible substructure of those galaxies.

\subsubsection*{Dwarf galaxies and linear scale counterparts ($10^{8}\,M_\odot \lesssim$\mvir$\lesssim 10^{11}\,M_\odot$)}

It is in this mass range where the most significant hints of problems with CDM occur, as described in Section~\ref{sec:hints}.  It is therefore in this mass range where observational progress is especially important and (hopefully) imminent \emph{if} there is sufficient effort by dark-matter-oriented (astro)physicists.  Progress is expected on at least three fronts: 
\begin{enumerate}
\item Robustly constraining the abundance, mass profiles, and orbits of Local Group dwarf galaxies. 
\item Matching galaxies with halos. 
\item Counting small galaxies (and their halos) at a variety of epochs.
\end{enumerate}
Significant progress here will go a long way to discovering the dark or baryonic origin of the hints described in Section~\ref{sec:hints}.

We consider the Milky Way satellites first.  The Milky Way's satellite dwarf galaxies are central to hints of new dark physics, and understanding them is essential to revealing the origin of the hints.  Their importance comes from the fact that we can study intrinsically faint systems best if they are close to us.  

In Section~\ref{sec:hints}, we argued that the missing satellites problem is considered solved, but there are still important related unresolved problems that will only be answered in the LSST era with a more complete survey of dwarf galaxies.  The most pressing issue is the precise number and distribution of satellite galaxies within the Milky Way, and field dwarfs in the Local Group \cite{tollerud2017,Kim:2017iwr}. Based on completeness corrections of the known Milky Way satellites, Ref.~\cite{Kim:2017iwr} predict a relatively low number of Milky Way satellites more luminous than Segue I ($\sim 100 - 150$) if the satellites are centrally concentrated, but up to 2000 if they are distributed closer to the prediction of Refs.~\cite{Garrison-Kimmel:2017zes,errani2017}.  If the latter is found to be the case, we may have a ``too many satellites problem'' --- that there are more satellite galaxies than we can comfortably accommodate with a CDM subhalo mass function without putting galaxies in truly tiny halos.  Observations with LSST are essential to matching galaxies with subhalos in the Milky Way.    

At present, there are two major sources of uncertainty about the kinematics of the Milky Way dwarfs, which arise from the fact that we have typically only line-of-sight velocities for bright stars in these galaxies (i.e., only one of the three velocity coordinates).  First, the mass of the Milky Way halo is uncertain to a factor of two.  Statements about how unusual (or not) the Milky Way's satellite population is (e.g., the presence of the Magellanic Clouds) are highly sensitive to the mass of the Milky Way halo.  With full three-dimensional bulk velocities for dwarf galaxies and globular clusters --- enabled in the future with a combination of line-of-sight velocities from ten- and thirty-meter-class telescopes and plane-of-sky proper motions with space-based astrometry (Table~\ref{tab:surveys}) --- it is possible to put significantly better constraints on the Milky Way halo mass \cite{BoylanKolchin:2012xy,Kallivayalil:2015xva}.  Stellar streams are also an essential tool in the Milky Way mass estimation toolkit, since stars in a stream have almost identical initial conditions, and map out a large fraction of the Milky Way potential \cite{bonaca2014,price-whelan2014,johnston2016}.  The discovery of new tidal streams with deep wide-field surveys, and their characterization in phase space with spectroscopy and proper motions, are important for maximizing their use in mapping the potential well of the Milky Way. 

Second, solutions to the TBTF and dwarf-scale cusp/core problems are hampered by the lack of three-dimensional velocity data for individual stars in galaxies.  There is a major degeneracy between the mass profile of dwarfs and the orbital structure of stars in those galaxies.  This degeneracy can be broken if the full three-dimensional stellar velocity vectors are measured \cite{strigari2007,Kallivayalil:2015xva}.  The uncertainties related to the internal velocities of satellites will be significantly ameliorated when it is possible to measure stellar motion in the plane of the sky, in addition to line-of-sight velocities.    Although much good work has been done with HST and will be done with JWST, the field will truly open up with the astrometry performed by the Gaia and WFIRST satellites \cite{Spergel:2015sza,sanderson2017a}.\footnote{R.~Sanderson, {\em WFIRST in the 2020's meeting}, \url{http://www.stsci.edu/~dlaw/WFIRST2020s/slides/sanderson.pdf}}  Interestingly, the Gaia satellite does not go deep enough to probe the internal velocities for most Milky Way satellites (although it does for bulk velocities \cite{2018arXiv180409381G,2018arXiv180501839M,2018arXiv180500908F,2018arXiv180410230S}), but it will play a critical role in anchoring the astrometric reference frame for WFIRST studies.  We strongly endorse the work of the WINGS science investigation team and the WFIRST astrometric working group's effort to explore the implications of WFIRST astrometry \cite{sanderson2017a} for Local Group dark matter science.  

While proper motions will enable incredible studies of the internal kinematics and structure of classical dwarfs ($M_* \gtrsim 10^5\,M_\odot$; see Section~\ref{sec:hints}) in the Local Group, we need new ideas to measure densities and density profiles of galaxies that are either small and/or far away.  The smallest galaxies, the ultrafaints, contain insufficient stars for the shape of their dark matter density profile to be measured, even with full three-dimensional stellar kinematics, although the observational depth of WFIRST will enable the use of some of the more abundant faint stars for proper motions. However these are the galaxies which are expected to have cusps in CDM regardless of baryonic physics, which increases the importance of observational tests of their central density profiles.  For small galaxies beyond the Milky Way, follow-up considerations are different depending on whether galaxies can be resolved into individual stars or not.  For both, even line-of-sight velocities will be a challenge---thirty-meter-class telescopes will be able to follow up only a small fraction of new discoveries.   With ten-meter-class telescopes, line-of-sight velocities for individual stars are obtained only for distances less than 1 Mpc.  Beyond that, we need light integrated over many stars and (especially) gas to get a signal in a decent amount of time.  The presence of nebular emission lines is essential to obtaining good velocity information in a non-prohibitive amount of time.  However, we expect the faintest satellites, the ultrafaints, to be devoid of ionized gas (although they might possibly have small reservoirs of neutral gas in the field \cite{jeon2017}).  Thus, the the only tool available to us is photometry.  We strongly endorse the development of novel ways to estimate the density profile of small galaxies with photometry alone, as proposed by Ref.~\cite{inoue2017}.  

The second important pathway to progress is determining how to match galaxies with halos, and to find ways to constrain the matter power spectrum on dwarf scales that do not necessarily depend on the exact mapping of galaxies to halos.  The former is essential for determining if baryons can solve the cusp/core problem in a CDM framework, and the latter is important to look for ``primordial" effects of dark matter on structure formation.  In both cases, we strongly encourage the cosmology community to extend their tools to low(er)-mass halo regimes if at all possible.  As discussed above, galaxy-galaxy weak lensing and two-point correlation functions can be used to match galaxies to halos statistically \cite{Mandelbaum:2005nx,Mandelbaum:2006pw,zehavi2011,Zu:2015xwa,Lan:2016cpm}.  Ref.~\cite{Guo:2017zyt} presented first results from a study of atomic hydrogen-selected galaxies at mass scales similar to the Magellanic Clouds.  Intriguingly, a recent study of cosmic shear with \textsc{CFHTLenS} suggests that the matter power spectrum is consistent with CDM down to scales of $M_{\rm halo} \approx 5\times 10^9\,M_\odot$ \cite{Jimenez:2017anl}.  With the flowering of wide-field surveys (Table~\ref{tab:surveys}), we expect constraints on the matter power spectrum and on the statistical matching of galaxies with halos, to become robust on dwarf scales, although there are important caveats \cite{leauthaud2017}.

Finally, we expect significant advances in the counting of individual halos by the galaxies (or gas clouds) they host in the near term, and at a wide range of redshifts.

At high ($z\sim 10$) redshifts, dwarf galaxies are expected to reionize the Universe; there are simply not enough big galaxies or quasars at early times to yield sufficient ionizing photons \cite{alvarez2012,robertson2013,boylan-kolchin2014,robertson2015,bouwens2015,song2016,livermore2017}.  Thus, dwarf searches are of high interest at high redshift.  The effect of the matter power spectrum on $k\sim 0.1$ Mpc$^{-1}$ scales (Figure~\ref{fig:halogalaxy}) may be probed at $z\sim 10-20$ 21-cm radiation \cite{mesinger2014} (Table 1), and are sensitive to the first stars rather than to the first galaxies \cite{iliev2003}.  At $z\sim 10$, dwarf searches are one of the primary drivers for the upcoming James Webb Space Telescope (JWST) \cite{stiavelli2009,windhorst2009}, and of major interest for WFIRST.\footnote{S.~Finkelstein, \emph{WFIRST in the 2020's meeting}, \url{http://www.stsci.edu/~dlaw/WFIRST2020s/slides/finkelstein.pdf}} The UV luminosity function of galaxies near the epoch of reionization can be an effective probe of power spectrum cutoffs (already excluding thermal relic WDM masses below 2.4 keV), with significant improvement expected in the JWST and WFIRST eras \cite{pacucci2013,menci2017}.  The star-formation histories of nearby galaxies, determined today with HST but in the future with JWST, can also be used as a ``time machine'' to constrain the reionization-era galaxy stellar mass function, and hence, the matter power spectrum \cite{Governato:2014gja,Chau:2016jzi,Boylan-Kolchin:2015kza,Boylan-Kolchin:2016lwt,Weisz:2017ema}.  Again, we encourage efforts to constrain using primordial dark matter physics using the galaxy luminosity or stellar mass function at early times.

Moving down in redshift, from measurements of the Lyman-$\alpha$ forest at $z\sim 3-6$ \cite{Croft:2000hs,Tegmark:2002cy,McDonald:2004eu,Seljak:2004xh,irsic2017}, the CDM paradigm is known to be consistent with observations down to $M_{\rm halo} \sim 10^{9}\,M_\odot$ \cite{Seljak:2006qw,irsic2017}.  The Lyman-$\alpha$ forest is a series of Lyman-$\alpha$ absorption features from gas clouds backlit by quasars.  Because gas density roughly traces the matter field, it can be used to constrain the matter power spectrum at moderate redshift, albeit with uncertainties related to the thermal history of the intergalactic medium (IGM) \cite{irsic2017}.  

The resulting information about the number of gas clouds at varying redshift can be used to constrain deviations from CDM. For example, free-streaming of ``warm'' (semi-relativistic) dark matter would suppress the power spectrum on scales probed by the Lyman-$\alpha$ forest. This constrains thermal relic dark matter to be heavier than $3.5-5.3$~keV, depending on assumptions about how smoothly the temperature of the IGM changes with time \cite{irsic2017}.  In the future, measurements from \textsc{X-Shooter} on the VLT, WEAVE-QSO on the William Herschel Telescope, and the DESI survey will expand the resolution, sightlights, and redshift of the Lyman-$\alpha$ forest dataset \cite{desi2016,pieri2016,rorai2017}, improving these limits.  We strongly support continued efforts to measure the Lyman-$\alpha$ forest and isolate systematic uncertainties in their interpretation.

Statistical samples of dwarf galaxies are being obtained in the local Universe using a variety of methods, the sample size of which will increase vastly with LSST in particular.  A galaxy luminosity function (or stellar mass function), robust to surface brightness detection limits, is a critically important input for abundance matching and two-point clustering statistics to associate galaxies with halos, as well as to disentangle baryonic from dark matter effects on the core/cusp problem, and to solve the ``TBTF in the field" problem (Section~\ref{sec:hints}).  The SHIELD survey focuses on finding field dwarfs in atomic hydrogen \cite{Cannon:2011fe}.  Other radio and intensity mapping surveys (Table~\ref{tab:surveys}) will or already have uncovered more.  Optical surveys --- either specialized like MADCASH \cite{carlin2016}, or general-purpose like the Dark Energy Survey \cite{Drlica-Wagner:2015ufc} (Table~\ref{tab:surveys}) --- will be able to reach ultrafaint, or nearly ultrafaint \cite{geha2017} scales in the Milky Way and beyond.  

As mentioned earlier in this section, the biggest challenge after finding these systems is to characterize their kinematics, internal and bulk, and measure their distances from us.  The low surface brightness of many of these objects and the paucity of ionized and neutral gas in the lowest-stellar-mass systems makes both extremely challenging.  Distances for nearby dwarfs are currently obtained either with variable stars or using the ``tip of the red giant branch" (TRGB) method \cite{lee1993,hatt2017}, both of which require resolved stellar populations.  Beyond a few megaparsecs, this requires space-based angular resolution, notably HST, JWST, and WFIRST.  We strongly endorse efforts to find and characterize dwarf galaxies using a variety of methods, in a variety of environments, and at extremely low surface brightnesses; and to develop new methods to characterize these galaxies.  We also endorse the application of statistical techniques developed for higher-redshift data \cite{Sales:2012ts,Nierenberg:2016nri} to the near field.

The broad message of this section is that enormous new and powerful data sets are imminent or already extant, and we must learn how to mine them for dark matter science.  We strongly encourage the use of ``standard'' tools in statistical cosmology to approach dwarf-scale problems, to complement the current approaches largely driven by the galaxy evolution community.  We anticipate many opportunities for important progress in resolving both the core/cusp and counting issues on dwarf scales in the next decade. However, work will be needed needed to propagate the astrophysical results out to theorists, as well as bringing new theoretical ideas to the simulators and observers to test.

\subsubsection*{Invisibles ($10^5\,M_\odot \lesssim$ \mvir$\lesssim 10^8\,M_\odot$)}

Below halo masses of $\sim 10^{8}\,M_\odot$, we expect dark matter halos to be largely devoid of baryons (\cite{Jethwa:2016gra,dooley2017,Munshi:2017xhq,Kim:2017iwr}).  Thus, measurements of dark matter halos on these scales must rely on gravity alone.  The two major methods under consideration on these scales are gravitational lensing and Milky Way stellar halo and stream perturbations (Section~\ref{sec:primer}).  We note that although on these scales dark matter searches are typically discussed in terms of halo counts or halo mass functions, the distribution of matter within halos can affect their observability.  Hence, it may be possible to unravel both primordial and evolutionary deviations from CDM, even without the presence of baryons. The most exciting aspect of tests in this mass scale is that the promise of lensing and stellar perturbations are about to be realized in new data sets.  Here, we describe why the data sets are about to grow exponentially in size, and what the remaining roadblocks are in mapping observations to \mhalo.

As discussed in the previous section on dwarfs, \emph{weak} lensing may possibly used to probe this regime \cite{Jimenez:2017anl}.  However, in this regime of halo mass, we usually discuss \emph{strong} gravitational lensing as a probe of the abundance of small halos \cite{2006glsw.conf...91K,2009arXiv0908.3001K,2010ARA&A..48...87T}.  As we showed in Section~\ref{sec:observations}, substructure along the line of sight connecting us to a galaxy- or galaxy-group-scale lens (including substructure in the lens and structure outside of it but along the line of sight) can perturb the apparent flux, position, and arrival time of light from strongly lensed background galaxy or active galactic nucleus (AGN).  Up until $\sim 2010$, most work focused on flux-ratio anomalies from seven radio-loud, quadruple-image AGN.  Excitingly, the flux-ratio anomalies were consistent with CDM, albeit with large uncertainties \cite{dalal2002}.  Since 2010, the method of ``gravitational imaging'' \cite{Koopmans:2005nr,vegetti2010} has gained traction, mapping out perturbations in a few dozen Einstein ring galaxies, which are also consistent with CDM but again with large uncertainties \cite{Lagattuta:2012gq,Vegetti:2012mc,Vegetti:2014lqa}.

Why so few targets, and what is the path to stronger tests of dark matter microphysics?  Radio-loud AGN were desired for flux-ratio anomalies because they are less prone to contamination by the microlensing induced by stars in the lens galaxy in optical bands, as the image size in the lens plane is comparable to the Einstein radius of stars as well as small halos.  The mid-IR and radio emission from AGN comes from a much larger region, and it much less susceptible to microlensing (see Ref.~\cite{nierenberg2017} for an illustration).  But such AGN are rare.  Up until recently, there were only seven known suitable systems.  In addition, a further complication is that for both flux-ratio anomalies and gravitational imaging, the angular size of systems is small (on the order of arcsecond(s)), meaning high-resolution imaging is required to study these systems.  

What is changing on the data side?  First, new wide-field surveys are finding more suitable candidates \cite{des2016,agnello2017,berghea2017} --- even SDSS is still being mined for lenses \cite{williams2017}.  Hundreds to thousands of quadruply lensed AGN are expected to be discovered, with commensurate numbers of Einstein ring systems \cite{oguri2010}.  Second, new facilities are enabling follow-up of these candidates.  New adaptive optics facilities provide the high-resolution imaging and spectroscopy required for these studies \cite{Lagattuta:2012gq,nierenberg2014}, the mid-infrared capabilities and grisms of JWST will enable many more quadruply-imaged AGN to be used for substructure studies, and the high-quality astrometry available in the radio will enable low-mass subhalo detections via gravitational imaging.  Third, new methods are being pioneered that allow most quadruply-lensed AGN to be used to measure flux-ratio anomalies without microlensing contamination with space telescopes (HST, JWST, and WFIRST \cite{nierenberg2014,nierenberg2017}), or enhance the science return on gravitational imaging surveys \cite{Rivero:2017mao,Daylan:2017kfh}.  Finally, Ref.~\cite{hezaveh2013} opened a window to study new sources with new facilities: clumpy, high-redshift galaxies with the ALMA interferometer \cite{Hezaveh:2014aoa}.  There is much to be done even with HST \cite{Dalal:2017lpy}.  The landscape of what is possible is changing rapidly.

Likewise, the prospects for finding small subhalos in the Milky Way halo are also changing with the advent of new data and new analysis methods.  As satellites of the Milky Way, especially the mysterious ultra-dense-in-stars globular clusters, are disrupted by tidal forces from the Galaxy, stars disperse along tidal stream.  Dark matter subhalos can ``punch'' through the streams, perturbing stellar orbits and leaving persistent gaps \cite{Carlberg:2009ae,yoon2011,Carlberg:2011xj,Carlberg:2013eya,Ngan:2013oga,erkal2014,erkal2015,erkal2016,Bovy:2015mda,Bovy:2016irg,johnston2016,stanford2017,erkal2017}. To probe the Galactic subhalo population and not the Galactic bar or molecular clouds, we require streams with large enough orbital pericenters so that they are insensitive to baryonic substructure, but small enough that the progenitor may be stripped \cite{Amorisco:2016evb,pearson2017}.  The discovery of suitable streams in new wide-field surveys, including Gaia, is already happening, and will continue into the LSST era \cite{Grillmair:2014bua,Grillmair:2015zra,grillmair2017,pearson2017,des2016,Abate:2012za,Myeong:2017skt,Shipp:2018yce,2018ApJ...860L..11K}.  A combination of deep imaging and proper motions will lead to a large number of exquisitely mapped stellar streams.
 It is expected that constraints on dark-matter models with primordial deviations from CDM can be significantly tighter than achieved with the Lyman-alpha forest or dwarf galaxy counts, with limits comparable to or better than from the Lyman-alpha forest anticipated \cite{banik2018}.  A new study suggests that halos as small as $10^6\,M_\odot-10^7\,M_\odot$ may be discovered by their effects of the smooth stellar halo as well \cite{Buschmann:2017ams,vantilburg2018}, or even at lower masses \cite{penarrubia2017}.  Both the theoretical modeling of the stellar halo and its streams, and the discovery of new streams, have advanced dramatically in just the past few years --- a trajectory whose derivative will only increase in the near future.  The upside of the Milky Way is that its halo can be mapped to high precision with current and future surveys; the downside is that it is a sample of one.  It will be extraordinarily powerful if a consistent signature of small halos is discovered in the Milky Way using stellar halo methods, and in other galaxies using other methods.  

Because of this new wealth of data, for both lensing and stellar halo methods, what is desperately needed is a clean mapping between dark matter theory and observables.  There is progress on both fronts, especially in the case of CDM \cite{Chen:2011wc,Despali:2016meh,Bovy:2016irg,Graus:2017rrr}.  But this is a specific instance where progress toward dark matter constraints will soon be theory-limited.  

\subsubsection*{Tests of CDM: the final frontier ($10^{-6}\,M_\odot\lesssim$\mvir$\lesssim 10^5\,M_\odot$)}
One of the strongest astrophysical predictions of the WIMP-$\Lambda$CDM paradigm is the hierarchy of halos down to Earth-mass scales.  A measurement of halos on these scales would be an astounding confirmation of this theory.  Although the predictions for the existence of substructure are strong for CDM, fluffy, extended, and small halos are difficult to detect.

A number of ideas have bubbled up in recent years to detect such small halos, but require additional validation.  A number of authors have suggested different ways to use \emph{time-domain} observations, in all of the strong, weak, and microlensing regimes, to detect halos in this mass regime \cite{Chen:2011wc,erickcek2011,garsden2012,Rahvar:2013xya,venumadhav2017,Boehm:2017wie}.  These ideas need to be validated in an observational context, but they point to a very important fact:  the current and next generation of optical wide-field surveys (Table~\ref{tab:surveys}) are time-domain surveys by design.  We encourage, in the strongest terms, scientists studying dark matter to think about the new axes of possibility, like the time axis, opened up by next-generation facilities.

Other attempts to discover dark matter on these small scales are model-dependent, but high-impact if those models for dark matter are correct. If dark matter consists of $\sim 100\,M_\odot$ primordial black holes, they can be detected with gravitational waves at LIGO, stellar microlensing, or stellar dynamics in ultrafaint dwarf galaxies \cite{Bird:2016dcv,Kovetz:2017rvv,Battaglieri:2017aum,Koushiappas:2017chw}.\footnote{W. Dawson, {\em WFIRST in the 2020's} Meeting, \url{http://www.stsci.edu/~dlaw/WFIRST2020s/slides/dawson.pdf}}  If dark matter consists of WIMPs, the substructure abundance can be backed out of indirect detection observations if a signal is seen in different types of cosmic environments \cite{Ng:2013xha}.

This is arguably the most challenging regime for astronomical dark matter searches, but is the one with potentially the largest payoff.  We strongly encourage new ideas in this regime, not just ones that exploit new observational capabilities, but ones that can be brought to fruition with dedicated new facilities.  

\subsubsection*{Dark matter vs. inflation}
Identifying a theory of dark matter from astronomical observations and experiments means seeing a consistent signature across \mvir\, scales and in the laboratory.  Figure~\ref{fig:fom} is our recommendation for how to begin characterizing consistency.  However, so far we discussed predictions in the paradigm of a primordial scale-free power spectrum from inflation.  Inflation can lead to a boost in small-scale power relative to the standard slow-roll case either ``primordially" or through altering the thermal history of the Universe \cite{Erickcek:2011us,Aslanyan:2015hmi}, or a suppression relative to the standard case \cite{Kamionkowski:1999vp}. There are a number of innovative methods to suss out the former, many which tie into either indirect dark matter searches (tiny compact structures can boost the signal) \cite{Scott:2009tu,Erickcek:2015jza} or gravitational lensing perturbations \cite{Bringmann:2011ut,Li:2012qha,Clark:2015sha}. The latter effect (modification of the thermal history) can easily be confused with a dark-matter-physics-induced truncation of the matter power spectrum.  A novel idea is to use spectral distortions of the CMB to distinguish between the two \cite{Chluba:2015bqa,Chluba:2016bvg,Nakama:2017ohe}.  This may be possible with the PIXIE satellite \cite{Kogut:2011xw,Cabass:2016giw,Chluba:2016bvg}. 

We endorse further attempts to distinguish between inflationary and dark-matter effects on the power spectrum, and in exploring the connection between inflation and the indirect detection of dark matter.

\subsubsection*{New ideas}
 In this section, we have laid out the prospects for both theoretical and observational progress to making an accurate map between dark-matter theory space and astronomical observables (i.e., a more accurate and nuanced version of Figure~\ref{fig:fom}), and to constraining the nature of dark matter more generally.  Our main message is that we expect potentially transformational progress in the next decade as theorists and observers come together to maximize the return on new observational and computational facilities and techniques.  However, there are some problems that require new ideas; and other problems where the current path to progress is straightforward but slow, and where new insights would be extraordinarily valuable.
 
 There are two major points we would like to make before we conclude.  
\begin{itemize}
\item First, arguably the strongest and cleanest predictions for CDM are on small scales.  In terms of the cusp/core problem, the cleanest systems to study are the ones where the baryons are the fewest and have the most boring behavior: in ultrafaint dwarf galaxies and small, dark subhalos.  But the smoking gun of CDM theory is the abundance of tiny, solar-system-sized dark matter halos.  A discovery of such small halos would be definitive proof of the CDM paradigm.  
 
However, it is extremely challenging to measure the abundance of and matter distribution within such small objects.  There simply are not enough stars in ultrafaint dwarfs to distinguish between core and cusp scenarios kinematically.  Finding dark subhalos relies on gravity, but halos are extended objects, which complicates searches (compare to primordial black holes, which would be relatively easier to discover on account of their compactness).  New ideas that will enable the discovery of the hierarchy of CDM-predicted subhalos with forthcoming or existing facilities, or which could be enabled by bespoke new facilities, are highly desirable.
 
\item Second, there are many new astronomical facilities coming on-line in the next decade, which we listed at the beginning of the section.  While dark matter science is not a primary driver for these facilities, we showed here that they have the potential to significantly advance this field.  Now is the time to think about and advocate for minor changes in survey design or guest-observer programs that have small marginal costs yet yield enormous progress in dark matter astrophysics. 
\end{itemize}

\subsubsection*{Astrophysics in the Nightmare Scenario }

We end this section with an investigation of the role astrophysics might play in the ``worst-case'' particle physics scenario: no dark matter signal is seen in any of the particle physics searches in the next decade.  In this scenario, there is no evidence found for physics beyond the Standard Model at the Large Hadron Collider; axion experiments (such as ADMX \cite{Asztalos:2009yp}) continue to report null results; no definitive signal of WIMP dark matter in direct-detection experiments; and no sign of dark-matter annihilation from gamma-ray and neutrino telescopes.  The lack of an axion signal would imply that QCD axions are excluded as dark matter unless the vacuum misalignment angle $\theta \ll 1$ or most axions cluster in Bose stars \cite{Guth:2014hsa}.  Would such null results imply that dark matter cannot largely consist of WIMPs or axions, or are we just sitting in an unfortunate (for detection) point in the parameter space?  Gravitational probes could help resolve the issue in the following ways:
\begin{itemize}
  \item If probes of the small-scale power spectrum at early times (via the  Lyman-$\alpha$ forest probing $z \sim 2$ or 21-cm fluctuations during reionization at $z \sim 10-20$) show indications of a small-scale cut-off in the power spectrum, and this is confirmed at low redshift by measurements of the subhalo mass function (e.g., with strong lensing or observations of Local Group satellite galaxies), then this would point towards dark matter paradigms with primordial deviations in $M_{\rm halo}$.  This could, for example, help motivate the development of X-ray satellites to look for line emission from sterile neutrinos in nearby galaxies, and the guide further theoretical work on non-trivial dark sectors. 
  
  \item If the matter power spectrum is consistent with CDM at early times (e.g., the CMB) but there are indications of late-time structure evolution (e.g., a  truncation in the halo or subhalo mass function where \mhalo~is larger today than indicated by the Lyman $\alpha$ forest), this would indicate that the dark matter decays or interacts with other dark-sector particles on cosmological time scales.  This would focus theoretical work on models that have evolutionary effects on $M_{\rm halo}$.  If there were no Standard Model particles emerging from dark matter particle interactions or decays, it would imply that the dark sector is complex but almost entirely decoupled from the Standard Model.
  
  \item Finally, if all gravitational probes of dark matter indicated that both large-scale and small-scale structure evolve as expected in the dark energy plus CDM cosmological framework, this would indicate that dark matter should be cold.  In this case, it is possible that dark matter could consist, say, of heavy (multi-TeV) WIMPs that are simply hard to create at colliders, due to the high energy scale; or detect via direct or indirect detection due to the fact that the dark-matter-particle number density scales inversely with mass.
\end{itemize}

In any of these three scenarios, astronomical observations play a key role in sharpening the allowed dark matter parameter space, and guiding particle physicists and experimentalists to new directions.

\section{Conclusions \label{sec:conclusion}}

As direct proof of physics beyond the Standard Model, dark matter is one of the most significant open questions in particle physics today. We know remarkably little about dark matter, and what little we do know has been discovered by astrophysics via its gravitational imprint on baryonic matter. 

These astrophysical measurements are indirectly sensitive to particle physics properties of dark matter, in particular those that modify the initial distribution of dark matter away from the predictions of cold dark matter, or provide mechanisms for energy exchange between dark matter particles at late times. These attributes are largely orthogonal to those accessible to particle physics experiments (e.g., direct detection, indirect detection, and collider production). Therefore, any comprehensive approach to the study of dark matter physics must incorporate results from both fields. 

Constructing a joint effort requires a common language and a clear sense of the open questions in both fields. To facilitate this, in this paper we have developed a simplified parameter space which allows theoretical models of dark matter to be characterized by the strength of their particle physics interaction $\Lambda^{-1}$ and the astrophysical scale at which non-trivial deviations from CDM would appear $M_{\rm halo}$. Our goal with this parameter space is that it will highlight and enable the advantages of interdisciplinary collaboration in dark matter physics. 

Having constructed this common dictionary between particle and astroparticle properties, we consider a number of hints in the existing data that are suggestive of new physics beyond pure CDM. These revolve around discrepancies between observation and simulations at mass scales of $10^9\,M_\odot \lesssim M_{\rm vir} \lesssim 10^{15}\,M_\odot$. Resolving the origin of these hints requires a more accurate consideration of baryonic effects, and is a necessary step in any effort to use astronomical observations to probe the particle physics of dark matter.

Regardless of the ultimate origin of these particular anomalies, the example of the ``Crisis in Small Scale Structure'' demonstrates the potential for collaboration between particle physicists and astrophysicists in unraveling the mysteries of dark matter. In Section~\ref{sec:future}, we outline some opportunities to use existing and near-future astronomical surveys to cover more of the \mhalo\, parameter space. We note that dark matter structures below $M_{\rm vir}\lesssim 10^7\,M_\odot$ are both largely unconstrained and of particular interest to a large class of theoretical models. New ideas are needed here.

Particle physics searches for dark matter are at a mature stage, and their sensitivity will only increase in the coming years. However, they cannot test all aspects of dark matter phenomenology, and astronomical observations have enormous discovery potential. The development of a comprehensive framework to use astronomical measurements to constrain dark matter theory, which can in turn motivate and guide future studies, will be {\em the} challenge to the field of dark matter science.

\section*{Acknowledgements}

We would like to thank Alyson Brooks, Keith Bechtol, Francis-Yan Cyr-Racine, Alex Drlica-Wagner, Manoj Kaplinghat, Stacy Kim, Leonidas Moustakas, Chanda Prescod-Weinstein, Kris Sigurdson, and David Weinberg for enlightening discussions.  This research has made use of the \texttt{matplotlib} python package \cite{hunter2007}, NASA's Astrophysics Data System, and INSPIRE.
M.R.B.~is supported by DOE grant DE-SC0017811.  A.H.G.P. is supported by the NSF Grant No.~AST-1615838.

It was brought to our attention that our choice of terminology used in Section II in an earlier version of this paper could be marginalizing to people of color. We sincerely apologize for this, and have changed the language in question. We thank those members of our community who raised this issue, and challenged us to do better.

\bibliographystyle{elsarticle-num}
\bibliography{DMastroparticle}

\end{document}